%% file: TESS_LCs.tex
\documentclass[fleqn,usenatbib]{mnras}

\usepackage{newtxtext,newtxmath}

\usepackage[T1]{fontenc}
\usepackage{ae,aecompl}

\usepackage{graphicx}	
\usepackage{amsmath}	
\usepackage{mathtools}
\usepackage{longtable}
\usepackage{subfigure}
\usepackage{natbib}
\usepackage{nccmath}
\usepackage{float}
\restylefloat{table}
\usepackage{multicol, blindtext}
\usepackage{caption}
\usepackage{booktabs}
\DeclareGraphicsExtensions{.png,.pdf}

\title[An Unsupervised Exploration of \textit{TESS}]{Linking Anomalous Behaviour with Stellar Properties: An Unsupervised Exploration of \textit{TESS} Light Curves}

\author[Crake \& Mart\'{i}nez-Galarza]{
Dennis A Crake,$^{1}$
Juan Rafael Mart\'{i}nez-Galarza,$^{2}$
\\
$^{1}$Institute for Astronomy, University of Edinburgh, Royal Observatory, Blackford Hill, Edinburgh, EH9 3HJ, UK\\
$^{2}$Center for Astrophysics | Harvard \& Smithsonian, 60 Garden Street, Cambridge, Massachusetts, 02144, US\\}

\date{Accepted XXX. Received YYY; in original form ZZZ}

\pubyear{2023}

\begin{document}
\label{firstpage}
\pagerange{\pageref{firstpage}--\pageref{lastpage}}
\maketitle

\begin{abstract}
With the upcoming plethora of astronomical time-domain datasets and surveys, anomaly detection as a way to discover new types of variable stars and transients has inspired a new wave of research. Yet, the fundamental definition of what constitutes an anomaly and how this depends on the overall properties of the population of light curves studied remains a discussed issue. Building on a previous study focused on \emph{Kepler} light curves, we present an analysis that uses the Unsupervised Random Forest to search for anomalies in \emph{TESS} light curves. We provide a catalogue of anomalous light curves, classify them according to their variability characteristics and associate their anomalous nature to any particular evolutionary stage or astrophysical configuration. For anomalies belonging to known classes (e.g. eclipsing binaries), we have investigated which physical parameters drive the anomaly score. We find a combination of unclassified anomalies and objects of a known class with outlying physical configurations, such as rapid pulsators, deep eclipsing binaries of long periods, and irregular light curves due to obscuration in YSOs. Remarkably, we find that the set of anomalous types differ between the \emph{Kepler} and \emph{TESS} datasets, indicating that the overall properties of the parent population are an important driver of anomalous behaviour.

\end{abstract}



\section{Introduction}

Recent, current, and upcoming sky surveys are transforming the field of time-domain astronomy by producing light curves for millions of astronomical objects, providing constraints to cosmological models, enabling the discovery of unexpected phenomena, and challenging our data storage and processing capabilities. The \textit{Sloan Digital Sky Survey} (SDSS) has provided insight into the large scale of the universe by providing millions of images and spectra of approximately one-third of the entire sky \citep{York00} and is about to become the most extensive time-domain spectral variability survey with the advent of SDSS-V \citep{Kollmeier2019}. As a prelude to the eagerly anticipated Vera C. Rubin Telescope's \textit{Legacy Survey of Space and Time} (LSST), which promises to be a generational step in the observations of the transient universe, the Zwicky Transient Facility (ZTF) is currently surveying the northern sky at a pace of more than 3750 square degrees per hour, discovering energetic transients every single night \citep{Bellm2014}. 

Space-based surveys have focused on stellar targets. The \emph{KEPLER/K2} mission observed more than 500,000 stellar objects across nine years of operation, increasing the number of confirmed transiting exoplanets from hundreds to just under three thousand. Additionally,  providing valuable information about intrinsic stellar variability \citep{Overbye2}. The next-generation exoplanet finding sky survey came online in April 2018, following the launch of the \textit{Transiting Exoplanet Sky Survey} (\emph{TESS}), with the primary goal of continuing the search for exoplanets, now focusing on smaller, rockier worlds in a much larger area of the sky with respect to \emph{Kepler} and looking at comparatively brighter stars. 

These and other similar surveys are producing an unprecedented onslaught of time-domain data. Yet, the challenge remains to effectively analyse the resulting plethora of data to accurately classify sources, select the most promising objects for spectroscopic follow-ups, and identify anomalies that require new theoretical frameworks, expanding our knowledge horizons. The latter is of particular interest within the exploration approach to astronomy research, in which we look for the \emph{unknown unknowns} by expanding the parameter space of observables, or by dissecting existing datasets in novel ways to find true astrophysical anomalies.

A rapidly increasing number of publications deals with the problem of anomaly detection in astronomical datasets, and in time-domain datasets in particular \textit{astronomaly} \citep{Lochner2021}, while others include; \cite{Giles19, Magalef2020, doorenbos2021, skoda2020, Baron17}. Anomaly detection is relevant because finding anomalous objects is a direct avenue to discovery, particularly when those discoveries challenge existing models, motivating the formulation of new paradigms. A recurring theme of this research relates to the very definition of what constitutes an anomaly. In general, the anomalous nature of a given object in a large dataset depends on both the features used to represent the objects and the method used to quantify the anomalies, and crucially, it also depends on the specific domain knowledge in a particular field, in that not all rare objects represent astrophysical anomalies from a purely scientific point of view.

In \cite{Martinez20} (MG21 hereafter), we have provided a plausible definition for anomalous behaviour in the context of \emph{Kepler} light curves: given a representation constructed by the light curve points (fluxes or magnitudes at each time snapshot) and its power spectrum (Lomb-Scargle periodogram). An anomalous light curve has features that appear to be sampled from a different distribution in the multi-dimensional space of those features compared to the bulk of the data. This definition is based on the results of applying a particular anomaly detection method (an adaptation of the Unsupervised Random Forest described in \cite{Baron17}) to the \emph{Kepler} light curves. We have demonstrated that the method successfully identifies \emph{bona-fide} anomalies, including Boyajian's star, as well as a variety of rare pulsators including $\delta$-Scuti stars, RR Lyrae stars, long period variables, among other types. 

To better understand anomalous variability in large time-domain surveys, one might ask if the anomalous nature of a given object is not only a function of the features and the method but also of the physical properties embedded in the dataset itself. Specifically, given a set of similar features from two similar datasets, are the anomalies similar and do they reveal any information about the physical properties of the objects in the dataset? For example, are RR Lyrae stars with larger periods more or less anomalous than their more rapid counterparts? One possible way to answer this question, in the context of regular, well-sampled light curves, is by applying the same anomaly detection method to both \emph{Kepler} and \emph{TESS} light curves and investigating the types of anomalies recovered in each case. By doing so, one can learn about the physical properties that result in anomalous variability in each case. In addition, by studying the properties of anomalous light curves, regardless of whether they belong to a known class, one can identify the relationship between the anomaly score and specific physical properties of the system.

In this paper, we apply the MG21 anomaly detection method to a set of SPOC-reduced \emph{TESS} light curves observed between July 2018 and May 2020 in the \emph{TESS} archive. Based on the results, we do the following: \emph{i)} We provide a catalogue of unreported anomalies for each of the \emph{TESS} sectors included in the analysis, classify them according to their variability trends, and provide anomaly statistics; \emph{ii)} We investigate how the anomaly score relates to basic stellar parameters given by the location of the objects in the Colour-Absolute Magnitude Diagram and how it relates to specific parameters of the system for anomalies of known class; \emph{iii)} We investigate the prevalence of instrumental and analysis artefacts in our list of anomalies; \emph{iv)} finally, we provide a discussion on how the overall properties of the population affect the selection of anomalies by comparing the anomalies found in this work versus those found in MG21 for \emph{Kepler} light curves.


In section 2, we investigate in detail the selected datasets and further information we use to analyse the results. Section 3 describes the fundamentals of the algorithm and the improvements made during this research. We also discuss how this is implemented with the selected databases. Section 4 explores what is identified by the algorithm and the implications of this. Section 5 explains the patterns identified within the data, linking this to astrophysical classes through the combination of anomaly score with \emph{Gaia} data to create a colour-coded Colour-Absolute Magnitude Diagram (CAMD). Section 6 explores the astrophysical process in detail behind the patterns identified. Section 7 concludes our results and the future potential of the algorithm presented.

\section{Datasets}

The two datasets that enable this research are presented in this section. The first dataset consists of the \emph{TESS} light curves, reduced using the \emph{TESS} Science Processing Operations Center (SPOC) pipeline. The second dataset consists of \emph{Gaia} DR2 photometric and astrometric data for our \emph{TESS}, where available.

\subsection{\emph{TESS} light curves}
\label{sec: lc_data}

\begin{figure}
    \includegraphics[width=(\columnwidth)]{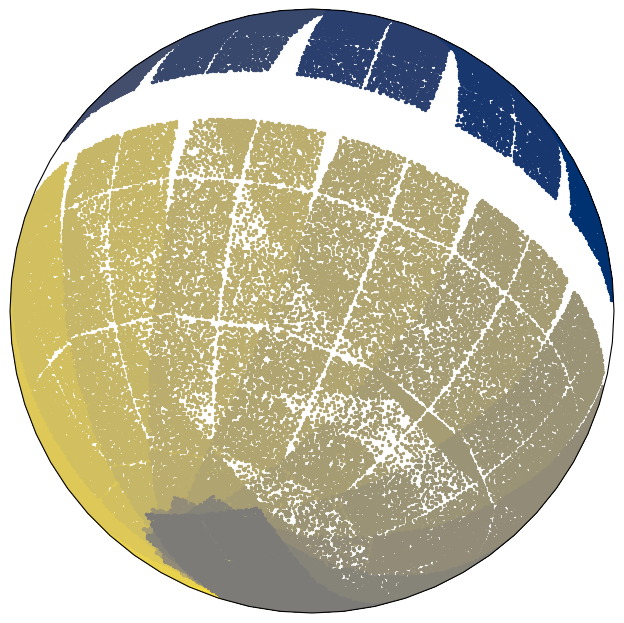}
    \caption{Visualisation of the observation segments in equatorial coordinates, each observing a patch of the sky for a duration of 27 days, with a cadence of 2 minutes. Due to overlaps at the poles, some targets are in multiple sectors and hence are observed for longer timescales, up to 351 days. This allows a cross-match to be made across sectors to evaluate the stability of our classification across datasets.}
    \label{fig: TESS_FOV}
\end{figure}

\emph{TESS} has an ambitious science case, with exoplanet detection being the main scientific goal. But many other astrophysical phenomena can be studied using the sensitivity and cadence of \emph{TESS} light curves, including astroseismology \citep{handberg21}, the dynamics of binary and multiple stars \citep{justesen21}, stellar rotational dynamics \citep{martins20}, and magnetic activity on the surface of main sequence stars \citep{cunha19}.

While both \emph{Kepler} and \emph{TESS} have similar science objectives, their input catalogue of targets differ in several ways. \emph{TESS} observes 85\% of the entire sky, an area that is 400 times larger than the \emph{Kepler} field. It aims at stars that are on average 10 times closer (and 30 to 100 times brighter) than those observed by \emph{Kepler}. To enhance the probability of observing small planets, the \emph{TESS} 2-minute cadence targets are specifically selected to be bright, cool dwarf stars \citep{Stassun18}. This implies that, on average, the \emph{TESS} targets analysed here are cooler, older, and less massive than the \emph{Kepler} targets, with populations of giants, such as the red clump, being much less prominent in \emph{TESS}, and white dwarfs more represented in comparison with \emph{Kepler} \citep{Berger20}. These astrophysical differences in the target list impact the anomaly detection results, as is later demonstrated in this paper. 

\emph{TESS} uses four identical cameras, each with four $2k \times 2k$ CCDs with a pixel scale of $21$ arcseconds, resulting in a total field of view of $24 \times 24$ degrees. The design results in the simultaneous observation of a $24 \times 90$ degree stripe that is scanned in steps over the sky to provide coverage of almost the entire celestial sphere, as can be seen in \autoref{fig: TESS_FOV}. Each observation stripe constitutes a sector and is typically observed repeatedly over approximately 27 days, producing full-frame images with a cadence of 30 minutes. Additionally, 2-minute cadence light curves are obtained for a subset of brighter and typically nearby stars. The \emph{TESS} light curves are processed by the Science Processing Operations Center (SPOC), including a pre-processing pipeline that corrects for systematics and performs photometric corrections. We use the light curves processed with the Presearch Data Conditioning (PDC) algorithm described in \cite{Fausnaugh18}. We further re-process the PDC light curves to remove single-pixel spikes unlikely to be of astrophysical origin and normalize the light curves by dividing the fluxes by the mean value.

Each \emph{TESS} sector contains about 20,000 PDC light curves with a 2-minute cadence. Objects located in the overlapping zones between sectors near the ecliptic poles, contribute more than one light curve to the dataset at different epochs. The \emph{TESS} database contains 2-minute cadence light curves associated with over 225,000 individual stars. For this work, we analyse the first 24 sectors, corresponding to approximately two years of observations. In each sector, there are gaps in the light curves where the spacecraft is offline for maintenance, with the largest intermissions caused by the inability to observe during data transmissions. \citep{Fausnaugh18}. These gaps typically span from one to a few days. In comparison with a \emph{Kepler} quarter, the total observation time in a \emph{TESS} sector is roughly three times longer (90 vs 27 days), the cadence is 15 times shorter (2 vs 30 minutes) and the number of targets is much smaller (20,000 vs 167,000), which is reflected in the data volume of a \emph{TESS} sector being about half of a single \textit{Kepler} quarter ($5-6$GB versus 11GB).

\subsection{\emph{Gaia} Photometry and Astrometry}
\label{sec: gaia_data}

The \emph{Gaia} data release 2 (DR2) consists primarily of astrometric (parallax and proper motions) and photometric (G, BP and RP bands) data for over 1 billion stars or about 1\% of the total Milky Way stellar population \citep{Perryman01}. The \emph{Gaia} DR2 all-sky survey provides photometric measurements for the majority of both \emph{Kepler} and \emph{TESS} databases. 

From the \textit{Gaia} DR2 catalogue, we collect the following values for matching TIC sources: \textit{phot\_G\_mean\_mag} ($92.5\%$ matching success with the TIC), \textit{bp\_rp} ($90.3\%$) and \textit{parallax} ($90.0\%$). The absolute $G$ magnitude is obtained from the apparent magnitude and the parallax-estimated distance to each source. \cite{Brown2018} provides a summary of all details within the database.
 
\section{Anomalous Time Series Detection Algorithm}

\begin{figure}
    \includegraphics[width=\columnwidth]{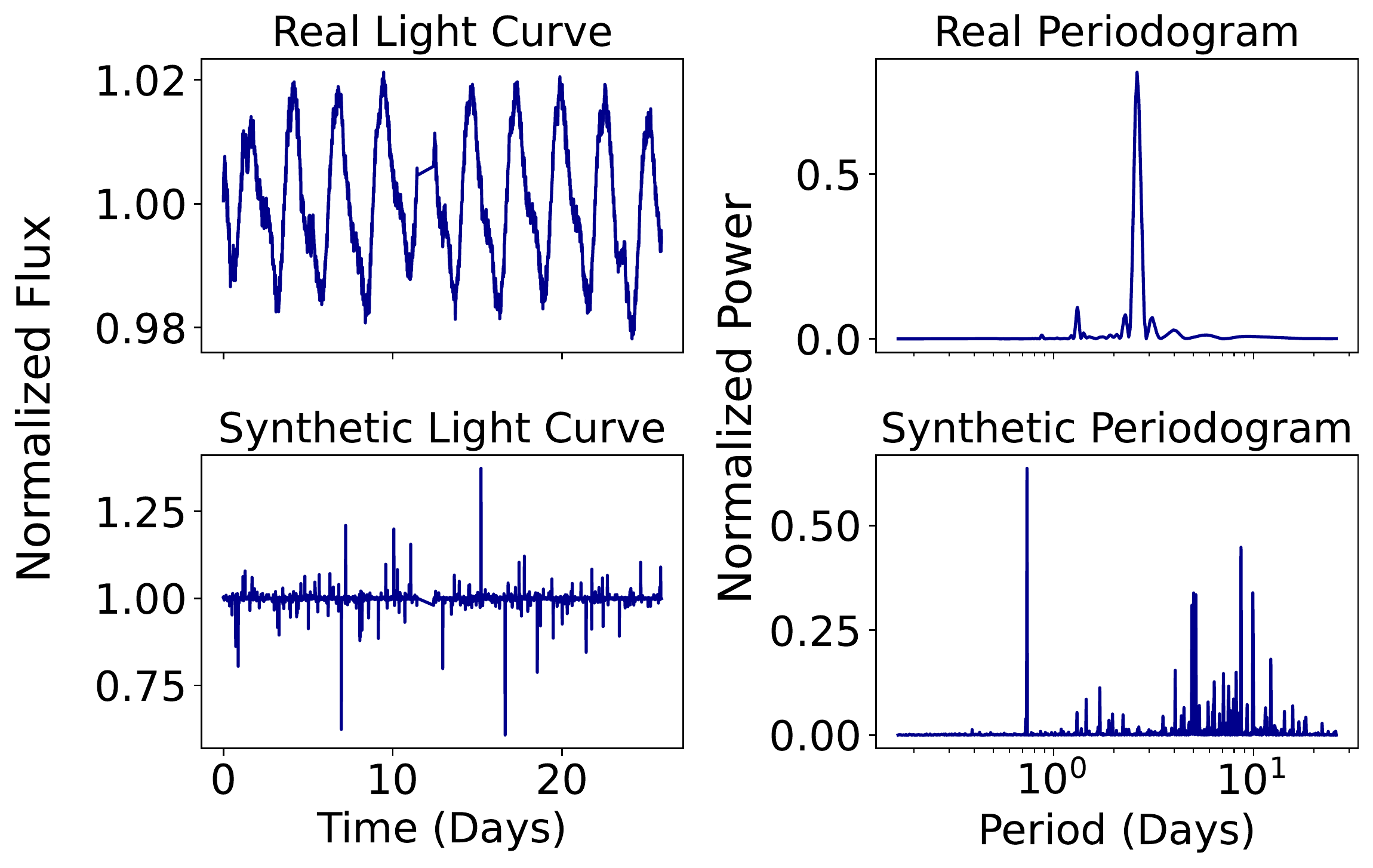}
    \caption{A visualisation of the \emph{real} against \emph{synthetic} datasets for both photometric flux and periodicity data. The non-linear sampling periods can be seen along the periodogram. Top Left: Example of an arbitrary real light curve. Bottom Left: An example of an arbitrary Synthetic light curve, notice the $\delta$-like spikes caused by the random selection of data from many separate light curves. Top Right: Example of a real periodogram for the light curve displayed Top Left. Bottom Right: An example of a Synthetic periodogram again selected at random. Notice the lack of trends or trends available within the \emph{synthetic} datasets.}
    \label{fig:Real_vs_Synthetic}
\end{figure}

We perform anomaly identification using the Unsupervised Random Forest (URF) algorithm, first used in astronomical datasets in \cite{Baron17} in the context of SDSS galaxy spectra, and previously explored by the authors in the context of \emph{Kepler} data in MG21. Given a set of features from the light curves the URF method assigns anomaly scores in two steps. First, using the data features as inputs, a random forest is trained to identify the original dataset from a synthetic dataset constructed by sampling the marginal distribution of the original features. Secondly, the population of terminal nodes is analysed for the original dataset to calculate an anomaly metric for each object. The origin of the anomaly score lies in \citet{Baron17}, but for this work, we use the modified version in MG21. The anomaly score is calculated by analysing the populations of each terminal node and creating the \textit{similarity score} as the average fraction of the total population in a given object's terminal node across all trees. The final \textit{Anomaly score} is defined as $1-S$, where $S =$ is the similarity score. The normalised anomaly score, therefore, reflects how many similar objects exist within the data set. A score of 1 signals that the object is the sole occupant of its terminal node across all trees, while 0 indicates the entire dataset occupies the same terminal node across all trees.

We now describe how the light curve features that act as inputs for the URF are extracted and the selection of the URF hyperparameters.

\subsection{Feature Extraction}
\label{sec:feature_extraction}

Following MG21, we construct the feature vector for the \emph{TESS} light curves by concatenating the vector of normalized light curve points and the vector of Lomb-Scargle spectral power values evaluated for a logarithmic range of frequencies (corresponding to periods spanning 4 hours to 27 days). The Lomb-Scargle method constructs the power spectrum of regular or irregular time series by using the equivalence between the least-square fit to a periodic signal using sinusoidal functions and the classical Fourier transform. Specifically, the power spectrum is a simple function of the $\chi^2$ value for each frequency of the sinusoidal model. A fit to the data is performed by adjusting the amplitude and the phase, resulting in an estimation of the harmonic content of the signal (the periodogram). This approach is computationally much less expensive than the Fourier transform \citep{VanderPlas2018}. The resulting vectors contain information on both the relative amplitude and the frequency properties of each light curve and are defined for the same set of times and frequencies for all light curves.

In MG21, we have thoroughly studied how the selection of these features affects the anomaly detection algorithm and have demonstrated that passing this set of feature vectors to the URF algorithm results in the successful identification of anomalies, including \emph{bona-fide} objects. In particular, we have demonstrated that the distribution of anomaly scores obtained from the analysis shows a clear, distinct peak of anomalous objects that are assigned a significantly higher URF score. In the \emph{Kepler} data, this population of anomalies is dominated by rare periodic and non-periodic variable stars, with large normalized amplitude variations, and characteristic timescales of either a few hours or a few months.

We now describe the method used to identify the optimal range of frequencies for the URF algorithm for the specific case of the \emph{TESS} light curves.

\subsection{Feature Optimisation}

\begin{figure}
    \centering
    \includegraphics[width=\columnwidth]{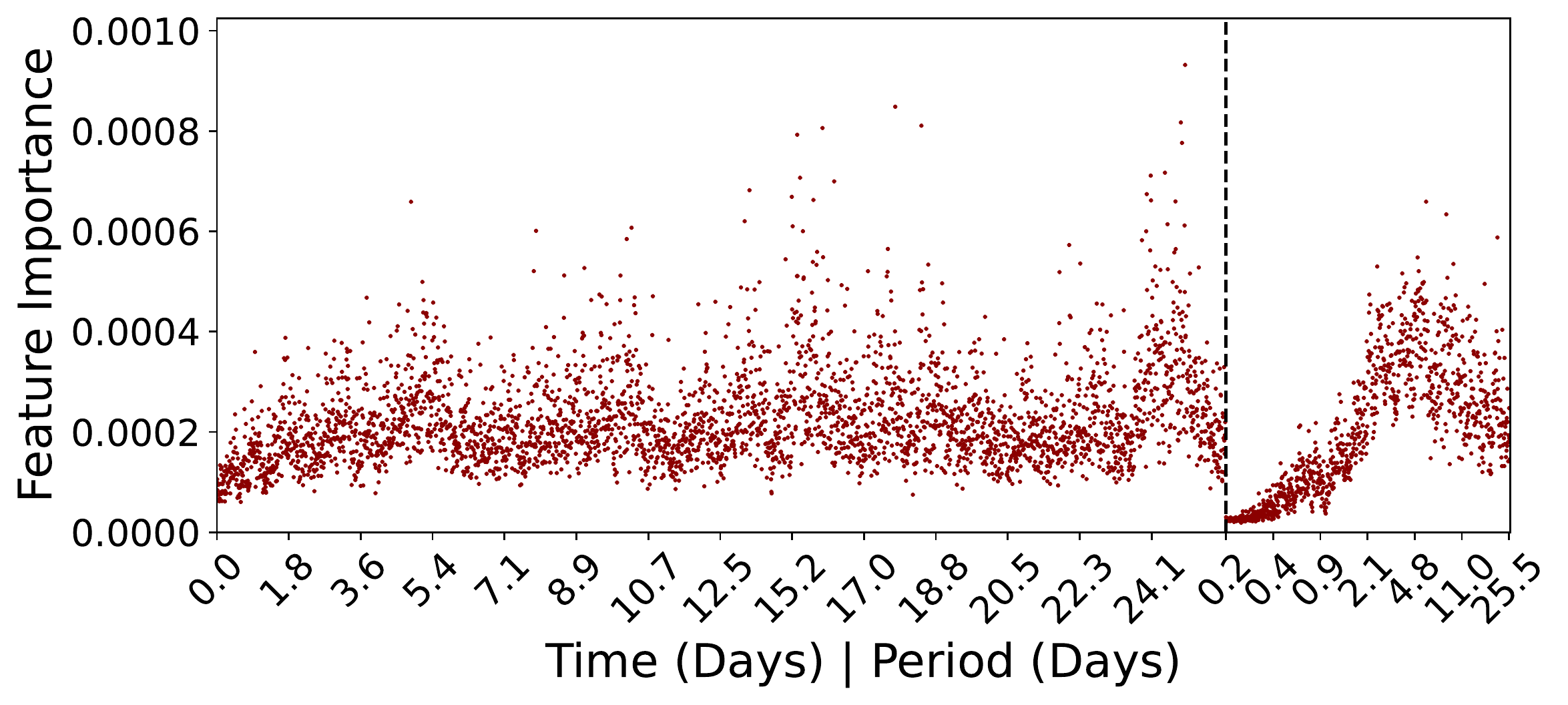}
    \caption{Example Importance scores for each feature during the training of the random forest using data from Sector 1. Data points to the left of the dotted line represent flux data, whereas those to the right represent the periodogram. As seen in the first periodogram data the lower frequencies hold negligible importance and why the minimum period is 4 hours rather than approaching the telescope cadence.}
    \label{fig: importances}
\end{figure}

At a 2-minute cadence over a range spanning 27 days, a typical \emph{TESS} short-cadence light curve contains over 19,000 points. Using the entire length of the light curves for this analysis is both scientifically unnecessary for our purposes and computationally prohibitive, as the computing time required by the URF scales linearly with the number of features used and quadratically with the number of objects considered. Consecutive light curve points at this time resolution tend to be highly correlated, and most stellar variability phenomena occur at timescales longer than 1 hour \citep{Eyer2008}. The correlation between points typically leads to overfitting when a random ensemble method, such as the URF, is applied to the data. Therefore, we opt for reducing the number of light curve points used in the analysis to 3000 points, which is achieved by performing a uniform and regular sampling of the original light curves. The resulting light curves have a time resolution of about 10-15 minutes or between a third and a half of the \emph{Kepler} cadence.

To investigate which frequencies contain information relevant to anomaly detection, we performed test runs of the URF algorithm using a different number of periodogram points, and frequency boundaries, while keeping the number of light curve points at 3000. The resulting URF scores did not change significantly for more than approximately 1000 elements in the periodogram, or for a lower bound of the frequencies corresponding to timescales shorter than a few hours. We, therefore, set the frequency boundaries to values corresponding to timescales between 4 hours and 27 days, which correspond to the duration of the observation for each sector. The frequencies are distributed logarithmically between these boundaries. We end up with a feature vector of length 4000, with the first 3000 points corresponding to the light curve points, and the last 1000 corresponding to the periodogram. Our Lomb-Scargle periodograms are computed before the subsampling to minimise the loss of information.

In \autoref{fig: importances}, we show the feature importance for each of the 4000 features, resulting from performing the random forest classification on the first step of the URF algorithm. We note that importance distributes more or less evenly among light curve points, while it peaks for the spectral power features at frequencies corresponding to timescales of about 1 day. This is as opposed to the importance distribution for the \emph{Kepler} light curves, where important timescales for anomaly detection were either of the order of a few hours or a few weeks. The distinctions here reflect the differences that we have discussed between the \emph{Kepler} and \emph{TESS} input catalogues, with dwarf stars overrepresented in the latter and giant stars underrepresented. Pulsation modes and variability types typical of giant stars are present in the \emph{Kepler} light curves and are unlikely to be found in the present dataset, whereas stellar flares, rotational patterns, and other features of dwarf stars are more likely to be found in the present study. 

\subsection{Optimisation of Hyperparameters}

The optimisation of the URF hyperparameters is done using k-fold cross-validation. Specifically, we employ the \emph{Random Search Cross Validation} implementation in \emph{scikit-learn} with 3-fold validation to maximise the accuracy of the random forest classifier. We randomise the following parameters: \begin{itemize}
    \item Number of Trees (n\_estimators): 10 evenly spaced options between 50 and 200 trees per iteration.
    \item Number of features at each split (max\_features): [sqrt, log2]
    \item Maximum depth of each Tree (max\_depth): [100, 300, 500, 700, 900, 1000, "None"]
    \item Minimum samples for Split (min\_samples\_split): ['None',2, 4, 7,10]
    \item Minimum samples in Node (min\_samples\_leaf): [1, 2]
    \item Bootstrap: [True, False]
    \item Warm Start: [True, False]
\end{itemize}

During the random search, we look for hyperparameters that maximise the validation accuracy of the RF classifier while avoiding over-fitting. Once the parameters that maximise this accuracy are identified, we fine-tune the specific hyperparameters near these values to increase the contrast in anomaly scores between objects. Fine-tuning is necessary because the parameters that maximise classification accuracy are not necessarily the same parameters that maximise the detection of anomalies. For example, by allowing a small amount of overfitting, we can improve the ability of the method to find anomalies as the more stringent isolation of light curves in the terminal nodes reduces the range of light curve shapes that are considered non-anomalous. We have therefore tuned the hyperparameters to obtain a sweet spot between classification accuracy and the resulting fraction of anomalies.

The final parameters used for this study are: n\_estimators=100, warm\_start=True, bootstrap=False, min\_samples\_leaf=2, max\_features='auto', min\_samples\_split=4, max\_depth=700.

\subsection{Implementation and Hardware}

The computation of the anomaly scores uses 96 GB of memory and approximately 16 hours of run time per sector. The periodograms, URF training and anomaly scores were calculated by Cuillin Computing Cluster at the Institute of Astronomy in the Royal Observatory Edinburgh. As described in MG21 and earlier in this paper, we use a modified version of the original URF algorithm that significantly reduces the computation time to calculate the anomaly score. Specifically, we do not use a pairwise match for each pair of light curves, as this is redundant. Instead, we shift to the analysis of the terminal node populations. The updated method scales linearly with the number of final nodes bringing increasing returns the larger the input catalogue.

The uncertainty in the anomaly scores originates from the random nature of the ensemble methods and the finite number of trees. This uncertainty is inversely related to the number of trees used and results in a variance associated with each computed anomaly score. The values provided here for the anomaly score of each light curve correspond to the average score over ten independent realisations of the method.

\section{Results}
\label{sec: results}

Results from the URF are published alongside this work, with \autoref{tab: datatable_cont} showing the format. Columns $1-5$ represent identifiers for the object, Column 6 represents the anomaly score from this work, and columns $7 - 10$ explain the sectors observed and the population of each light curve, e.g. the anomalous, bulk or intermediate populations. Columns 11 \& 12 are observations also from this work and products of upcoming analysis in \S~\ref{sec: objects_found} \& \S~\ref{sec: high_sigma}.

\input{Tables/results_split_1.tex}
\input{Tables/results_split_2.tex}

\begin{figure*}
    \includegraphics[width=\textwidth,height=\textheight,keepaspectratio]{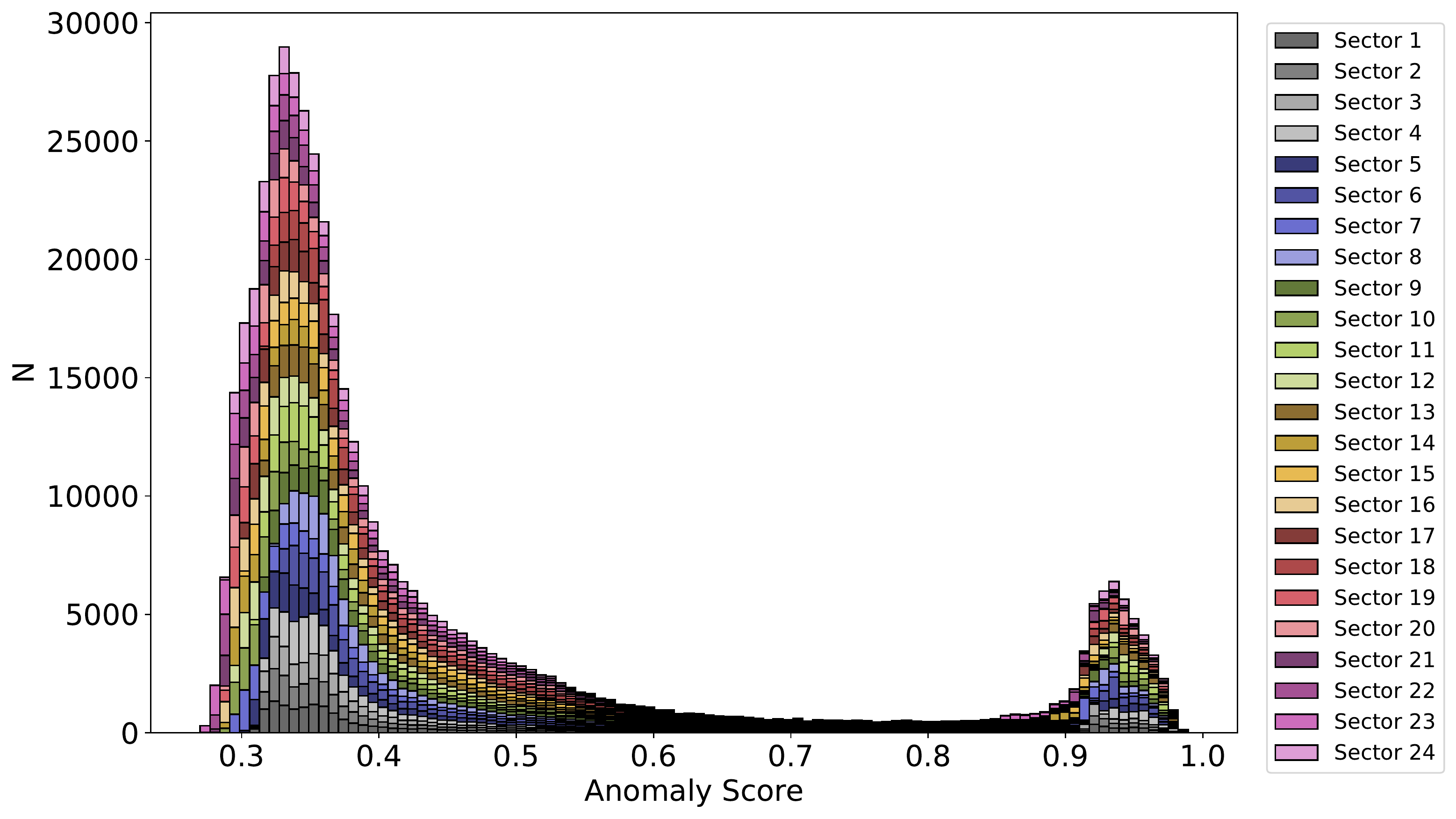}
    \caption{Stacked histogram of all 24 Sectors colour-coded by sector. The overall distribution can be seen to be consistent across all sectors despite independent analysis. Notice $\sim 85\%$ of all objects have an anomaly score less than 0.6. With no further analysis, this represents an order of magnitude reduction in size for the catalogue of anomalous stars to classify.}
    \label{fig: stacked_hist}
\end{figure*}

\subsection{The emergence of distinct populations}

\autoref{fig: stacked_hist} shows the distribution of URF anomaly scores for the 24 \emph{TESS} sectors considered in this work. The distribution consists of two distinct peaks, one centred at a score of about 0.35, and another at about 0.93. A sparsely populated region of objects with intermediate scores lies between them. The bulk of the population (objects with anomaly scores $\lesssim 0.6$) corresponds to what we can consider \emph{normal} light curves, that is, objects whose light curve features are sampled from a parent population that represents the most common forms of variability (or lack thereof). Containing objects with anomaly scores $\gtrsim 0.9$, on the other hand, represent light curves with outlying amplitude and frequency properties sampled from a different distribution with respect to the bulk of the objects. These are the ones that we will consider anomalies in this work. Up to this point, they only represent anomalies from a purely data-centred perspective. Whether they constitute truly novel astrophysical objects, observational and/or pipeline artefacts, or only rarer classes that are underrepresented in the dataset, is the subject of our investigation in the upcoming sections.

Overall, about 82\% of the \emph{TESS} targets have an average anomaly score $<0.6$; 7\% of them have an intermediate average score between 0.6 and 0.9, and 11\% have average anomaly scores higher than 0.9. The average is calculated over multiple observations of the same object, which is more likely to occur for targets located at high ecliptic latitudes. The distribution of anomaly scores is very similar from sector to sector, apart from a different normalisation in each case, due to differences in the total number of targets detected in each sector.  In particular, the population of outliers with scores $>0.9$ represents a similar fraction of the cases in all sectors, with small variations. For example, in Southern Hemisphere sectors, the fraction of anomalous light curves is $0.11^{+0.03}_{-0.02}$, whereas, for Northern Hemisphere sectors, it is $0.08^{+0.007}_{-0.018}$. Sector 23 is an outlier with a fraction of anomalous objects of only 0.02. The variation in the fraction of outliers between hemispheres is understood to be due to different fractions of the sector field containing the galactic plane. The average value of the anomaly score is very similar for all sectors, with a standard deviation of only 0.02 for the 24 averages. This all suggests that each sector is representative of the entire dataset and that the anomalous nature of the objects is related to astrophysical or observational phenomena that are common to all sectors, which favours an interpretation according to which most of the anomalies - but not all of them - are objects of a known class with particularly outlying variability parameters.

In our results (An example of which can be found at \autoref{tab: datatable_cont}), we present the average anomaly score for each anomalous object in our dataset, computed across all sectors in which the object is observed. For this work, we define anomalous objects are those for which the average score is higher than 0.9. To indicate the dispersion of each object's score across sectors, we also list sector IDs where the object is detected and the sectors where the score is higher than 0.9 or below 0.6. We justify providing the average score as most objects show nominal variance between sectors. Additionally, we provide a flag (\textit{high variance}) for the objects that contradict this assumption. We dedicate part \S~\ref{sec: high_sigma} to discuss the prominent cases where a large dispersion in anomaly scores is observed across sectors, as those objects represent anomalies that are not persistent in time.

\subsection{Consistency Across Sectors}
\label{sec: sigma}

\begin{figure}
    \includegraphics[width=\columnwidth]{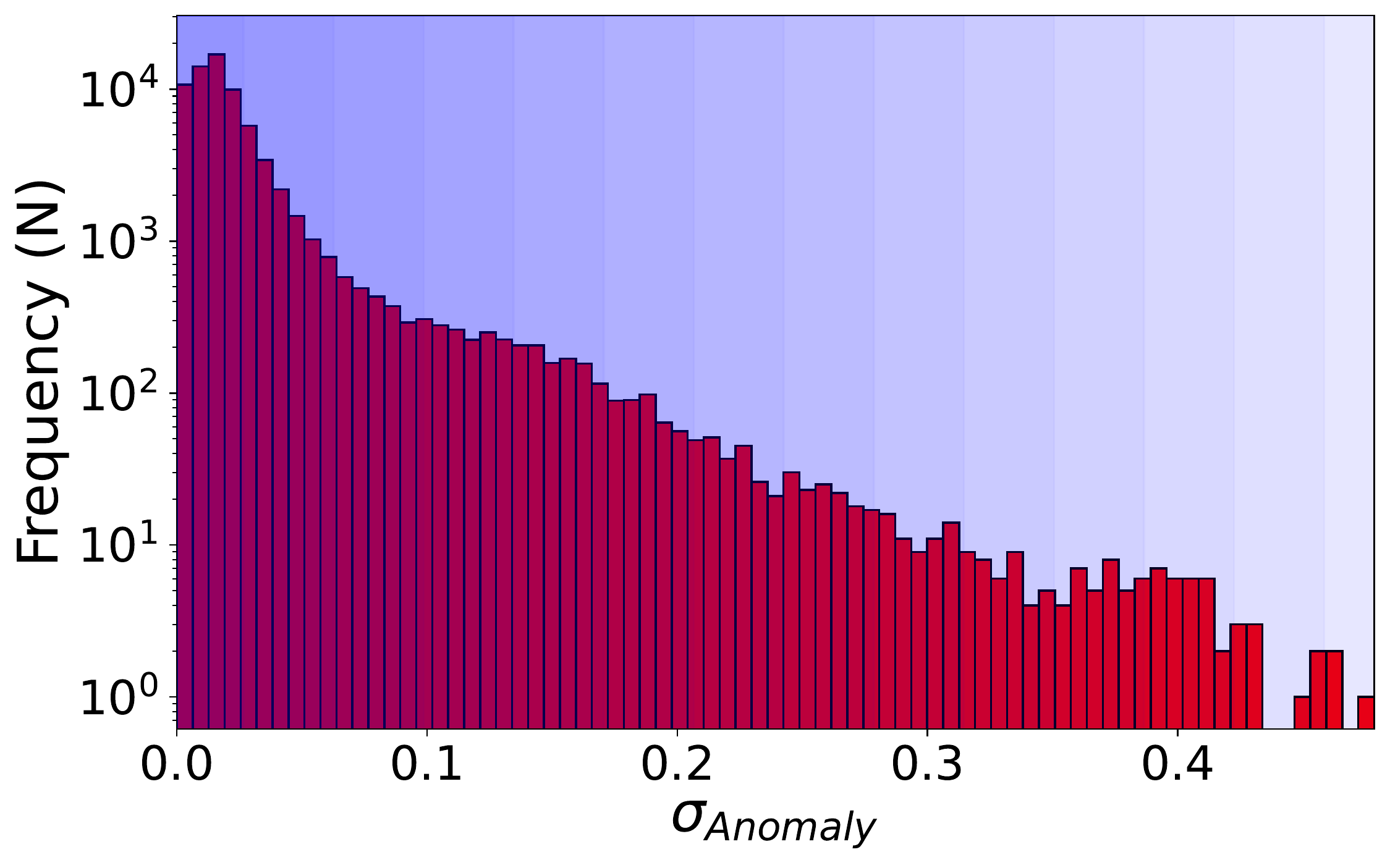}
    \caption{Distribution of the standard deviation of anomaly scores for objects observed in multiple sectors. The majority of objects show minimal deviation between sectors, with a median standard deviation of 0.0166. The steps in gradient, shown in blue, represent one sigma of the shown distribution ($\sigma_{\sigma_{Anomaly}} = 0.036$).}
    \label{fig: anomaly_variance}
\end{figure}

Our dataset contains over 400,000 light curves for more than 200,000 unique targets. Despite an average of 2 light curves per object, less than half ($\sim70,000$) are observed in multiple sectors. The majority of these objects have similar anomaly scores across all observations. However, there is a small fraction where the anomalous variability pattern occurs during a particular observation. This event leads to an increased anomaly score in that sector, with the other sectors remaining unaffected. Naturally, this leads to an increased variance for that object, indicating the anomalous light curve may be an exceptional or rare behaviour that is not persistent in time, for example, a transient event, ASASSN-21lw \citep{Shappee2014}.

Nonetheless, we compute the standard deviation of the anomaly scores for all objects observed more than once. We show the resulting distribution of standard deviations in \autoref{fig: anomaly_variance}. We find consistency between the scores across observations, with 95\% of the objects having $\sigma_{Anomaly} < 0.1$. The mean and median of the $\sigma_{Anomaly}$ distribution are 0.027 and 0.017 respectively. Implying the average anomaly score is a good diagnostic for the anomalous nature for the vast majority of the sources. We now discuss the cases where this is not true.

Only about 0.38\% of the objects have a standard deviation higher than $\sigma > 0.25$ between sectors. These might indicate short-lived transient events, or periodic signals such as dips with a period significantly longer than the duration of the \emph{TESS} light curves. Astrophysical transients such as cataclysmic variables, stellar flares, microlensing events and other unknown explosive events could also be part of this group. Additionally, data processing or instrumental artefacts, such as sudden drops in the baseline level of the light curve, can also be identified in this fashion. We discuss this group of anomalies in \S~\ref{sec: high_sigma}.

A total of 44,889 light curves in our sample have anomaly scores greater than 0.9, representing 25,858 unique objects. As is usually the case in astronomical anomaly detection, the list of anomalies is too large to allow for the characterisation of each light curve. Our approach for prioritisation here is to explore the anomalous objects to understand the astrophysical mechanisms driving the anomaly score, followed by methods probing beyond the anomaly metric alone. Such methods potentially isolate the rarest or most extreme anomalous behaviours, such as objects likely to be non-persistent anomalies which indicate the observed anomalous behaviour is restricted to a single sector, as opposed to behaviour that results from repeating or periodic patterns. Nevertheless, we provide the list of anomalies and provide a description based on the overall light curve type, determined using visual inspection for sectors 13 onward. Additionally, in the next section, we incorporate information related to the astrophysical properties of these sources, such as luminosities and effective temperatures, to assess the relationship between anomalous nature and evolutionary stage for stars in the \emph{TESS} input catalogue.

\section{Linking anomaly scores to astrophysical properties: the Colour-Absolute Magnitude diagram}
\label{sec: hr_diagram}

\begin{figure*}
   \includegraphics[width=\textwidth]{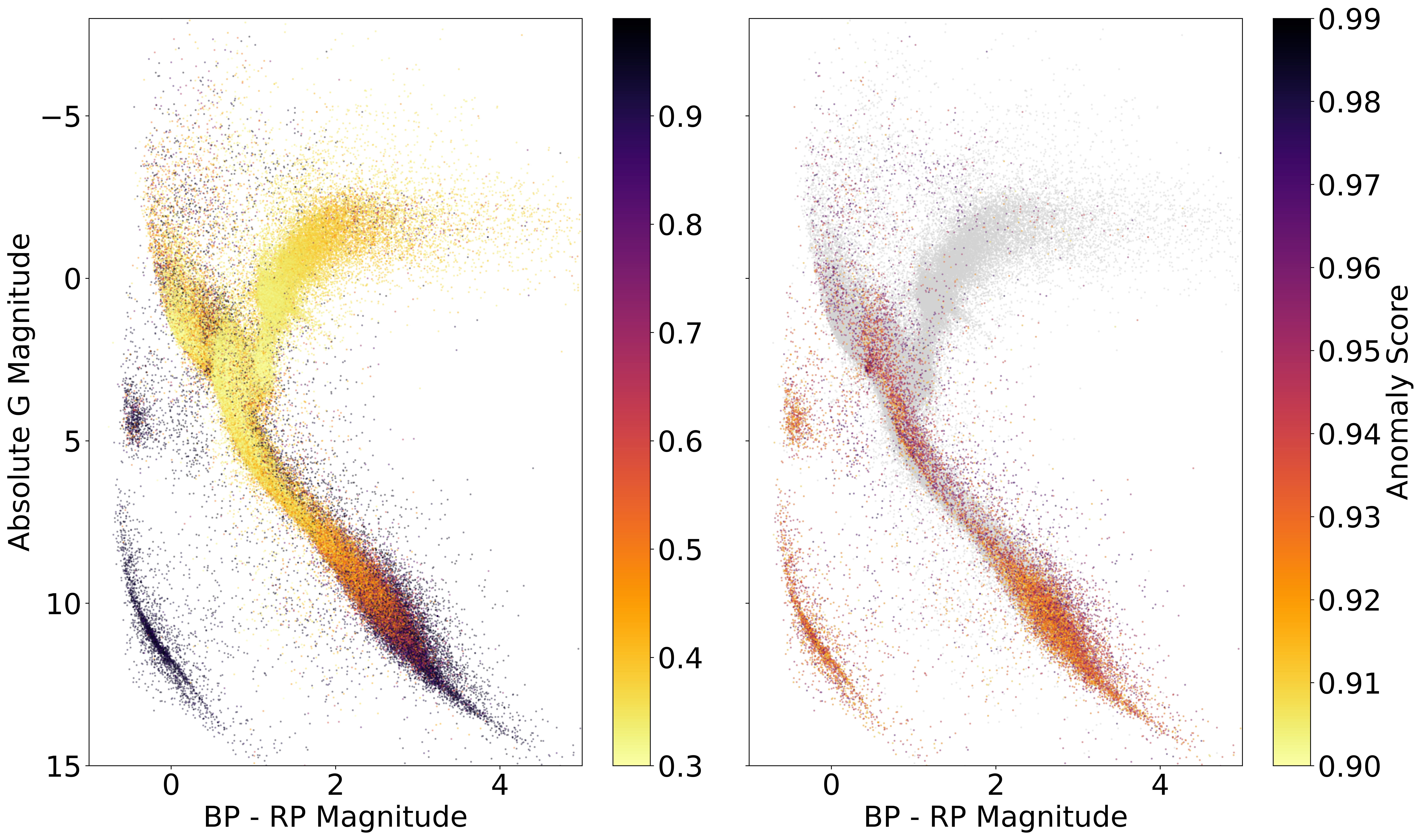}
   \caption{\textit{Left:} Colour-Absolute Magnitude diagram of \emph{TESS'} all objects, colour-coded by their average anomaly score across all sectors. Distinct regions with a high anomaly score can be seen, meaning the algorithm is classifying intrinsically similar objects with a similar anomaly score across the diagram. \textit{Right:} Colour Magnitude Diagram of the outlier population colour coded by anomaly score, with the remaining population shown in grey for comparison.}
   \label{fig: HR_full}
\end{figure*}

The discussion of whether a particular light curve represents a true astrophysical novelty or relates to an understood phenomenon within a particular data set is non-trivial due to the required domain knowledge that may require modelling or other tools of inference. An initial step involves identifying the objects with a known class. Only about 30\% of the anomalies identified in this work have a known class described in independent studies. Since the vast majority of the TIC objects are stars, we can rely on additional information from their \emph{Gaia} observables to constrain their physical properties. In \autoref{fig: HR_full} we show the \emph{Gaia}-generated Colour-Absolute Magnitude diagram (CAMD) for our targets, colour-coded by the average anomaly score across all sectors. This diagram allows us to relate the anomaly scores to particular spectral types or evolutionary stages, setting a general framework for the interpretation of our results. Our cross-match indicates that $\sim90\%$ of all targets considered here have reliable distance and photometry measurements from \emph{Gaia}. This fraction is similar in all sectors.

There are several interesting trends in \autoref{fig: HR_full}. A prominent region of the CAMD with overall high anomaly scores is the white dwarf branch. The instability strip is also a prominent region of relatively high anomaly scores, potentially driven by the radial pulsations of starts occupying this part of the diagram. Other regions with relatively high anomaly scores include; supergiants, subdwarf B stars, M-dwarfs and brown dwarfs. There are also objects whose optical variability is likely to be extrinsic, rather than intrinsic, or a combination of both. These include young stellar objects (YSOs), which show variation due to internal stabilisation and obstructions from dusty halos. Additionally, objects that occupy sparsely populated regions of the CAMD, e.g. the coolest region of the M-dwarf branch, are often classified as an outlier. Our results demonstrate the isolation observed in the CAMD, due to the extreme luminosities or temperatures, is reflected in their variability patterns.

We observe an increase in the anomaly score as we move into cooler, low-mass dwarf stars, from convective envelopes into fully convective M-type stars. In this regime, it has been proposed as part of basic dynamo theory that stellar magnetic activity should increase inversely to mass, accounting for the proportionally deeper convective envelope \citep{Radick1992, Brun2017}. A significant part of the variability in these stars is the result of chromospheric activity, such as sunspots \citep{Baliunas1995, Schuessler1996, Barnes2001, Barnes2011}. 

Focusing now on the right panel of \autoref{fig: HR_full}, where we have limited the anomaly score range to only the most anomalous scores, we note that the highest anomaly scores are distributed among stars in different regions of the CAMD, with an increased density of anomalous objects near the instability strip where the MS meets the giant branch. Low-mass stars and white dwarfs consistently have lower anomaly scores, whereas stars in the giant branch generally do not make the $>0.9$ score mark to be included in the anomalous group, which was not the case for \emph{Kepler} targets. In the particular case of white dwarfs, variability is dominated by rapid flux variations that deviate between 1\% and 2\% from the mean flux. In \S~\ref{sec: types_of_object} we describe the variability trends in each case and investigate potential physical mechanisms for the anomalous behaviour. 

The MS remains, at best, sparsely populated by anomalous light curves, likely the result of the relatively stable hydrogen-burning cycle of MS stars. In this region, the observed anomalous time-domain behaviour is caused predominantly by extrinsic phenomena, such as eclipses. As for RGB stars, they are largely stable throughout this phase of evolution at the timescales probed by the \emph{TESS data}, and they are sufficiently represented in the TIC not to be considered a rare population. The timescales of their spectrally dominant variability modes are longer(hundreds and thousands of days) than the timescales of \emph{TESS} observations. The assumption of variability over long timescales does not hold for all types of supergiant stars, with some of them undergoing different stages of evolution, particularly members of the blue supergiant population. We expand our exploration of giant stars further in \S~\ref{sec: giants}. We note here, however, that there are signs of increased anomaly score in the cool extremes of the red giant branch, where we observe variability with periods between $\sim30 - 80$ days which can be considered "long period variability" compared to the observation timescales.

In what follows, we explore the set of anomalous light curves, both in terms of their value (i.e., are there any true astrophysical one-offs?) and how the anomaly score relates to specific physical parameters for objects belonging to known classes. This will allow an investigation of the distribution of physical parameters for objects of specific types (e.g. eclipsing binaries) and the most extreme among those. Our methodology is as follows: first, we perform a visual inspection of the anomalous light curves and categorise them according to their overall variability pattern. We then use additional metrics, such as the standard deviation in the anomaly score across observations, to search for non-persistent anomalies. Finally, we study anomalies of specific know classes (e.g. white dwarfs, eclipsing binaries) and investigate what the anomaly score tells us about their specific configurations.

\section{Most Anomalous Objects Identified}
\label{sec: anomalous_objects}

\subsection{Visual Inspection of Light Curves and Classification}
\label{sec: objects_found}

We perform a visual inspection of all anomalous (score $\geq 0.9$) light curves in sectors 13 through 24 that do not have an unambiguous classification in the SIMBAD database. This includes objects that only had uninformative labels in SIMBAD, such as "Star". Rather than identifying all potential unique objects, our initial goal is to understand the diversity and composition of the landscape of \emph{TESS} anomalous light curves.

We inspect each anomalous light curve individually, regardless of whether the associated object has multiple light curves. This work has analysed over 12,500 light curves. As a result of the inspection, we exclude sectors 22 and 23 from any further analysis due to a disproportionate population of artefacts not corrected by the PDC pipeline, an example given in \autoref{tab: my_labels}, leading to inconsistent outlier populations. Sector 23 is the most affected with only 0.8\% of light curves classified as anomalous, compared to the typical 5-10\% in other sectors. Although we exclude them from the analysis, we still include results from these sectors in our catalogue of anomalies for completeness.

Our classification is based on the morphological properties of the light curves rather than on astrophysical types for two main reasons: first, we would require a vast amount of domain knowledge and consideration to correctly assign accurate astrophysical classes based on the light curve alone. In addition, since we expect some anomalies do not belong to any known class, our best efforts can only describe their morphology. We converged to the final classification in \autoref{tab: my_labels} iteratively by initially setting a set of labels based on Sector 24 alone. These were updated as more analysis revealed slightly different patterns in different sectors. Finally, we reclassified sector 24 to reduce potential inconsistencies from inexperience and missing labels during the initial classification.

As an example, RR-Lyrae stars are classified as \textit{"sinusoidal"} due to their periodic pattern resembling a sinusoidal function (see \autoref{tab: my_labels}). When appropriate, additional labels are included to refine the class, so "sinusoidal asymmetrical" indicates a distinctive pattern in which there is a sinusoidal pattern with asymmetrical peaks or troughs indicating more rapid increases in brightness than decreases or vice versa. Objects that do not fit in any morphological types are left unlabelled. If the astrophysical nature of the variability pattern is clear from the morphology, this is reflected in the label, as is the case for \textit{"transit events"}, which primarily include eclipsing binary systems. Overall, descriptive labels are used in ambiguous cases to minimise mislabelling and class overlap. In \autoref{tab: my_labels}, we compile a summary of the most common labels and provide light curve examples. The only exception to this morphological system of classification is white dwarfs, which we label according to the external catalogue provided by \cite{Fusillo2019}. This divergence in approach is required as the white dwarf population systematically returns higher than average anomaly scores that are difficult to determine visually. We believe this is due to rapid pulsations described in \S~\ref{sec: WDs}.

\input{Tables/class_labels}

\subsection{Composition of Outlying Population}
\label{sec: types_of_object}

\begin{figure}
    \includegraphics[width=\columnwidth]{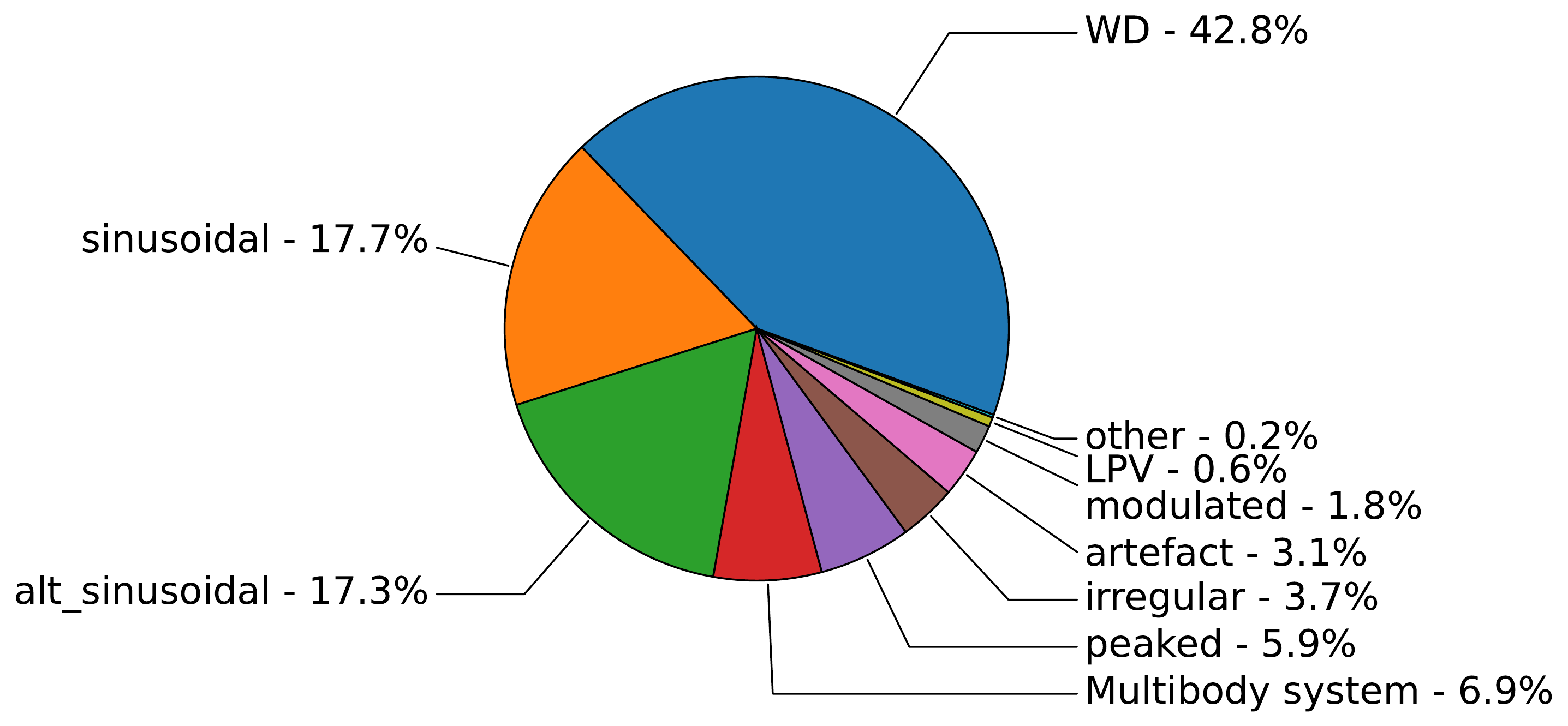}
    \caption{The distribution of descriptions for objects with anomaly scores above 0.9 and lacking prior classification. The figure includes $\sim 30\%$ of objects in sectors 13 - 21 and 24, with the remaining objects lacking defining features upon visual inspection or catalogue comparisons.}
    \label{fig: weirdest_pie}
\end{figure}

Out of the 12,500 light curves inspected, 4,128 are classified using the morphological system summarised in \autoref{tab: my_labels}. An additional 3063 objects have been previously classified as white dwarfs (WDs) by \citet{Fusillo2019}, with only $\sim 1\%$ falling in one of the morphological categories. We also identify several objects with minor artefacts, e.g. single-pixel spikes and trends in small sections of the data. Since these minor features are not likely to dominate the anomaly score, we classify these objects based on the morphology of unaffected parts of the light curve.

In what follows, we refer to light curves with a repeating pattern of short but statistically significant dips, including transits and suspected eclipses, as "multi-body" systems. The breakdown of light curve classification is as follows: 3063 are classified as white dwarfs, 1275 as sinusoidal, with an additional 1244 showing additional patterns to the sinusoidal baseline (alt\_sinusoidal), such as modulation (see \autoref{fig: example_modulated}), short periods (less than $\sim8$ hours) or asymmetrical peaks. There are 500 multi-body systems, 423 periodically peaked light curves showing repeated peaks of emission, 264 irregular light curves and 237 dominated by pipeline artefacts. In addition, 129 are modulated patterns and 43 show long-period variability (LPV), defined in this work as a pattern with an apparent period $\geq 26$ days. 13 light curves do not fit any of the morphological classes but have a unique description. \autoref{fig: weirdest_pie} provides a visualisation of this breakdown.

\begin{figure*}
  \includegraphics[width=\textwidth, keepaspectratio]{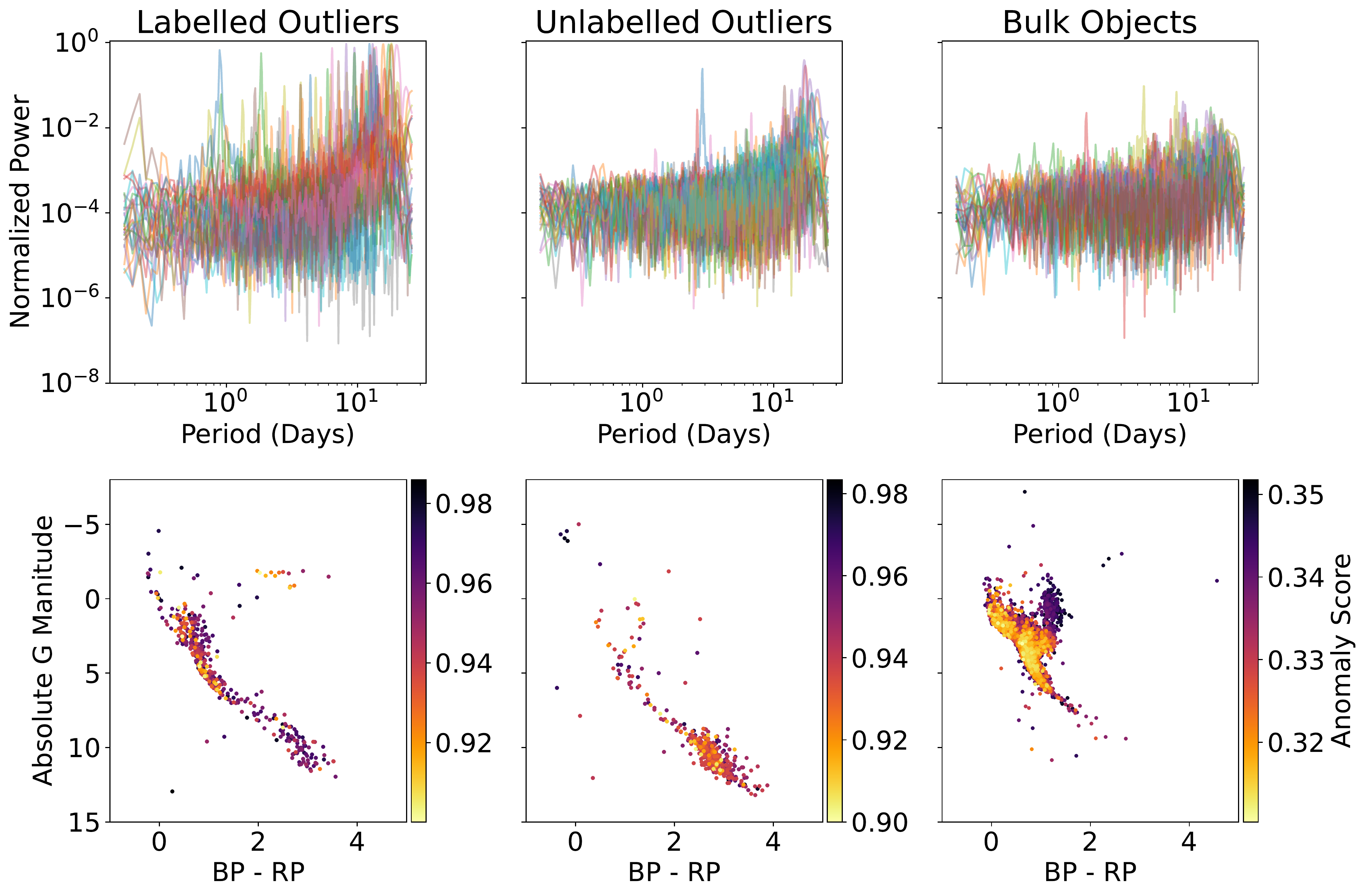}
  \caption{\textit{Top:} A random sample of $\sim30$ periodograms from the labelled outliers, unlabelled outliers that show an increased variability magnitude, and objects with anomaly score less than the peak of the bulk population (labelled "Bulk") from Sector 13. \textit{Bottom:} CAMDs for the same example periodograms, highlighting the type of objects in each class. White Dwarf stars are deliberately removed from the sample as these are classified using an alternative approach.}
  \label{fig: periodograms_comparison}
\end{figure*}

There is a group of light curves for which the only noticeable feature is a relatively higher amplitude than a non-variable noise-dominated light curve that populates the low-anomaly end of the distribution. The majority of these higher amplitude light curves are associated with low-luminosity, cool M-dwarf stars. As the light curve shows little evidence of anomalous behaviour, we must explore the frequency space and investigate their power spectrum. \autoref{fig: periodograms_comparison} shows the periodograms and location in the CAMD for three types of objects: anomalies that fit in any of our morphological types (left), anomalies that show the higher amplitude feature (centre), and non-anomalous light objects (right).

We note that the periodograms of M-dwarf anomalies are relatively featureless compared to the morphologically classified objects, which indicates that they do not have well-defined periodic features. However, similar to the morphological anomalies, they have a greater fraction of the power contained at low-frequency modes than the bulk objects. This pattern indicates dominating variability at relatively long (weeks) timescales. Evidently, these objects are anomalous due to both higher amplitude and in their frequency spectrum, supporting results from MG21, where we found low-frequency modes of variability are associated with anomalous light curves in \emph{Kepler} data.

The stellar rotation might be related to this anomalous behaviour. As demonstrated in the observational study by \citet{Popinchalk2021}, M-dwarfs of earlier spectral types (M4-M6) are rapid rotators, with rotational periods of less than one day to as short as a few hours. M-dwarfs become cooler as they evolve, causing their rotational period decreases, possibly due to magnetic effects \citep{Garraffo2018}, until it stalls for late types. The reduction in rotation explains why the majority of field M-dwarfs with measured rotation periods are greater than 10 days, with non-periodic variability timescales typically a few weeks to months in the later types. This non-periodic excess of spectral power at longer timescales is what makes their light curve anomalous. The fact that we observe a gradient in the anomaly score along the main M-dwarf portion of the main sequence, with cooler M-dwarf stars being more anomalous, is indicative that our method is sensitive to the disappearing rotation signature. These results, once more, suggest a correlation between the anomaly scores and the object's evolutionary stage, at least for certain types of objects. 

\begin{figure*}
  \includegraphics[width=0.95\textwidth, height=0.95\textheight, keepaspectratio]{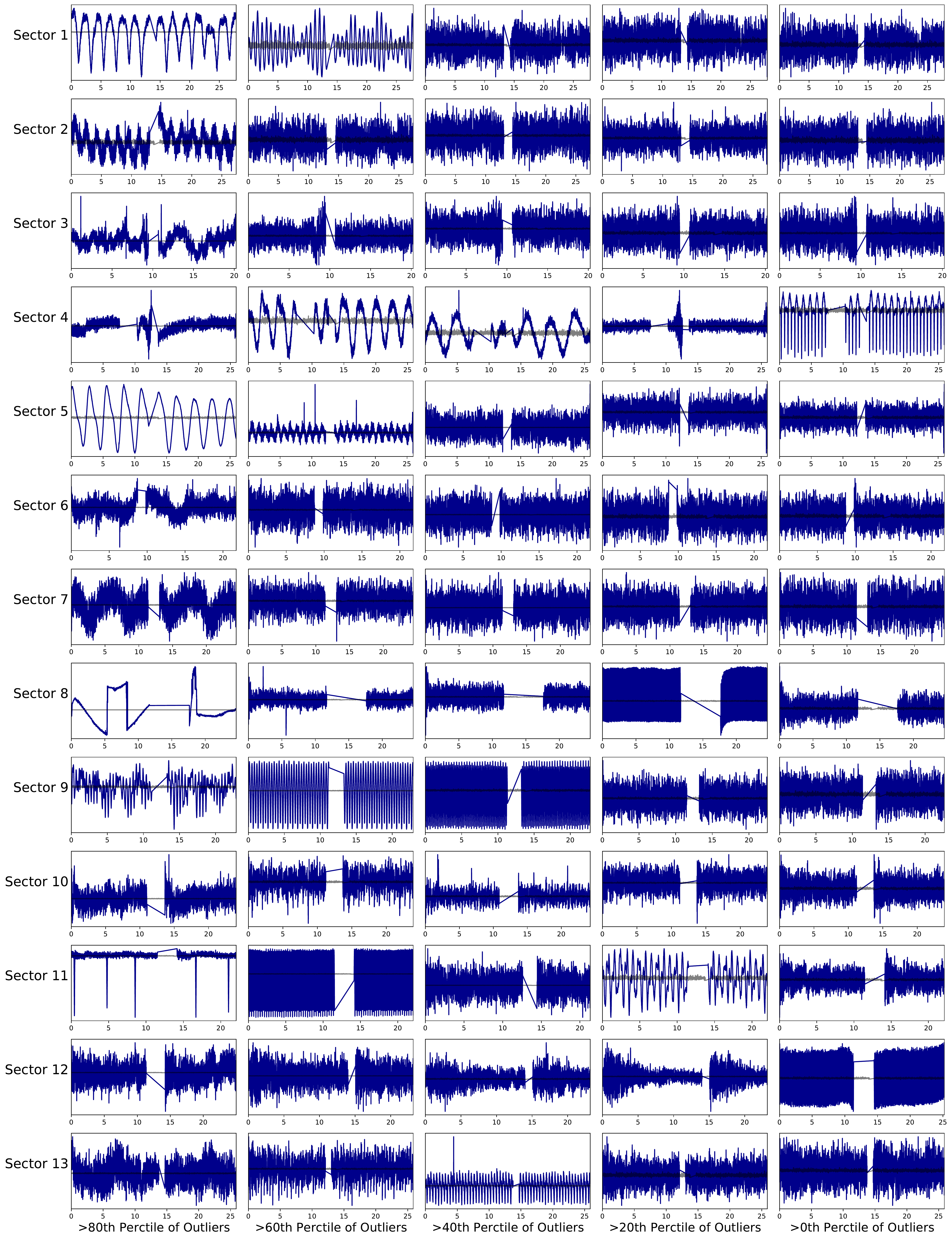}
  \caption{Examples of outliers identified in Sectors 1 - 13 without well-defined classification labels. Anomalous behaviour decreases from left to right, with a random sample taken from each 20th percentile. The light curves are in blue, with the black light curve representing a very non-anomalous light curve, TIC 261337074, for reference on magnitude variations.}
  \label{fig: outliers_13}
\end{figure*}

\begin{figure*}
  \includegraphics[width=\textwidth, height=\textheight, keepaspectratio]{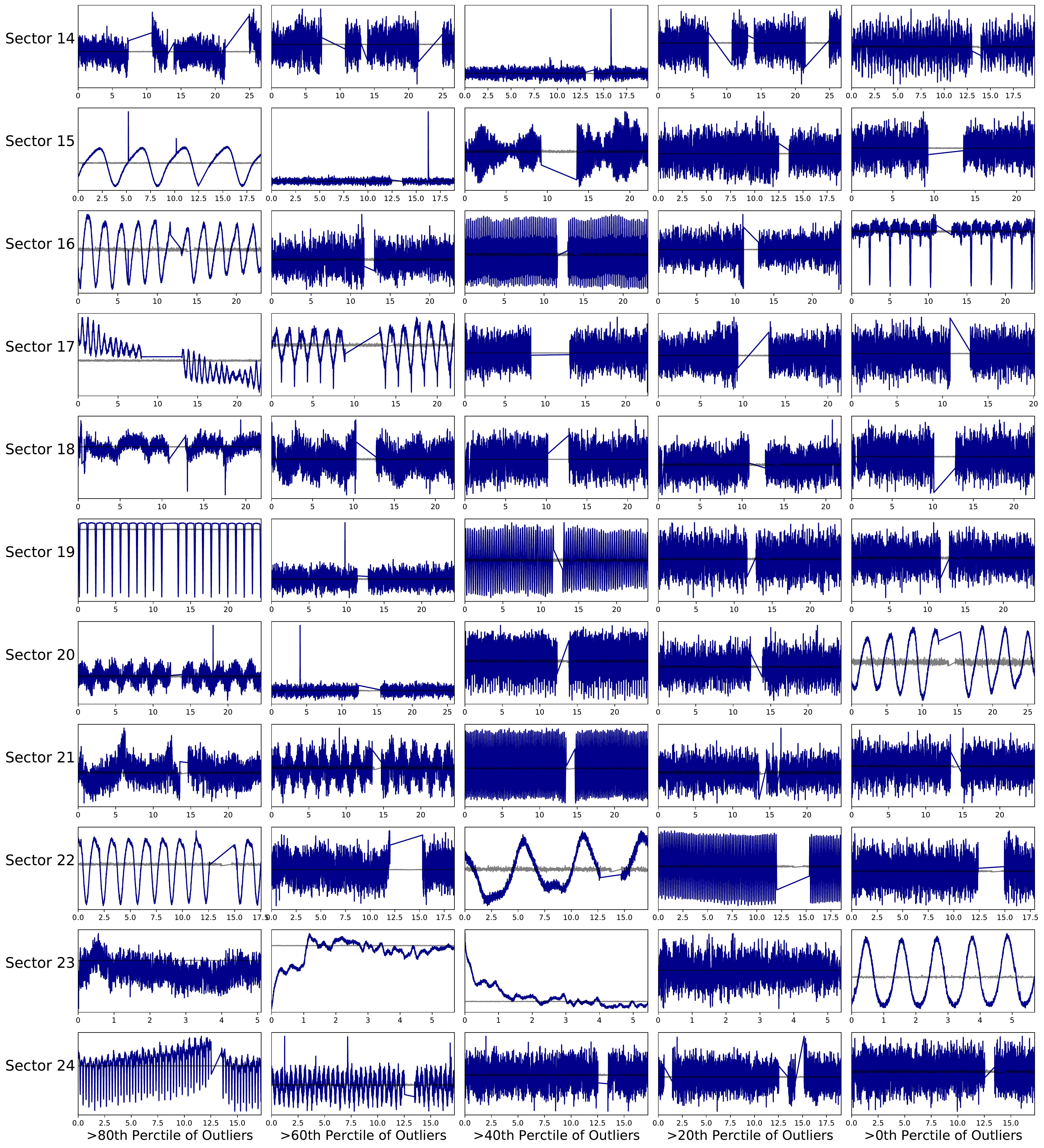}
  \caption{Examples of outliers identified in Sectors 14 - 24 without well-defined classification labels. Anomalous behaviour decreases from left to right, with a random sample taken from each 20th percentile. The reference light curve in black is the same as in Figure \ref{fig: outliers_13}. Plots from Sector 23 reveal the nature of the detrimental trends identified}
  \label{fig: outliers_24}
\end{figure*}

\subsection{Non-persistent anomalies}
\label{sec: high_sigma}

\begin{figure}
  \includegraphics[width=\columnwidth]{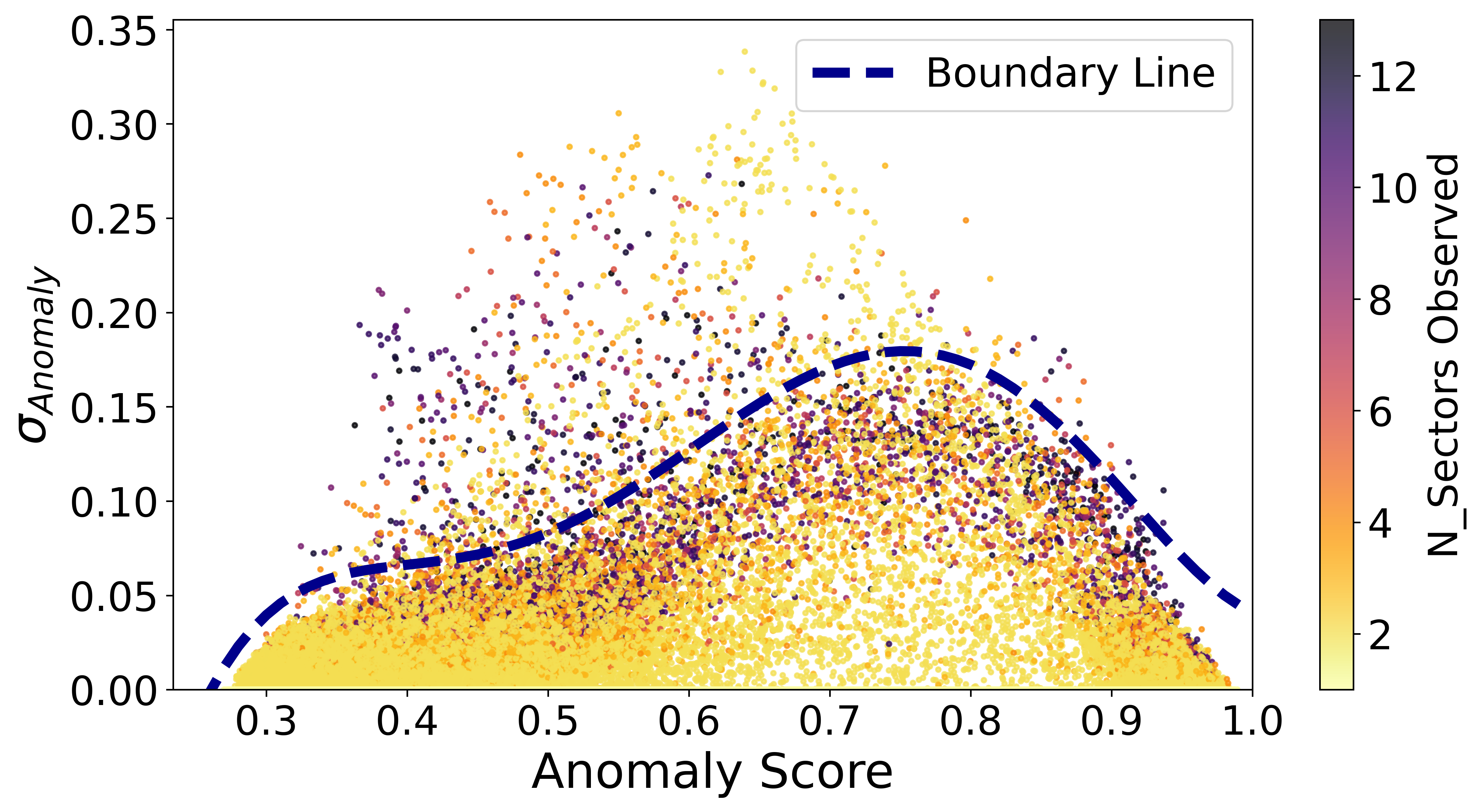}
  \caption{A visualisation of the standard deviation of anomaly score between sectors vs the average anomaly score. The colour bar represents the number of sectors observed. As observed, we expect the intermediate population to show an average higher standard deviation due to the smaller population within this region. Nonetheless, the objects that occupy the region above the bulk are of the greatest interest. The boundary line represents the boundary between the bulk and the high standard deviation sources for this section of analysis.}
  \label{fig: avgWeird_std}
\end{figure}

A small fraction of the objects show high variance in the anomaly score between sectors, which may indicate a particularly elevated score in only one of the observed epochs. These are candidate non-persistent anomalies, i.e., anomalous behaviour that is non-repeating and could signal single-event transients. About 95\% of the objects show a standard deviation below 0.1 in their scores, and just 0.38\% of them have a standard deviation above 0.25 (see \autoref{fig: anomaly_variance}). In this subsection, we investigate these high-variance objects.

In \autoref{fig: avgWeird_std}, we plot the average anomaly score against the score's standard deviation for each object. Symbols are colour-coded by the number of observations. Most objects populate a relatively dense region below the polynomial dashed line, with those above the dashed line considered "high variance objects" in this work. In total, we find a population of 1789 high variance objects, with the most extreme variances ($\sigma_{\rm{Anomaly}}$) reaching 0.35. The population generally shows a higher variance in the intermediate region of anomaly scores (0.6-0.9), possibly due to the reduced population in this area compared to the extremes of the anomaly range. We now investigate the differences in variability in the high variance population to the outliers discussed in \S~\ref{sec: types_of_object}.

\begin{figure}
    \includegraphics[width=\columnwidth]{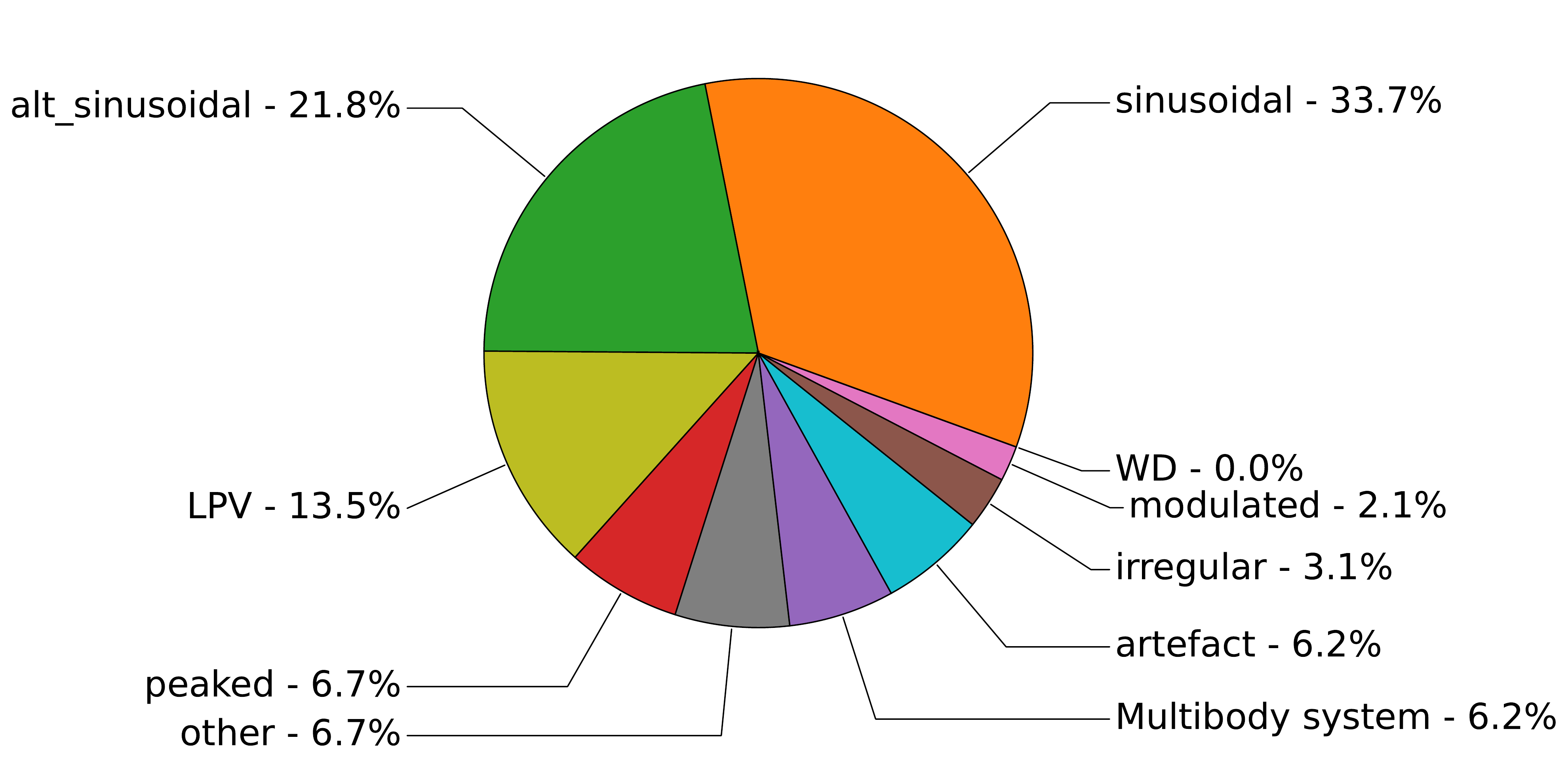}
    \caption{Distribution of descriptions for objects with anomaly score above 0.9 in the labelled sector and a standard deviation that lies above the curve in Figure \ref{fig: avgWeird_std}. The figure includes objects from sectors 13 - 21 \& 24 from the classifications in Figure \ref{fig: weirdest_pie}.}
    \label{fig: high_stdev_pie}
\end{figure}

\begin{figure}
    \includegraphics[width=\columnwidth]{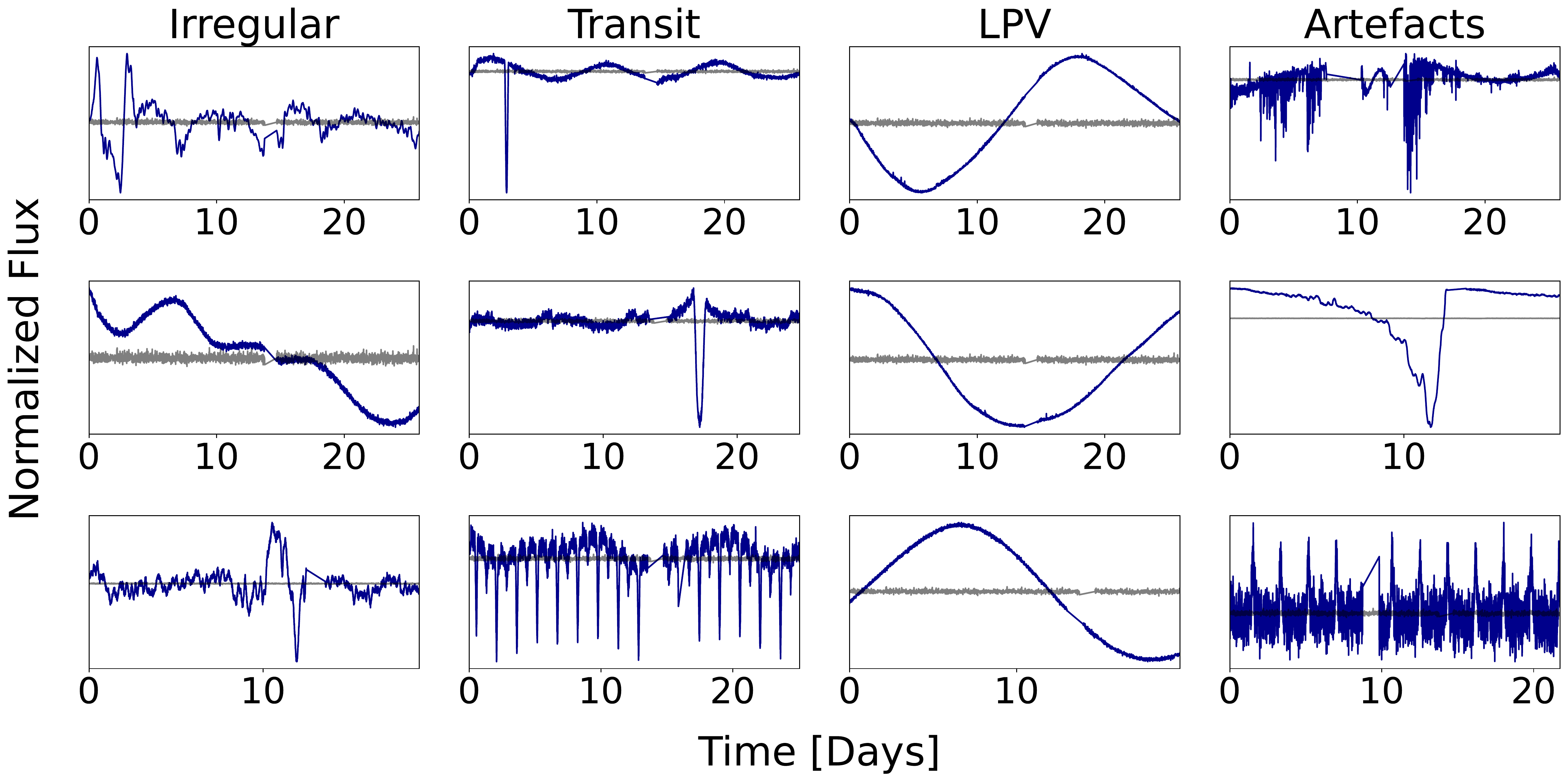}
    \caption{Examples of the most interesting object types within the high standard deviation population. From left to right, we see examples of \textit{irregular} light curves showing peculiar patterns. The second column shows transit events, with the top two examples showing periods exceeding the length of observations. The third column shows examples of long-period variables. Finally, we have a few examples of typical artefacts, including trends, frequent dips and calibration star variability.}
    \label{fig: high_stdev_examples}
\end{figure}

We concentrate here on objects with high variance for which at least one of the individual scores is higher than 0.9. We identify 636 such light curves, corresponding to 410 unique objects. We examine the recorded classifications, similarly to the approach in \S~\ref{sec: objects_found}, and inspect their light curves to evaluate their variability properties. Compared to the total outlier population, we find a greater fraction of these objects classified as "other" in the population of non-persistent outliers. Sinusoidal types still dominate, but there is also an increased fraction of "artefacts" and "Long Period Variables". The fraction of objects with multiple transits remains approximately the same. Remarkably, white dwarfs are not represented in this group at all. \autoref{fig: high_stdev_pie} visualises the breakdown of classes identified within the high variance population.

Further investigation reveals that the sinusoidal light curves found in this group have longer periods when compared to the sinusoidal anomalies in the general outlier group, with practically no light curve showing periods shorter than one day. Perhaps the most interesting objects in this group belong to the "irregular" and "multi-body" system categories. In \autoref{fig: high_stdev_examples}, we show examples of light curves in this group. The upper left panel shows an irregular light curve corresponding to high proper motion star TIC 370327409, which has been classified as a binary \citep{Eggleton2008} and has spectral type K2. The lower left light curve corresponds to TIC 156462093) and shows a light curve similar to those of heartbeat stars previously found by \emph{Kepler} \citep{Thompson2012}

In the second column of \autoref{fig: high_stdev_examples} we identify two transit events of interest (\textit{Top:} TIC 230125546, \textit{Middle:} TIC 350298314). These are likely binaries with orbital periods long enough to be captured during transit only once during the \emph{TESS}, with the transit visible in the plotted light curves. In fact, both objects are identified in \citet{Prsa2022}, a dedicated search for binary stars. TIC 350298314 is measured to have a period of $47.72\pm3.516$ days, placing it in the top 20 longest periods identified out of 4584 confirmed systems. Because of the limitations on inclination required for a transit to occur at such long orbital periods, it is evident this represents an extreme example of an outlying class within our sample. The identification of this extreme system confirms the effectiveness of exploring non-persistent outliers in the case of transits with large member separation.

Other examples in \autoref{fig: high_stdev_examples} include LPVs and artefacts. We note that in the case of artefacts, there is a subclass of objects that look like repeating, regular flaring events (bottom right panel). We have included them as artefacts because we have determined to be, in the majority of cases, they are the result of a background subtraction in which the background area contained a source with repeating transits. However, gravitational lensing is one potential astrophysical phenomenon that could theoretically cause this behaviour. If a compact, undetected object orbits a star, the resulting lensing pattern can look like the bottom right plot in \autoref{fig: high_stdev_examples}. We have listed all artefacts, including inverted transit candidates, in \autoref{tab: datatable_cont}.

\section{Analysing Evolutionary Stages Across the Observational Hertzsprung-Russel Diagram}
\label{sec: object_class_analysis}

In this subsection, we investigate the relationship between the anomaly score and the physical properties of anomalies of known class. The motivation is that even if the general astrophysical type is known, there can be members of a particular group with outlying physical parameters that could indicate anomalous configurations. Examples include orbital periods that are either too short, too long or very deep transiting events. Below we perform this analysis for \emph{TESS} objects of interest, stellar systems with multiplicity greater than 1, such as eclipsing binaries, pulsating stars, young stellar objects, giants, and white dwarfs.

\subsection{TESS Objects of Interest}
\label{sec: TOI}

\begin{figure}
    \includegraphics[width=\columnwidth]{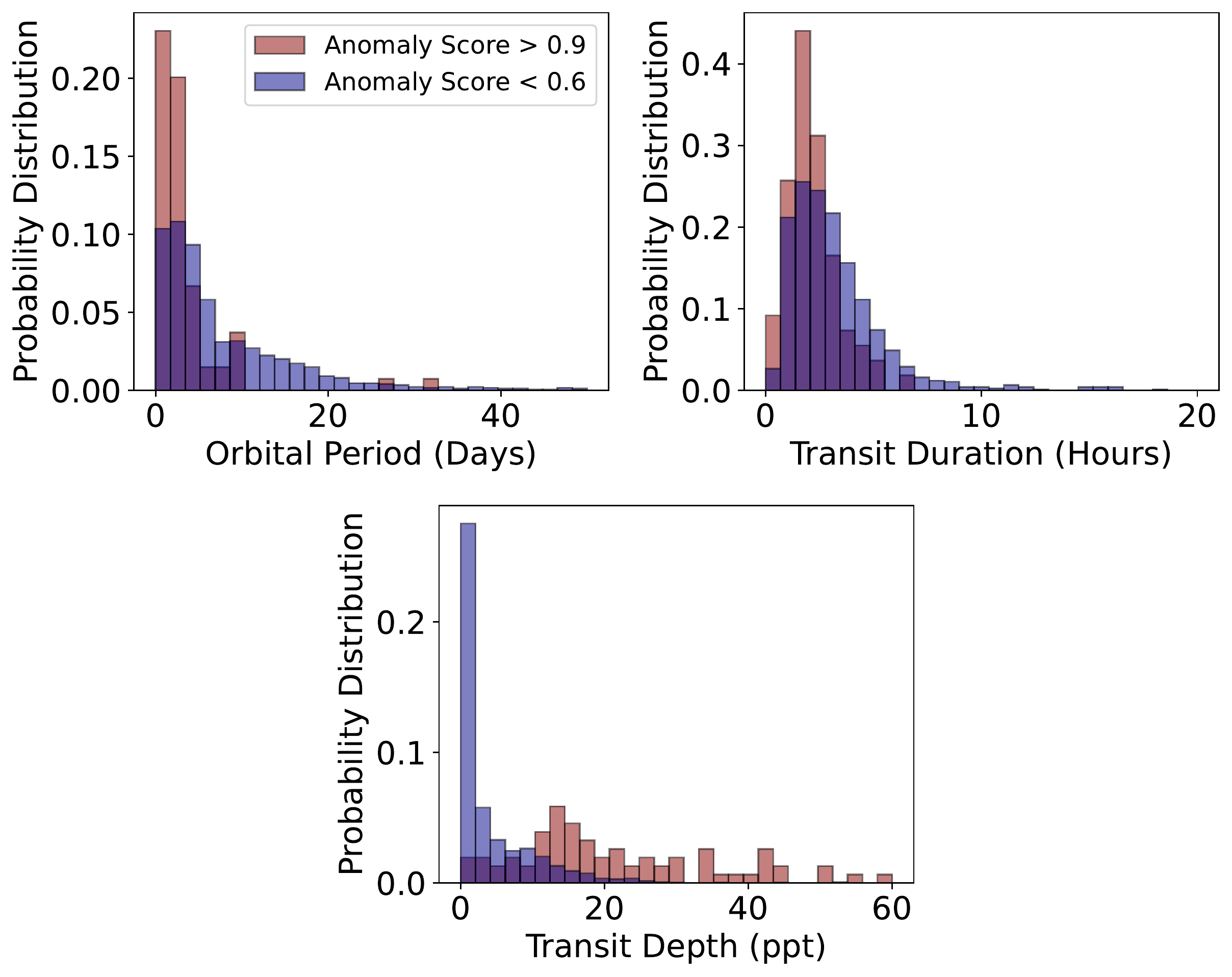}
    \caption{Set of normalised histograms comparing orbital parameters for exoplanet systems from the bulk and outlying populations. \emph{Top Left:} Orbital period in Days, \emph{Top Right:} Transit duration in Hours, \emph{Bottom:} Transit Depth in parts per thousand (ppt).}
    \label{fig: exoplanet}
\end{figure}

The \emph{TESS} mission team regularly releases a list of \textit{TESS objects of interest} (TOI's hereafter) containing the most promising exoplanet candidates. For detailed follow-ups with several transits per object, candidates with estimated orbital periods of less than ten days orbiting bright host stars have priority \citep{Guerrero2021}. At the time of download, the list of TOIs contained 2,647 exoplanet candidates\footnote{\hyperlink{https://tess.mit.edu/toi-releases/}{https://tess.mit.edu/toi-releases/}}. We now turn to the question of whether transiting exoplanets show up as anomalies in our analysis. Since we do not analyse folded or stacked light curves, it is unlikely that this particular anomaly detection method will detect most of the candidates. 

We look at the 1,262 exoplanet candidates from the TOI list that overlap with the sectors analysed in this paper. For TOIs, 87\% are bulk objects, with anomaly scores less than 0.6, $6.3\%$ are anomalous objects, with anomaly scores above 0.9, and the mean anomaly score is 0.44, with a standard deviation of 0.16. As we suspected, these values are not significantly dissimilar to the general population, indicating that exoplanet transits do not dominate the anomaly score of the stellar system in this work. Nevertheless, we have the data to explore how the orbital period, transit duration and transit depth affect the anomaly score. This analysis is relevant in the study of the most prominent exoplanet transits and population studies of eclipsing binary stars.

In \autoref{fig: exoplanet}, we show the distributions of transit parameters for TOIs in two different ranges of anomaly scores: anomalous with an anomaly score above 0.9 and non-anomalous objects with an anomaly score below 0.6. We note that anomalies clearly show deeper transits than the bulk objects, with anomalous transits being close to 15-20 ppt in depth versus $<1$ for the majority of the bulk population. Furthermore, there are indications of a difference in the distribution of orbital periods, with exoplanet anomalies generally having orbital periods shorter than the bulk objects, with the outlier orbital periods primarily remaining below five days. Although in this case, the difference is not as prominent as with the transit depths.

In the upper panel of \autoref{fig: binary_hist}, we compare the anomaly score distributions for TOIs and eclipsing binaries. We note that, at just $6.3\%$, anomalies represent a smaller fraction of the total in the case of TOIs than binary stars, for which the vast majority of light curves are anomalous. The distribution suggests that dim exoplanet transits are best inferred from phased, stacked light curves and are not picked up as anomalies when individual, unphased light curves are considered. Overall, it appears that in transits, regardless of whether they are produced by eclipsing stellar companions or exoplanets, anomalies are selected to be deeper eclipses with somewhat shorter orbital periods. This has implications for the typical exoplanet size. For example, for fixed semi-major axes, inclination and stellar properties for a K4V star, the radius of the exoplanet would need to be about three times larger to create a transit with the average depth of the anomalous objects, with respect to the size of the typical "normal transit" (see https://ccnmtl.github.io/astro-simulations/exoplanet-transit-simulator/). As deeper transits are more anomalous in this work, we conclude transits produced by larger planets will be considered more anomalous than those of smaller planets in the \emph{TESS} dataset.

The main effect of the orbital period on the shape of the light curve is the number of transits that occur during a single observation. At approximately 27 days, the data is optimised for orbital periods less than roughly ten days. The effects would most prominently reflect on harmonics shown in the light curve periodogram for exoplanet transits. On the other hand, the transit duration as a fraction of the orbital period is more informative of the system. For a given size planet, the transit duration depends on multiple factors including inclination and orbital distance. Yet, in the upper right panel of \autoref{fig:  exoplanet}, there is not a major difference observed in the distribution of transit durations between the anomalous and non-anomalous populations. This may be due to the selected features or, most likely a strong selection effect in the orbital parameter space of TIC objects.

\subsection{Multi-Body Systems}
\label{sec: binaries}

\begin{figure}
    \includegraphics[width=\columnwidth]{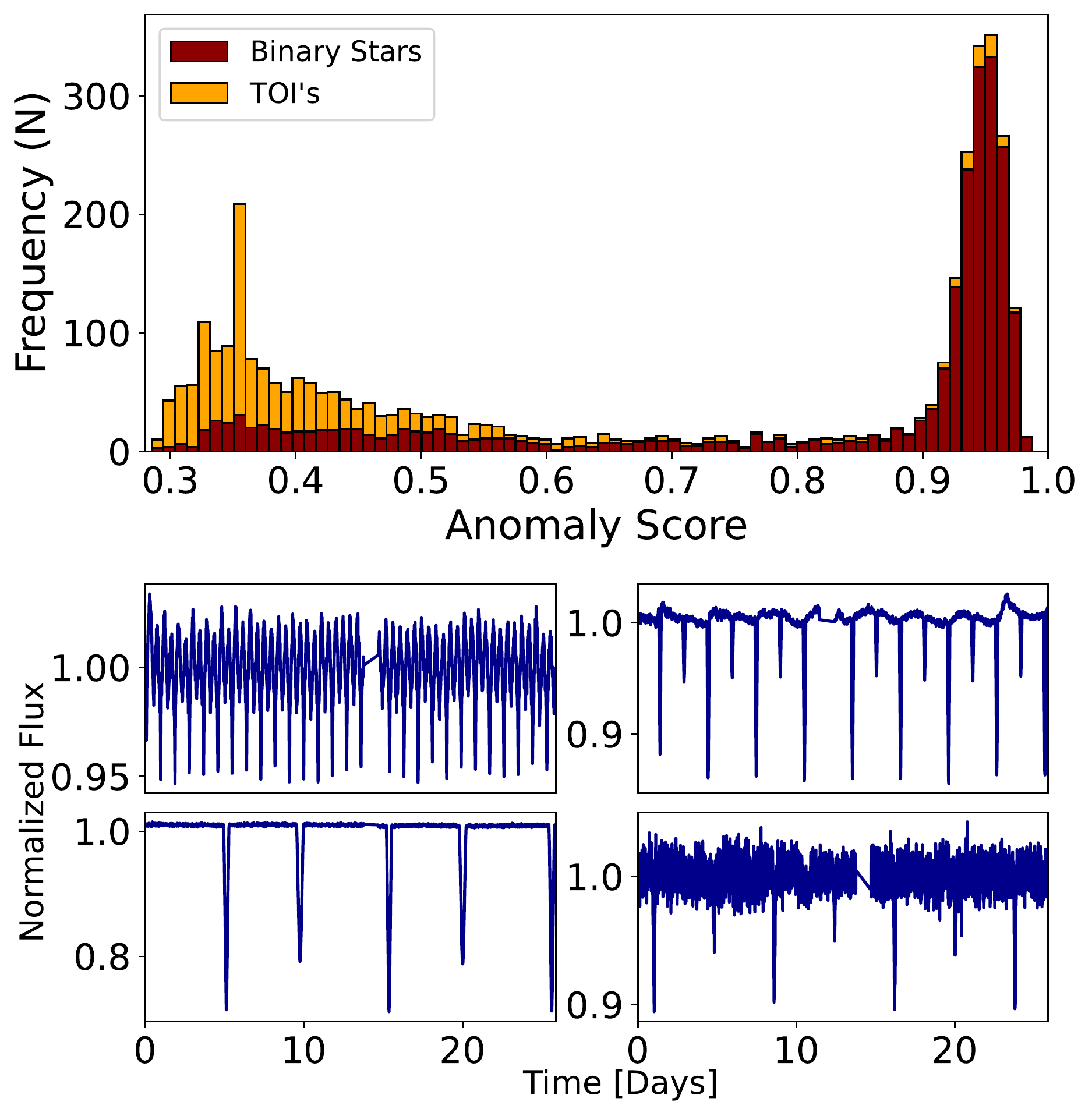}
    \caption{\textit{Top Panel:} Stacked histogram of average anomaly scores for Binary stars (red) compared to TOI's (yellow). These distributions highlight the effect of different transit depths.\textit{Bottom four panels:} Example light curves of binary systems showing an anomaly score above 0.9, showcasing the range of transit depths and periodicity of the anomalous systems.}
    \label{fig: binary_hist}
\end{figure}

\begin{figure}
    \includegraphics[width=\columnwidth]{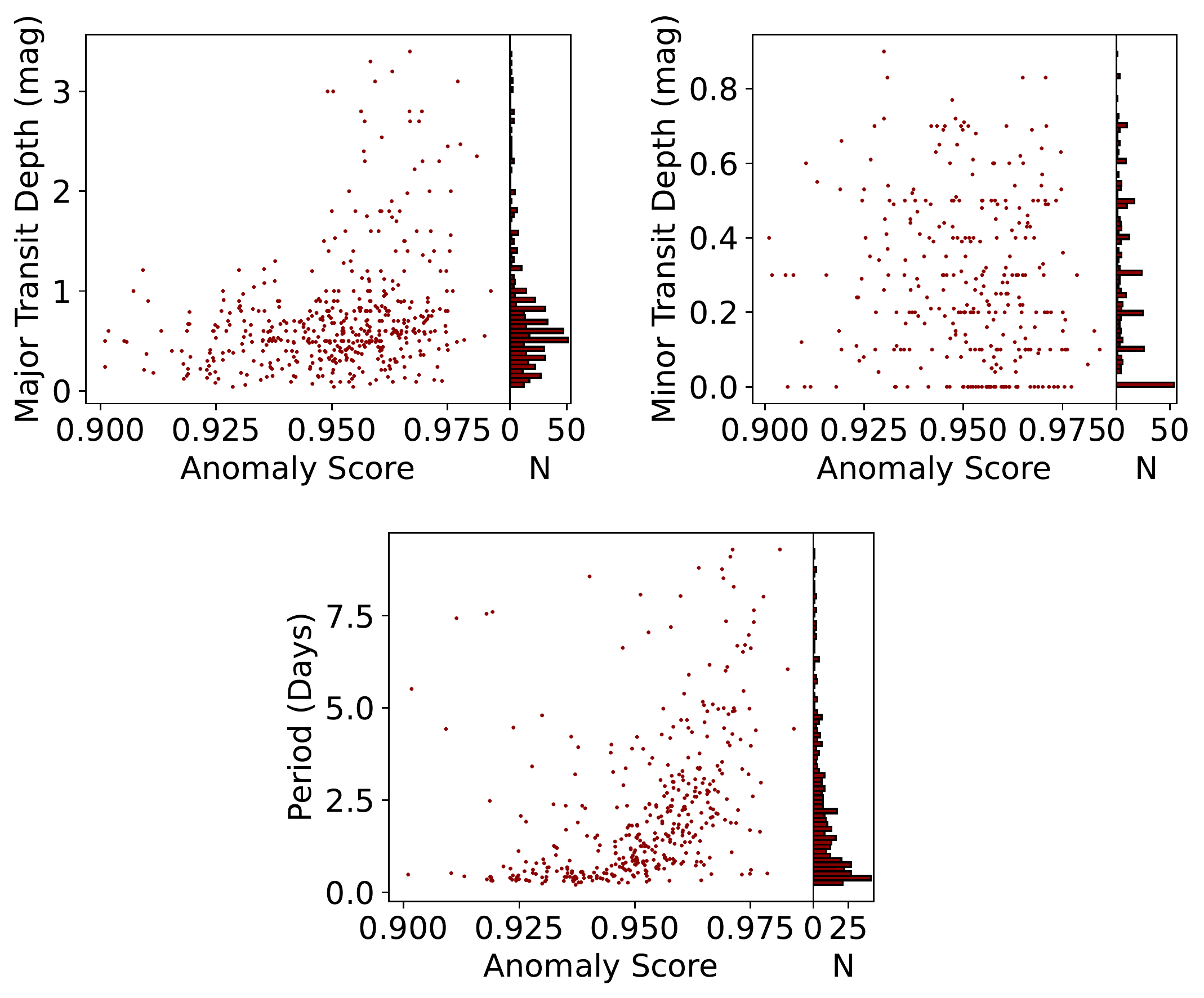}
    \caption{Representation of candidate features which drive anomaly scores in binary systems. \textit{Top Left:} Anomaly score against primary transit depth, \textit{Top Right:} Anomaly score against secondary transit depth, \textit{Bottom:} Anomaly score against orbital period zoomed into the area with correlation, with the histogram of binary periods shown to the right.}
    \label{fig: binary_scatter}
\end{figure}

We call \emph{multi-body systems} \emph{TESS} sources whose light curves show regular dips that can be associated with eclipses. They are predominantly eclipsing binaries or single exoplanet transits, but systems with multiplicity greater than two that show more than one frequency in the eclipses are also included in this category. Between $20 - 80\%$ of all stars are predicted to have companions, with the actual fraction depending on the spectral type \citep{Traven2020, Duchene2013}. However, to show up as eclipsing light curves in \emph{TESS} data, the orbital inclination and periods must be within specific narrow ranges. As a result, eclipsing light curves are rare, accounting for about 1\% of the entire dataset. This results in the majority of them (67\%) being identified as anomalies by our method.

In \S~\ref{sec: TOI}, we have reported a correlation between the depth of the transit in exoplanet light curves and the anomaly score. We now investigate if this correlation extends to eclipsing binary stars. The top panel of \autoref{fig: binary_hist} shows that eclipsing binaries are labelled as anomalous at a significantly higher rate than exoplanets, with the vast majority of binaries having scores higher than 0.9. Presumably, this is because they result in deeper transits. In \autoref{fig: binary_scatter}, we compare the anomaly score to key transit parameters from \cite{Avvakumova2013}: primary transit depth, secondary transit depth and period. 

The results indicate that the primary transit depth is a driver of the anomaly score. Practically all the transits with depths larger than 1.5 mag have anomaly scores higher than $0.95$ as opposed to shallower primary transits, for which none has an anomaly score $< 0.95$. Secondary transit depths, on the other hand, have no discernible effect on the anomaly score, likely because of their shallower profile. A much better predictor of the anomaly score is the binary orbital period, with longer periods having a clear tendency to be more anomalous. The fact that longer orbital periods that result in fewer transits during the \emph{TESS} observation are rare in the TIC is a selection effect due to the mission specifications and not necessarily an astrophysical trend \citep{Prsa2022}. Nevertheless, our results imply that deep transits observed fewer times during a \emph{TESS} light curve are among the rarest variability type in this particular dataset. Our catalogue of anomalies contains previously unidentified objects of this rare class. In the lower panel of \autoref{fig: binary_hist}, we show examples of binary transits with various orbital periods and transit depths.

The anomaly score can therefore be used as a probe of the physical parameters of binary and eclipsing systems, and in general, as a detector of these transits in a \emph{TESS}-like dataset. In our catalogue of anomalies, \autoref{tab: datatable_cont}, these systems are labelled in column 12 as either \textit{"multi-body system"}, or \textit{"Transit"}.

\subsection{Pulsating Variable Stars}
\label{sec: instability_strip}

Pulsating variables show periodic intrinsic variability that result from the expansion and contraction of their surface layers as they transition out of the main sequence. The specifics of those oscillatory modes, as well as their periods and amplitudes, depending on the underlying physics. They occupy a specific region of the CAMD, known as the instability strip. \citet{Gautschy96} present a review of stellar pulsations along the CAMD, emphasising that pulsations can occur at any mass and in various evolutionary stages. Cepheids and RR-Lyrae stars are among the most represented types within this class, of which \citet{Clementini2019} identify 140,874 Cepheids and 9,575 RR-Lyrae stars in Gaia data. These objects represent less than 0.01\% of the entire Gaia DR2 catalogue. To assess what physical mechanisms drive the anomaly score in pulsating stars, we rely on the independently determined classifications documented in SIMBAD. The resulting catalogue of known types of pulsating stars contains 156 RR-Lyrae stars, 468 Cepheid variables, and an additional 887 $\delta$-Scuti and $\gamma$-Dor stars. Also included are all other stellar types whose labels contain \textit{Pulsating Variable Star}, even if the specific types are unknown.

\begin{figure}
    \includegraphics[width=\columnwidth]{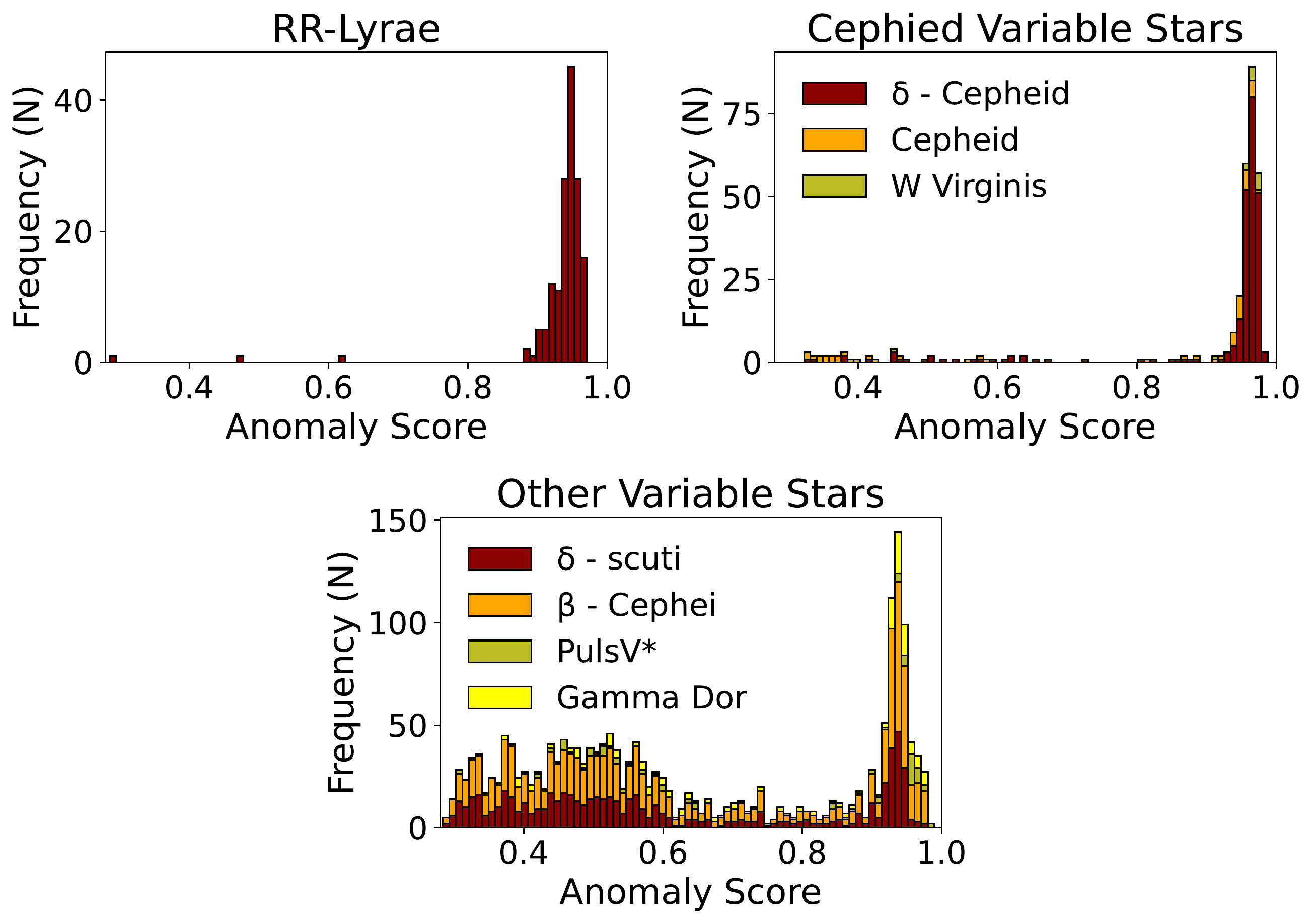}
    \caption{\emph{Top Left:} Anomaly score of RR-Lyrae stars across all sectors. \emph{Top Right:} Anomaly score of Cepheid Variable Stars in all sectors. \emph{Bottom:} Anomaly score of other variable stars across all sectors. (e.g. $\delta$-Scuti type variable.)}
    \label{fig: pulsating_hist}
\end{figure}

\autoref{fig: pulsating_hist} shows the distribution of anomaly scores for the most prominent pulsating types. The top left panel shows the distribution of scores for RR-Lyrae stars. 96\% of the members of this class have an anomaly score greater than 0.9, with 98.1\% of them having a score above 0.8. The classification is due to the RR-Lyrae variability pattern being uncommon in the TIC with a highly variable light curve, meaning our method will flag RR-Lyrae stars by assigning them a high anomaly score. Their pulsation periods range from $\sim$0.2 days to 1 day \citep{Soszynski2014, Soszynski2019}, which corresponds to the range of frequencies extracted for the periodogram features. The few RR Lyrae stars with an anomaly score below 0.8 lack the expected RR-Lyrae pattern in their light curves, indicating a potential data issue most likely caused by either an incorrect classification or a cross-match error between catalogues. In the context of automatic source classification, the fact that false classifications can be identified based on their outlying anomaly score can be used to improve existing training sets, increasing their purity.

The top right panel of \autoref{fig: pulsating_hist} shows the anomaly score distribution for Cepheid variables, split into the following sub-classes: $\delta$-Cepheid, W-Virginis and those classified simply as Cepheids. $\delta$-type and W-Virginis types are identified as anomalies in about 90\% of the cases. On the other hand, those classified as Cepheids are identified as anomalies only $\sim50\%$ of the time. As we had discussed previously (see \autoref{fig: periodograms_comparison}), anomalies show more structure in their periodograms, with clearly indicated peaks in the power spectrum. In particular, this holds for pulsating stars, with different types having unique characteristic frequencies. Cepheid stars with comparatively lower anomaly scores do not show a distinct spectral signature in the periodogram or higher power contained in long timescale ($>1$~day) variability. Cepheids have periods that can exceed 100 days, whereas $delta$ and W-Virginis periods are not known to exceed $\sim20-30$ days \citep{Soszynski2017}, and this can be another reason why a smaller fraction of Cepheids are detected as anomalous since their periods are outside the range of frequencies probed here. Nevertheless, the fraction of anomalous Cepheid variables ($\sim50$\%) remains significantly higher than the fraction of stars identified as anomalous in the entire catalogue ($\sim10$\%).

\begin{figure}
    \includegraphics[width=\columnwidth]{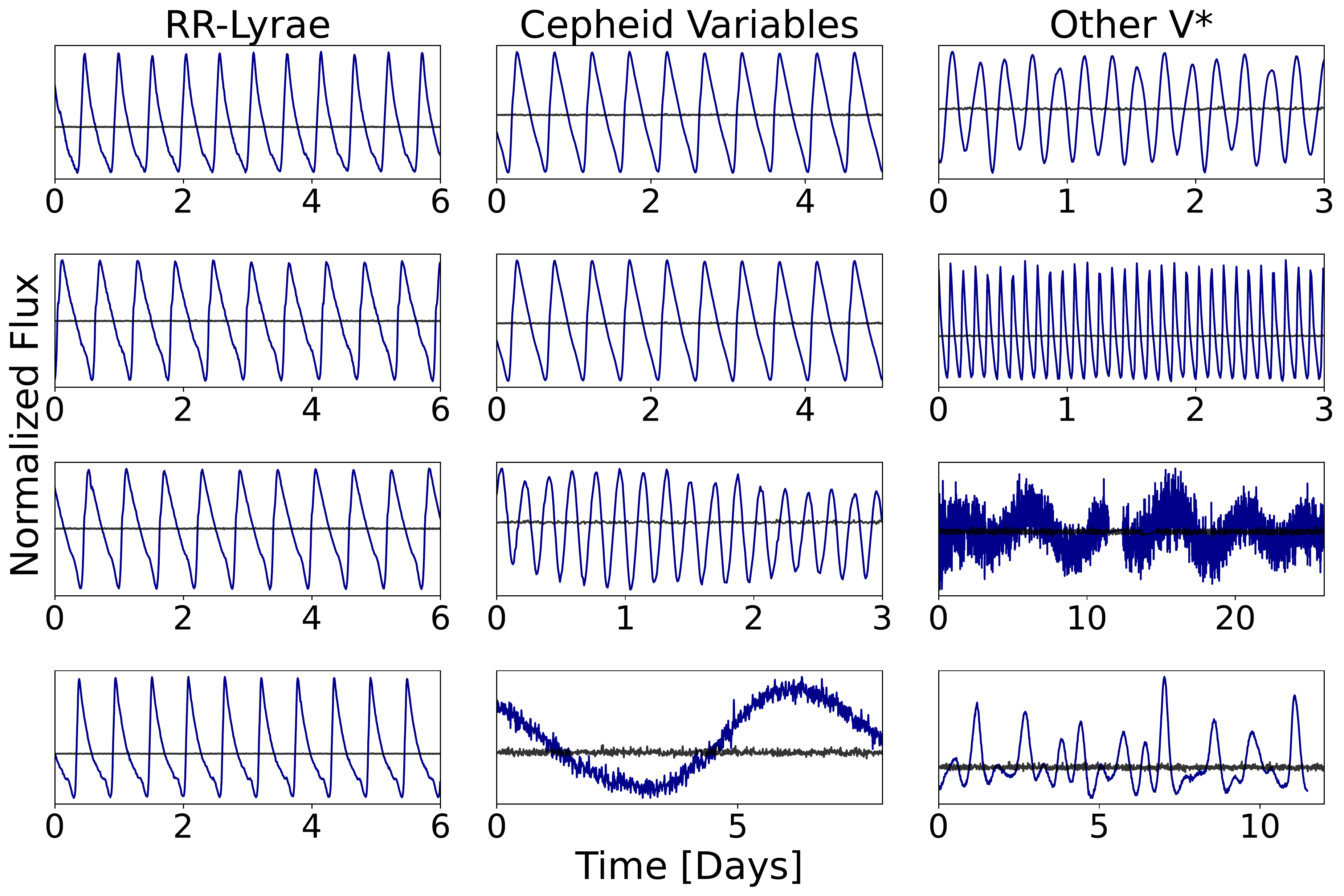}
    \caption{Visualisation of objects identified as outliers in the pulsating type stars with their distributions found in \autoref{fig: pulsating_hist}. Target light curves are shown in blue, with the black light curve representing a non-anomalous light curve, TIC 261337074, for reference on magnitude variations. From the top down, other V* types are: delta-Scuti, delta-Scuti, Pulsating Variable and Gamma Dor type Variables.}
    \label{fig: pulsating_LCs}
\end{figure}

The bottom panel of \autoref{fig: pulsating_hist} shows the distribution of anomaly scores for the remaining types of known pulsators, including the ambiguous \textit{pulsating variable} class\footnote{Classifications defined by the \href{https://simbad.u-strasbg.fr/simbad/sim-display?data=otypes}{SIMBAD Astronomical Database}}. Compared to RR-Lyrae and Cepheids, a significantly smaller fraction of these objects corresponds to anomalous light curves. For $\delta$-Scuti stars, the fraction is 24.2\%, whereas for plain pulsating stars, $\beta$-Cephei and $\gamma$-Dor stars the fractions are respectively 21.3\%, 42.7\%, and 45.2\%. For $\delta$-Scuti and $\beta$-Cephei stars, this is somewhat due to their short pulsation periods, ranging from 0.02 to 0.25 days \citep{Ziaali2019}, placing the majority of these objects outside the range of frequencies probed by our periodogram features. $\gamma$-Dor stars show multiple oscillation modes with periods ranging between 0.5 and 3 days, resulting in the multi-modal pattern seen in the bottom right panel of \autoref{fig: pulsating_LCs} \citep{Pietrukowicz2020, Tkachenko2013}. The pulsating stars without a specific class have a distribution of anomaly scores similar to the distribution for $\delta$-Scuti stars, suggesting a similar distribution of their frequency properties. The periodograms of those with anomaly scores below 0.7 show little to no structure across the entire range and therefore have no discernible oscillatory modes within the range of periods probed here (4 hours to 27 days). While a verification by anomaly score can not be used \emph{in lieu} of a classification for these objects, it at least suggests that unclassified variable pulsating objects with anomaly scores below 0.9 are more likely to be similar to $\delta$-Scuti or $\gamma$-Dor than RR-Lyrae or Cepheid variables.

We conclude that periodic pulsations are a driving factor behind the anomalous nature of \emph{TESS} light curves when both the light curves and the power spectrum are considered together. A similar behaviour is observed for the \emph{Kepler} light curves in MG21, where we also found pulsations to be among the main drivers of anomaly score. However, a smaller fraction of $\delta$-Scuti stars are anomalous in \emph{TESS} than in \emph{Kepler}, where practically all $\delta$-Scuti stars were anomalous. The discrepancy between the two datasets is most likely due to a difference in the parent population of stars in both surveys, with M dwarfs being much more represented in both the general population of \emph{TESS} targets and the group of anomalous objects. The anomaly score alone is not enough to distinguish between anomalous pulsators. The typical $\kappa$-mechanism driven stars show a similar distribution of anomaly scores to the non-radial, gravity mode pulsations in $\gamma$-Dor stars \cite{Kaye1999}. In \autoref{tab: datatable_cont}, the majority of these pulsating sources are noted as \textit{sinusoidal}, with additional notes including \textit{asymmetrical} or \textit{rapid} depending on the specific shape. $\gamma$-Dor stars show a distinct pattern predominantly noted as \textit{irregular peaked}.

\subsection{Young Stellar Objects}
\label{sec: YSO}

\begin{figure}
    \includegraphics[width=\columnwidth]{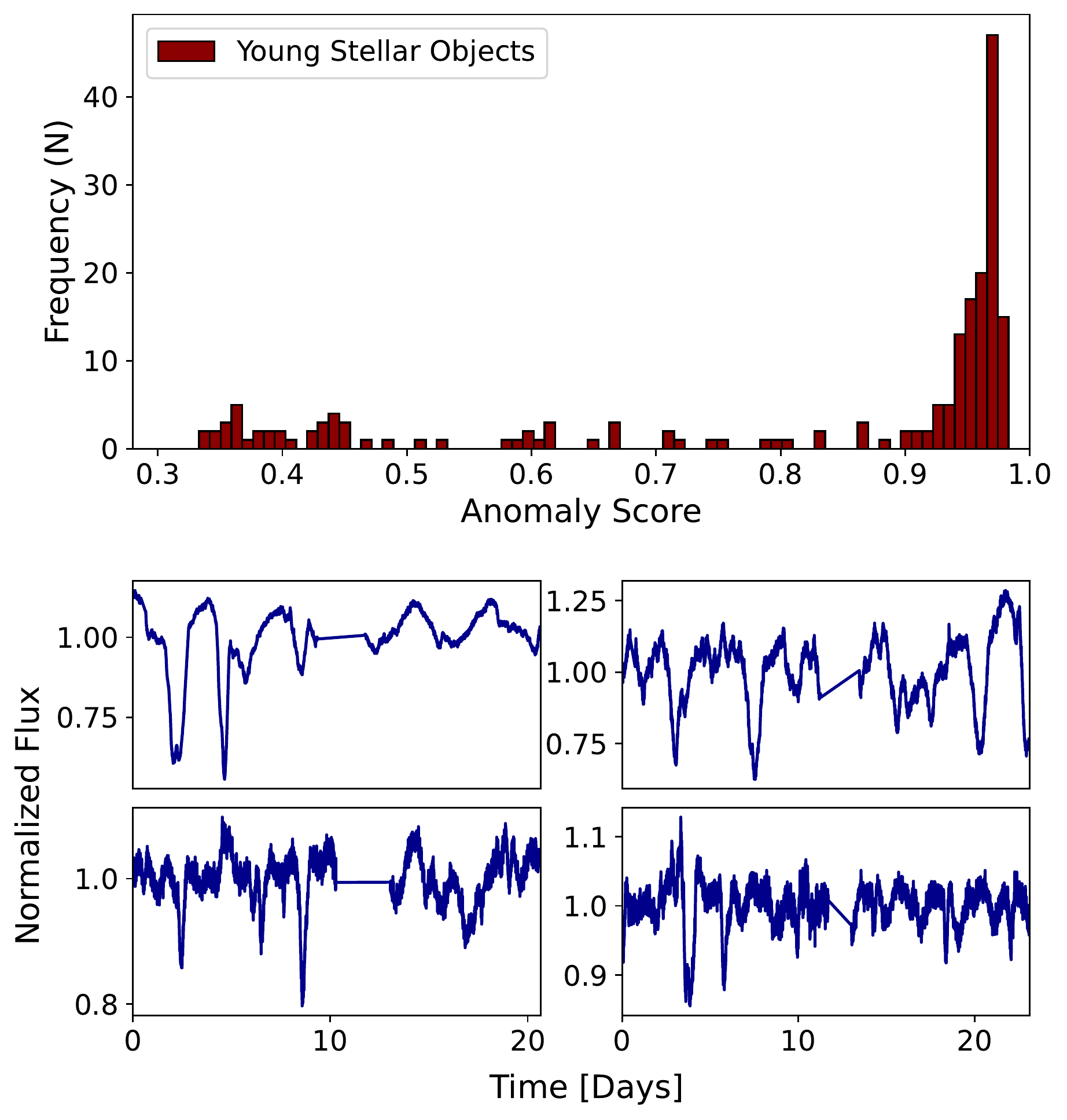}
    \caption{\textit{Top Panel:} Histogram of anomaly scores for young stellar objects. Note that the scores are from individual sectors. \textit{Bottom four panels:} Example light curves of young stellar objects showing a weirdness score above 0.9.}
    \label{fig: example_YSOs}
\end{figure}

Young stellar objects (YSOs) are stars in their earliest evolutionary stage, post-cloud collapse, but before they enter the Main Sequence. The youngest protostars are within a thick envelope of gas and dust arising from the original birth cloud, and the resulting obscuration manifests in the light curve of these young stars as an irregular variability pattern. As the primordial cloud dissipates, magnetic activity intensifies, producing extreme stellar flares as dissipated magnetic energy converts into kinetic energy. These flares often interact with the remaining circumstellar disk. In addition, a non-uniform distribution of starspots \citep{Bhardwaj2019} can cover up to 40\% of their surface \citep{Herbst1994, Kesseli2016, Guo2021}. These conditions all add to the stochasticity of the light curves of young stars. Furthermore, their relatively short rotation periods (typically 1 to a few 10's of days \citep{Froebrich2021}) allow semi-periodic signals to be present.

The top panel of \autoref{fig: example_YSOs} presents the distribution of anomaly scores for previously known YSOs in our dataset. We note that YSO light curves tend to rank highly in the anomaly score distribution, with most objects making the 0.9 cut considered an anomaly. The four lower panels of \autoref{fig: example_YSOs} show representative YSO light curves. The irregular pattern, with relatively long-term baseline variability, as well as the presence of dips are likely contributors to the anomaly score. In fact, among the types of light curves evaluated as part of this work, the YSO light curves are the closes analogues of a remarkable light curve, namely Boyajian's star. While most of the objects classified in our nomenclature as 'irregular' probably are unclassified YSOs, we believe that if any true analogues of Boyajian's star have been detected, they would belong to this group.

Out of a total of 190 YSOs, 67\% have anomaly scores above 0.9 and 81\% above 0.6. The remaining objects, which are part of the bulk population, show significantly less structure in their light curves. As low-mass YSOs evolve towards the main sequence, both the accretion rate and the size of the disk reduce, causing less variability observed in the corresponding light curve (Chapter 6 in \cite{Schulz2005} provides a detailed explanation of YSO evolution). Hence, the reduction in variability observed in the YSOs in the bulk population indicates a more evolved system than their anomalous counterparts, with less obscuration, accretion, or magnetism-related variability imprinted in their light curves. Furthermore, when studying a CAMD, these \emph{normal} YSOs also align closer to the main sequence than the anomalous YSOs, further indicating they correspond to a more evolved system. These observed young stars are expected to be of a low mass, as the timescales and radiative feedback in high-mass young stars generally prevent the formation of a long-lasting accretion disk \citep{Shepherd2003}. Therefore, our method consistently flags the irregular light curves that result from a combination of accretion activity and obscuration corresponding to young low-mass stars as anomalies. These are the closest analogues we find to the unique behaviour shown by Boyajian's star.

In \autoref{tab: datatable_cont}, these systems are predominantly classified as \textit{Irregular Dips}, with those previously classified as YSOs marked as such in the \textit{main\_type} column.

\subsection{Giants}
\label{sec: giants}

\begin{figure}
    \includegraphics[width=\columnwidth]{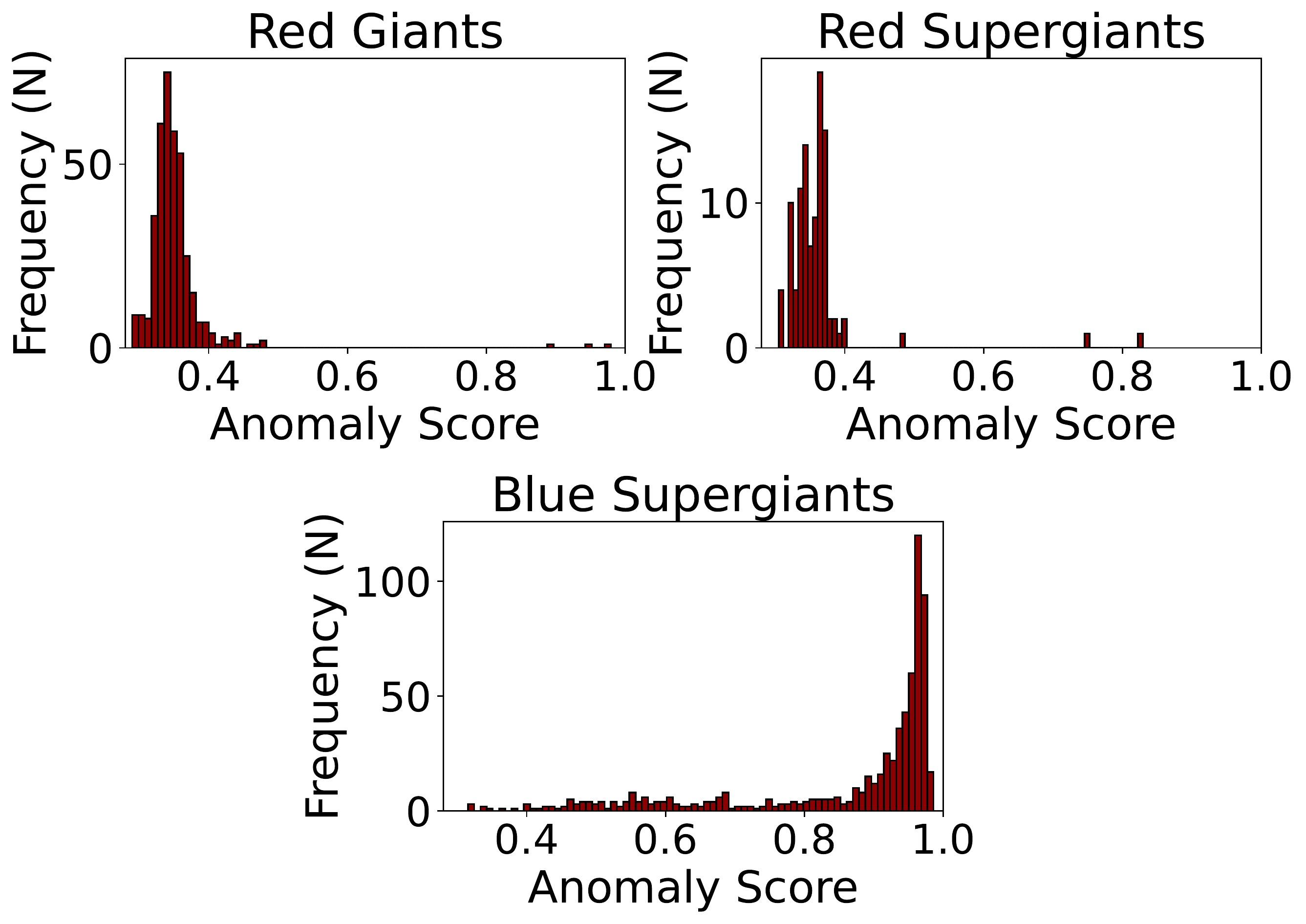}
    \caption{Histograms of anomaly scores for different stages of giant stars. Note the dominant spike for the class in all cases and a small number of outliers in each plot, demonstrating an alternative approach to identifying outliers than simply a high anomaly score.}
    \label{fig: giants_hist}
\end{figure}

\begin{figure}
    \includegraphics[width=\columnwidth]{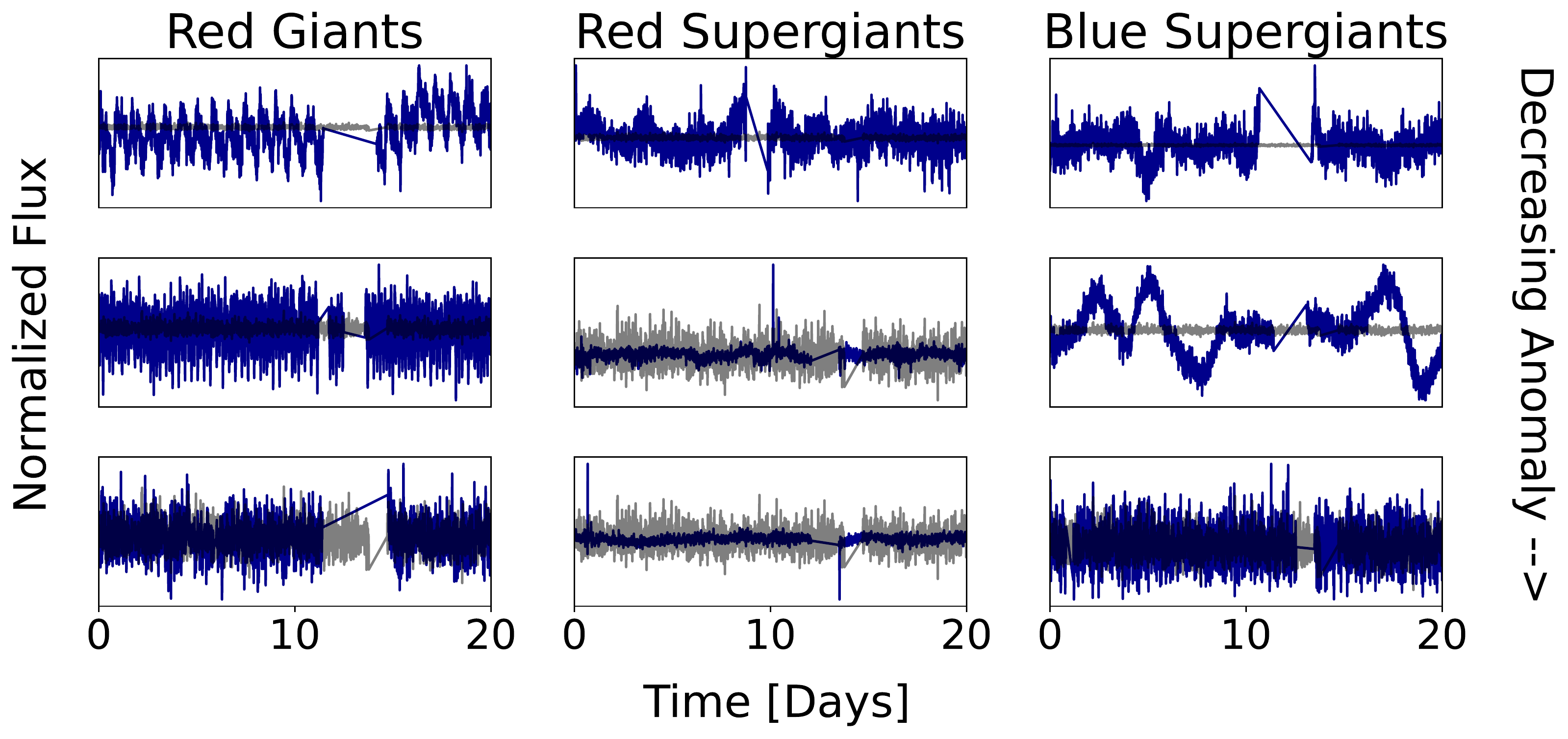}
    \caption{Example light curves from giant stars in blue, with a non-anomalous base light curve from TIC 261337074 shown in grey to highlight the magnitude of variations. The light curves cover a range of weirdness scores, with the selection reflecting the distribution seen in \autoref{fig: giants_hist}. From top to bottom, Red giant weirdness = [0.9, 0.4, 0.3], red supergiants weirdness = [0.8, 0.4, 0.3], and blue supergiant weirdness = [0.9, 0.8, 0.3].}
    \label{fig: example_giants}
\end{figure}

Up to this point, we have focused on identifying objects of known classes with anomalous light curves and understanding the underlying mechanisms that drive their classification as anomalies. However, we might miss additional insightful information by solely focusing on this anomalous/non-anomalous dichotomy. For instance, the anomaly score of predominantly non-anomalous classes may reveal extreme examples or variability patterns of interest that do not quite reach the anomalous threshold. Furthermore, the anomaly score may still correlate to physical processes within these classes. In this subsection, we explore the variability properties and distribution of anomaly scores for giant stars, namely the post-main sequence Red Giants and Red Supergiants, as well as the main sequence Blue Supergiants. These giants show mixed behaviour in terms of their mean anomaly score and are, therefore, a good target for understanding the relationship between the score and light curve attributes.

Stars occupying the giant branch are post-main-sequence stars with large surface areas resulting in an increased brightness. We focus here on previously identified red giants, red supergiants and blue supergiant stars. \autoref{fig: giants_hist} shows the distribution of anomaly scores for Giant stars. Red Giants and Red Supergiants are predominantly classified as non-anomalous, whereas the majority of Blue Supergiants have high anomaly scores, with most having a score higher than 0.9. This is also apparent in the CAMD of \autoref{fig: HR_full}, where the RGB stars prominently show low anomaly scores.

Anomalous Blue Supergiants, the majority of stars of this type, are characterised by shorter variability timescales and larger variability amplitudes compared to most red giants (which have low anomaly scores), as shown in the right-hand column of \autoref{fig: example_giants}. They can also show a combination or rapid variability modulated by high amplitude variations in intermediate timescales (centre right panel). As the anomaly score decreases, they start to resemble any other giants with a low score, and the light curve becomes dominated by a constant, low amplitude noisy continuum. These Blue Supergiants are undergoing an unstable stage in their evolution; they are young high-mass stars with high circumstellar dust fractions and low-frequency gravity waves \citep{deWit2014, Bowman2019}. Red Giants and Supergiants, on the other hand, are intrinsically more stable, with variability patterns observed over months to decades dictated by slow pulsation mechanisms \citep{Kiss2006}. Anomalous luminous stars in the TIC are, therefore, likely to be Blue Supergiants.

Starting in late 2019, a Red Supergiant star, Betelgeuse, experienced a well-documented period of instability. This period consisted of a significant dimming spanning several months, leading to several proposed hypotheses to explain the cause of the dimming \citep{Joyce2020, castelvecchi_2020, Levesque2020}. Remarkably, one of the two light curves of red supergiants that appear as anomalous outliers in the distribution shown in the upper right panel in \autoref{fig: giants_hist} corresponds to an observation of Betelgeuse in Sector 6 (the remaining outlier being an artefact) and has an anomaly score of 0.82. Betelgeuse's light curve shows regular increases in brightness, as shown in the middle top panel of \autoref{fig: example_giants}. The observation predates the great dimming event by approximately ten months.

Unfortunately, Betelgeuse is not observed again by \emph{TESS} until Sector 33, which occurs after the dimming event has concluded and the unusual periodic increases in brightness have subsided. No observations with different facilities and similar cadence or sensitivity are available during the time preceding and following the dimming event, and therefore we cannot verify when this anomalous pattern began. We note that the regular increases in luminosity prior to the dimming event suggest an intrinsic variability, which contradicts the dusty veil model presented in \cite{Montarges2021}. Determining if the anomalous behaviour we have identified here is related to the major dimming event would require additional observations and comparisons to evolutionary models, which is outside the scope of this work. It is nonetheless remarkable that we can detect anomalous behaviour in the light curve of Betelgeuse shortly before the major dimming event.


\subsection{White Dwarf Stars}
\label{sec: WDs}

\begin{figure}
    \includegraphics[width=\columnwidth]{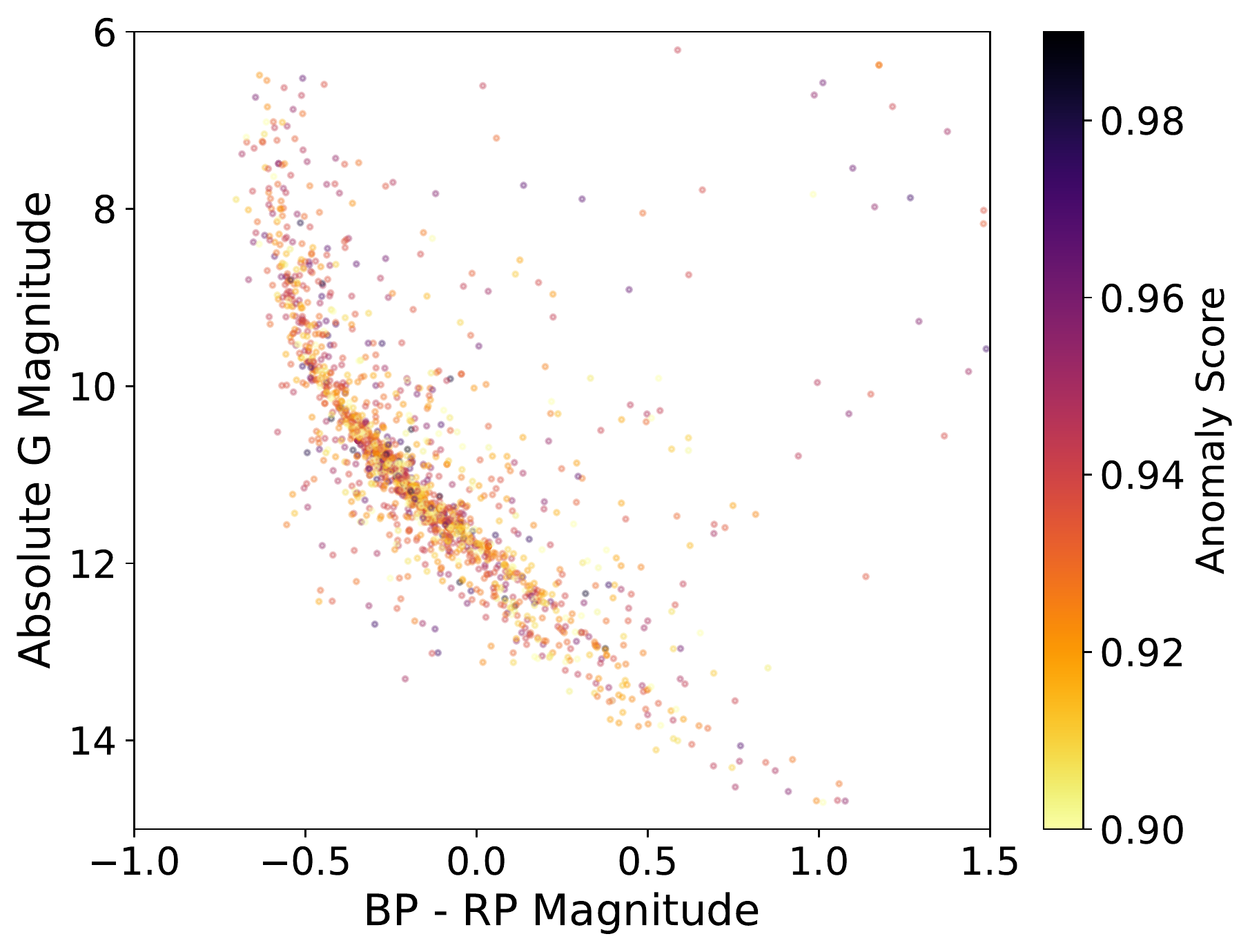}
    \caption{CAMD colour coded by anomaly score, zoomed in on the white dwarf stars. There appears to be little to no correlation along the length of the cooldown track, indicating no correlation between the cooling stage and anomaly score.}
    \label{fig: WD_hr_zoomed}
\end{figure}

\begin{figure}
    \includegraphics[width=\columnwidth]{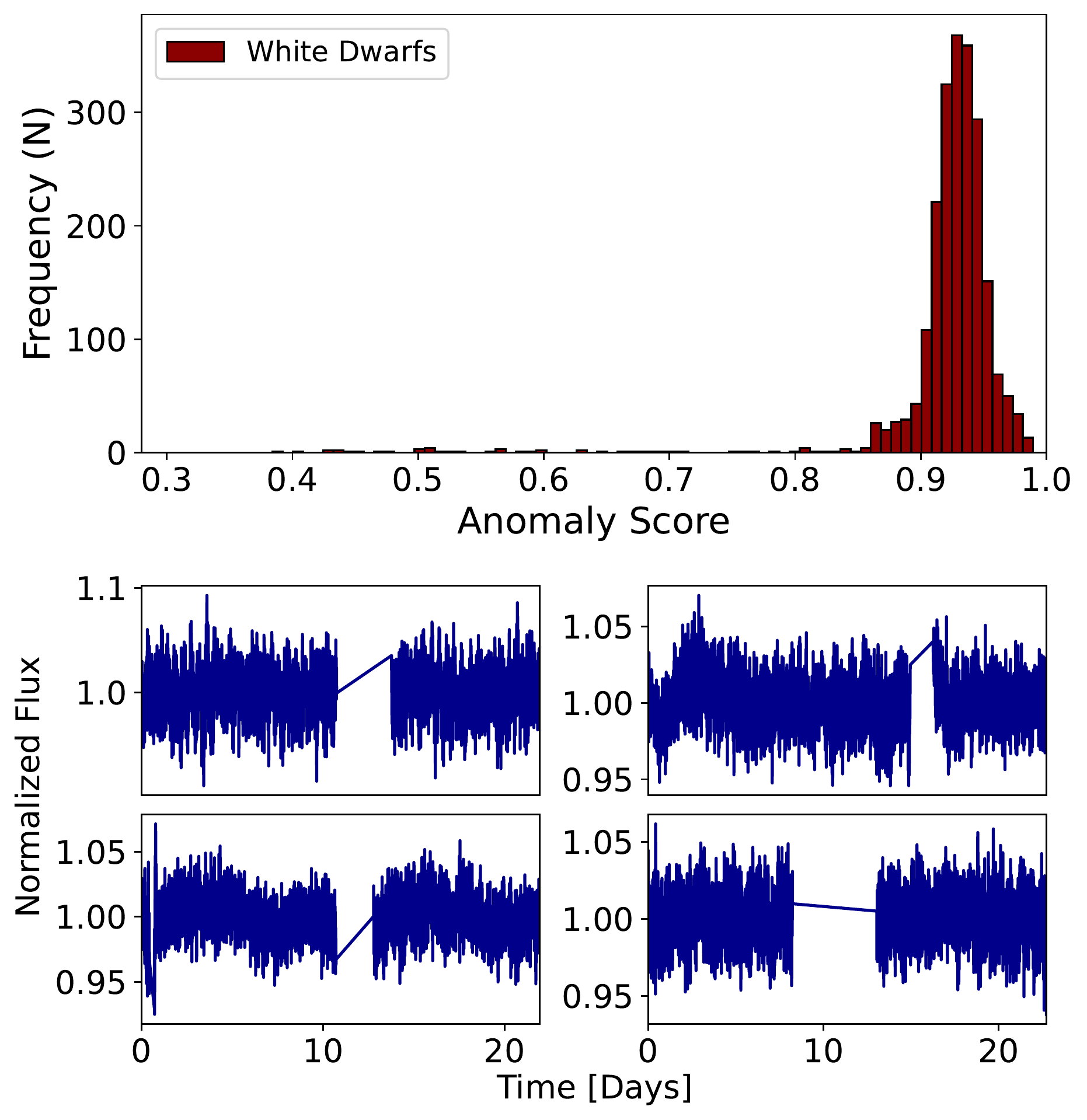}
    \caption{\textit{Top Panel:} Histogram of average anomaly scores for White dwarf stars found in \citet{Fusillo2019}. \textit{Bottom four panels:} Example light curves of white dwarf stars showing an anomaly score above 0.9 despite a lack of visually defining features.}
    \label{fig: example_WDs}
\end{figure}

White dwarfs are one of the individual groups of stars with consistently high anomaly scores. We have cross-matched our list of anomalies against the catalogue of white dwarfs in \citet{Fusillo2019}, identifying 2195 examples with a high ($>0.7$) probability of being a true white dwarf. This population represents less than 1\% of the objects in our dataset. Of these, $90.8\%$ (1995 examples) have an anomaly score above 0.9, and $98.1\%$ (2153 examples) have a score above 0.8. The high prevalence of anomalous light curves among white dwarfs suggests a prevalent physical mechanism driving a unique type of variability among them. 

In \autoref{fig: WD_hr_zoomed}, we show an inset of the CAMD only including the WD branch, colour coded by anomaly scores higher than 0.9. There is no evidence of increasing anomaly with temperature or luminosity. White dwarf stars evolve through cooling, becoming less bright and redder over timescales comparable to the age of the universe. The lack of a correlation along the cooling track suggests that the mechanism causing an anomalous variability is not related to evolutionary changes in surface temperature. Amplitude variations are likely to be part of the reason for the WD anomaly, with variations of up to 10\% in the normalized flux, as shown in the lower panel of \autoref{fig: example_WDs}. These amplitude variations are significantly higher than those observed in MS stars of similar luminosities.

The lower panels of \autoref{fig: example_WDs} show anomalous white dwarf light curves. They reveal a rapid variability pattern of high amplitude and lack of periodicity, which is also confirmed by the periodograms. \cite{Althaus2010} explain in detail the pulsation periods of white dwarf stars, and \cite{Corsico2019} further classifies a range of white dwarf pulsation types. These pulsations typically have short periods spanning 2-100 minutes, with amplitudes between 0.4 mmag and 0.3 mag. The periodogram shows little structure or clear power peaks, which indicates that the power spectrum is not sensitive to these short-period pulsations at the \emph{TESS} cadence. However, the amplitude of the magnitude variations in these relatively dim stars are consistent with the pulsation types described in \cite{Corsico2019}, representing a likely candidate for the mechanism driving high anomaly scores in White Dwarf Stars.

The above discussion and the fact that white dwarfs are not present in the high variance population, described in \S~\ref{sec: high_sigma}, suggests that, at any luminosity or absolute magnitude, our method can identify additional white dwarf candidates. We can define white dwarf candidates as those with a high anomaly score ($>0.9$), BP-RP colour smaller than 1, and do not return a high variance between observations. We also refer the reader to \S~5.8 in \cite{GCNS2021} for an alternative machine-learning method to identify white dwarfs.

\subsection{Modulated light curves}

\begin{figure}
    \includegraphics[width=\columnwidth]{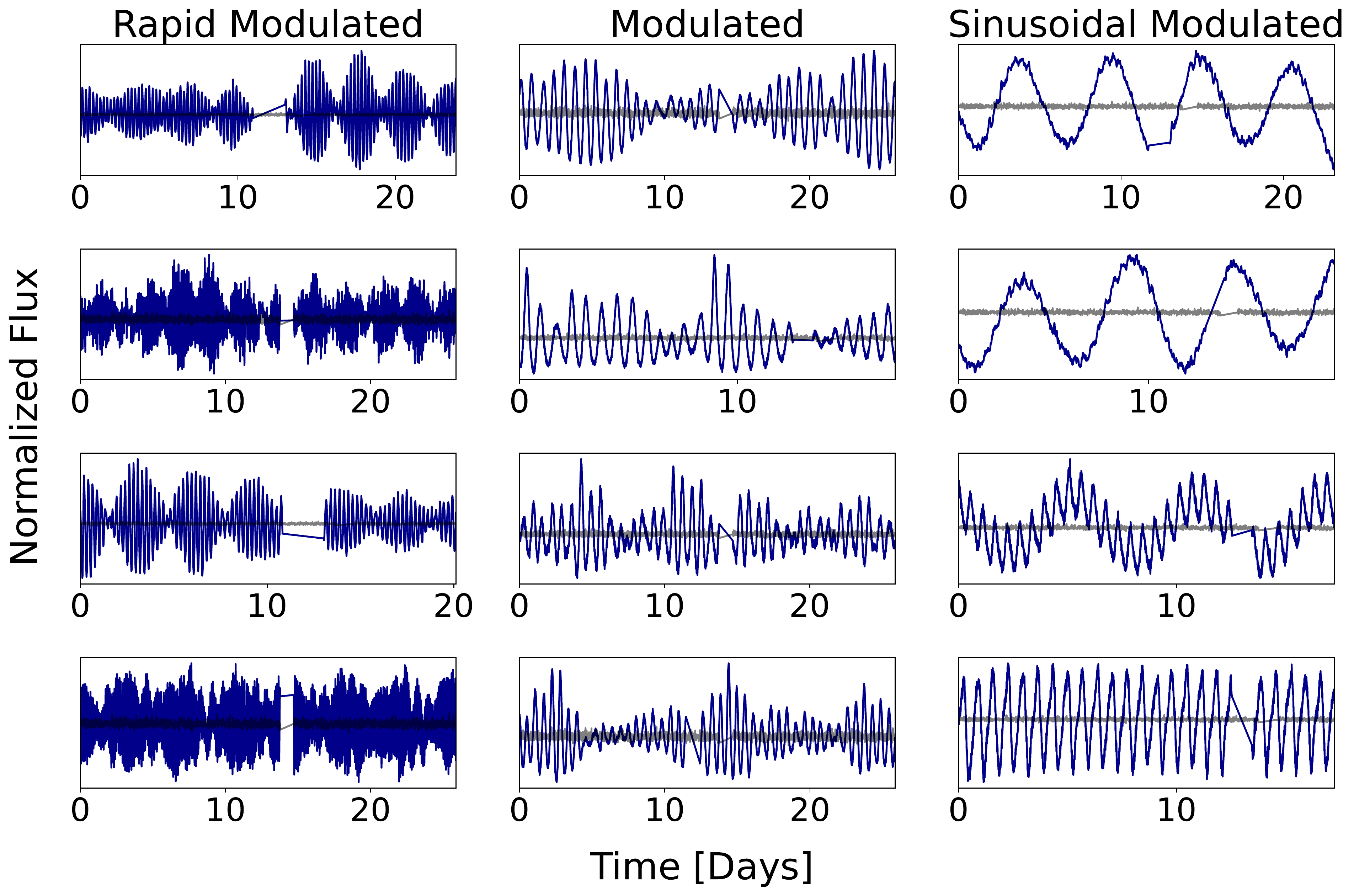}
    \caption{Example light curves of modulated light curves. These show a range of patterns with variations in both frequency and magnitude effects, suggesting multiple mechanisms driving the variability.}
    \label{fig: example_modulated}
\end{figure}

\begin{figure}
    \includegraphics[width=\columnwidth]{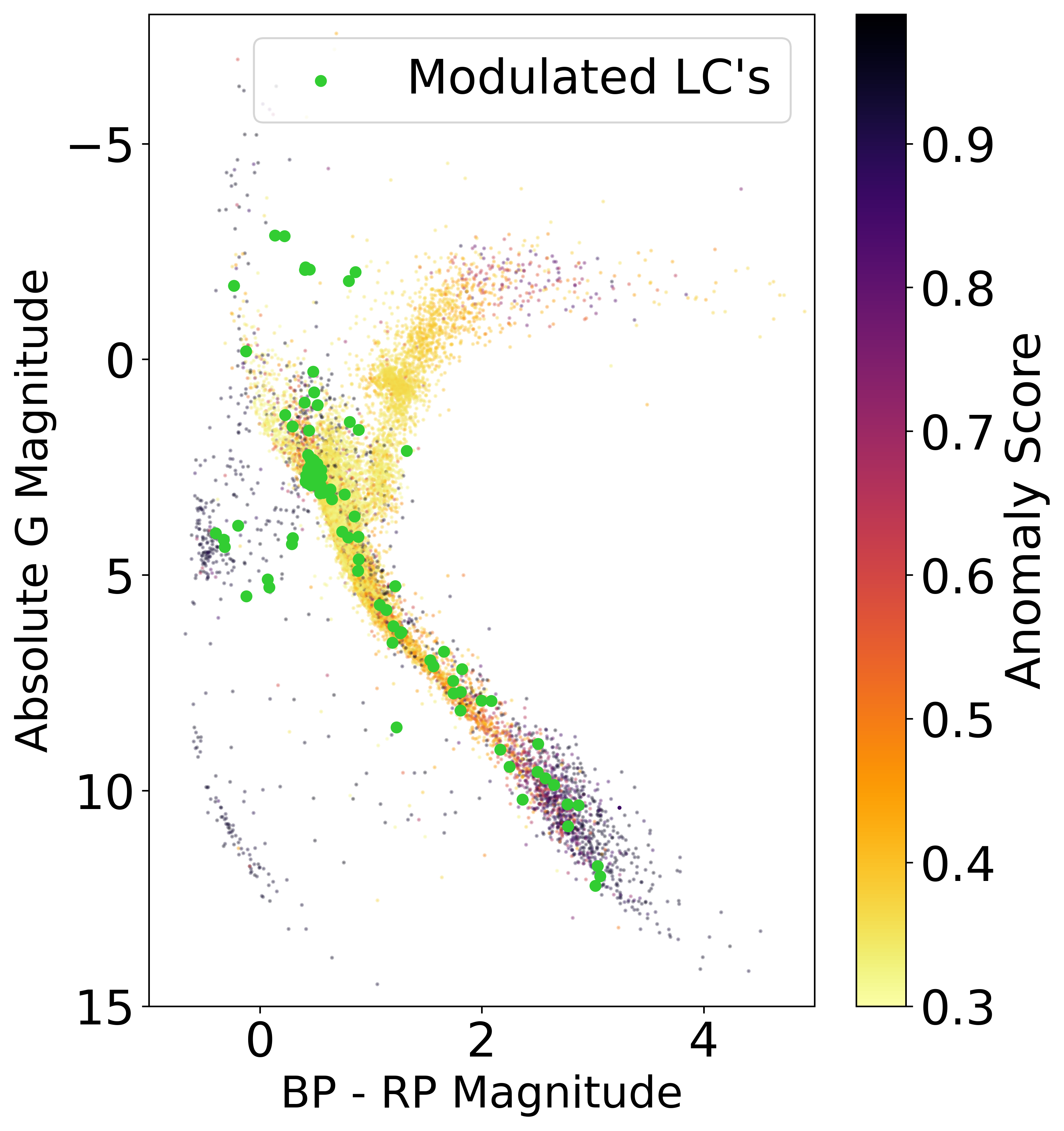}
    \caption{A CAMD colour-coded by average anomaly scores with the position of modulated light curves highlighted in green. Evidently, modulation patterns are present in many evolutionary stages and mass ranges, signalling multiple mechanisms causing such a variability pattern.}
    \label{fig: modulated_CAMD}
\end{figure}

A fraction of the light curves in our sample show a combination of a higher frequency pulsation modulated by a lower frequency pulsation, as in the examples shown in \autoref{fig: example_modulated}. We refer to these as modulated light curves. In total, we have identified 132 of these objects. There seems to be no preferential location for these objects in the CAMD, except that no examples are found along the giant branch, as shown in \autoref{fig: modulated_CAMD}.

In \autoref{fig: example_modulated}, we group objects in this class in sub-categories, as follows: the \textit{rapid} sub-class shows fast variations within a broad modulation envelope; the \textit{modulated} sub-class is similar to the previous sub-class, but shows a longer period ($\sim$10 days) for the lower of the two frequencies; finally, the \textit{sinusoidal} sub-class shows a dominant, long-period sinusoidal pattern combined with a higher frequency variation of low amplitude. These sub-classes have somewhat blurry boundaries, with some of the light curves fitting in none or more than one sub-class. If more than one sub-class is suitable to describe an object,  we label it with multiple types, e.g. \textit{rapid modulated sinusoidal}.

Given that objects with this modulated pattern are widespread over most of the CAMD, as well as the substantial differences between the sub-classes, it is unlikely that a single, intrinsic physical mechanism is behind the observed variability. These sources have a wide range of masses and evolutionary stages. Those which lie on the instability strip are likely RR-Lyrae stars undergoing the Bla\v{z}kho effect \citep{Blazko1907}, a long-term modulation effect first observed in \textit{RW Draconis} that have since been observed in other RR-Lyrae type stars. Despite being discovered over a century ago, this effect is poorly understood, with available models unable to fully explain the observed behaviour \citep{Benko2011, Sharka2020}.

In \cite{Saylor2018}, the authors search for low mass modulators in the SUPERBLINK catalogue observed during the \emph{Kepler K2} mission. The modulation pattern of GKM-dwarf-type stars is understood to be caused by the presence of starspots on fast rotators. This effect is well-modelled, but a limited number of examples have been found, with \cite{Saylor2018} identifying only 508 candidates from a catalogue containing 3 million objects. This study highlights the rareness of these types of phenomena and the importance of additional methods and catalogues like the ones presented here to identify further candidates. Other studies also suggest sunspots as a possible cause for modulation in higher mass main sequence stars. See \cite{Balona2011} for a detailed discussion.

While we do not explore other sources of modulation that certainly exist, we have highlighted that frequency modulation is related to astrophysical phenomena of interest. We also identify many candidates in the TIC using our unsupervised method, starting from a very general collection of light curves. In our catalogue (\autoref{tab: my_labels}), we assign the \textit{modulated} label to these candidates, with sub-class labels \textit{sinusoidal} or \textit{rapid} when appropriate.

\subsection{Artefacts}
\label{sec: artefacts}

\begin{figure}
    \includegraphics[width=\columnwidth]{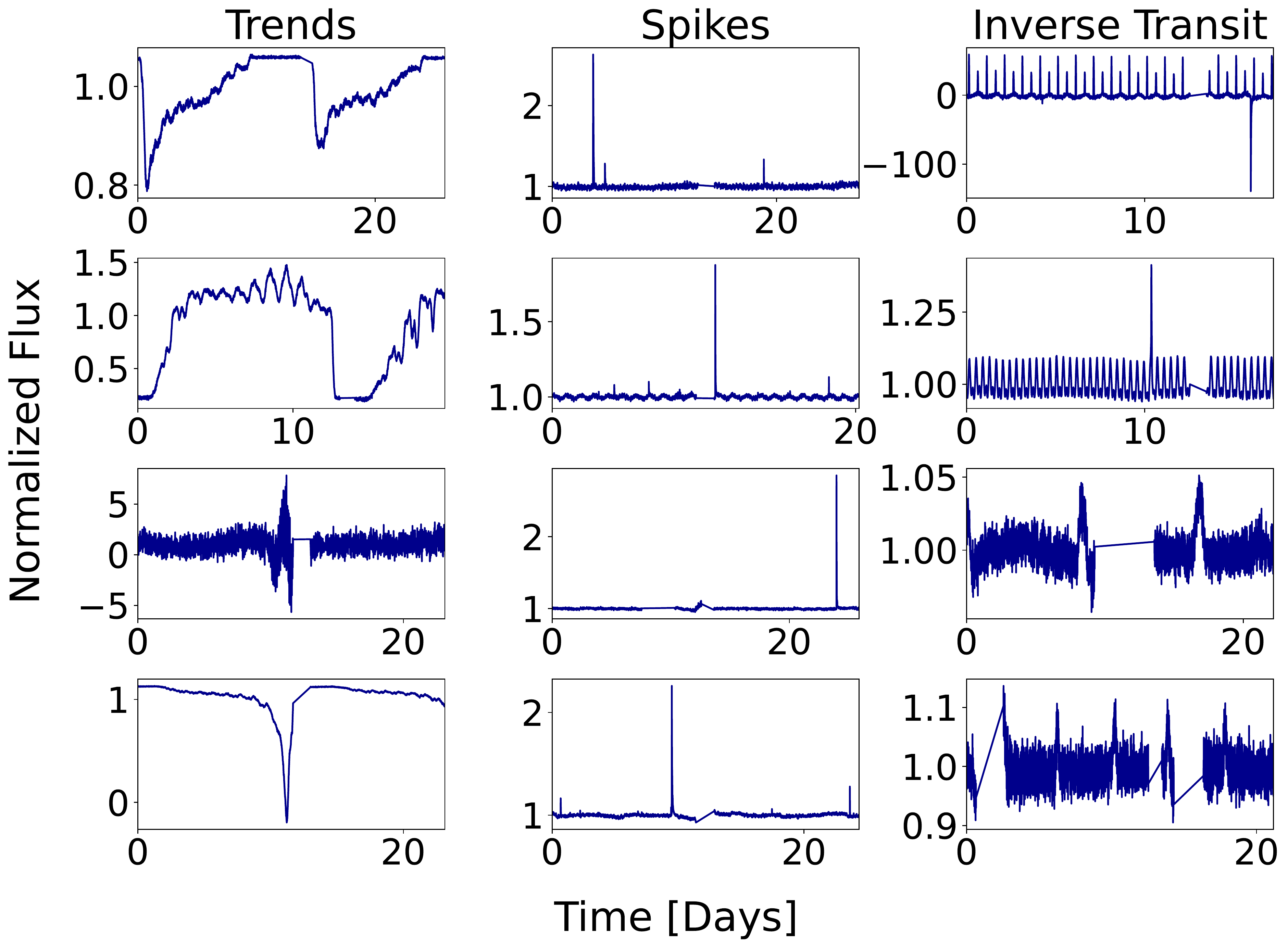}
    \caption{Light curve examples showing the most common types of artefact found, including trends, spikes and calibration star issues causing false inverse variability patterns.}
    \label{fig: example_errors}
\end{figure}

Artefacts are rare features in light curves that do not have an astrophysical origin and are instead related to unintended instrumental or processing effects. Examples include hot pixels, baseline trends, detection gaps, or pipeline errors. Although artefacts do not represent astrophysical events, their identification is crucial in assessing the quality of catalogue and data releases \citep{Lochner2021}. An unsupervised anomaly detection method cannot independently differentiate an artefact from a rare astrophysical phenomenon. To better characterise the anomalies, it, therefore, becomes necessary to understand the artefacts associated with the data in question. \emph{TESS} light curves are subject to several sources of artefacts, ranging from single pixel spikes to calibration issues that span significant fractions of the light curve. Discontinuity in the light curve due to a sudden change in the detector throughput is an example of the latter. The PDC pipeline produces the released \emph{TESS} light curves by correcting for the typical sensor response curve, yet some artefacts inevitably remain. Despite our best efforts to filter out single-pixel spikes, some remain in the final light curves that we have analysed. In \autoref{fig: example_errors}, we show examples of common artefacts in \emph{TESS} light curves.

Due to their often extreme effects and rarity within the dataset, artefacts can be assigned a very high anomaly score. This is particularly true when they result in discontinuities with timescales comparable with the duration of the observation, such as those seen in the left panels of \autoref{fig: example_errors}. Due to their distinct features (discontinuities, sudden variations in flux, etc.), these dominating patterns are identifiable by a visual inspection. On the other hand, single-pixel events do not dominate the anomaly score, as they usually account for a single feature. In fact, in \S~\ref{sec: objects_found}, we have identified that light curves that resemble each other have very similar scores, even when one is affected by a single pixel spike.

Finally, some artefacts result from imperfect background subtraction in crowded fields. When a crowded field contains a variable star, the background subtraction process can be adversely affected. The background subtraction in such a field results in an inverted imprint of the variability onto the target light curve. For example, transiting systems (such as binaries or exoplanets) incorrectly subtracted with the background will appear as repeating flare-like features in the target light curve. With the correct symmetry, these artefacts can be indistinguishable from genuine astrophysical novelties, such as gravitational lensing events. Therefore, they need to be identified and individually assessed. The final column of \autoref{fig: example_errors} contains examples of this phenomenon.

Fortunately, most artefacts are sector-dependent, particularly those associated with the detector response. Since artefacts result in a high anomaly score, they increase the standard deviation in the score for observations of the same object. It is, therefore, wise to identify artefacts from searches of anomalies that occur in single sectors, as we have discussed in \S~\ref{sec: high_sigma}. In \autoref{tab: datatable_cont}, we have labelled identified artefacts in the \textit{Notes} column.

\section{Discussion}

\begin{table*}[t]
\begin{center}
    \begin{tabular}{cccccc}
    \toprule
        TIC ID & Anomaly Score & Notes & Reference \\
        (1) & (2) & (3) & (4) \\
    \midrule
 TIC 21505340  & 0.96 & Quasi-periodic oscillations in WD star.        & \cite{Littlefield2021} \\
 TIC 157376469 & 0.96 & Binary system of K-type main sequence stars.   & \cite{Pan2021}\\
 TIC 63328020  & 0.95 & dipole pulsation mode in the eclipsing binary. & \cite{Rappaport2021}\\
 TIC 264509538 & 0.50 & Pulsations of the Rapidly Oscillating Ap Star. & \cite{Shi2020}\\
 TIC 229804573 & 0.95 & compact hierarchical quadruple system.         & \cite{Borkovits2021}\\
 TIC 234523599 & 0.96 & A giant planet, transiting an M3 dwarf star.    & \cite{Bakos2018}\\
 TIC 257459955 & 0.91 & Pulsating helium-atmosphere white dwarf.       & \cite{Bell2019}\\
 TIC 278659026 & 0.73 & g-mode hot B subdwarf pulsator.                & \cite{Charpinet2019}\\
 TIC 38586082  & 0.86 & Peculiar variable star of alpha2 CVn type      & \cite{Khalack2019}\\
    \bottomrule
    \end{tabular}
\caption{A list of known light curves of interest within the \textit{Tess} literature.\newline
(1) - TIC ID from \textit{Tess} catalogue \newline
(2) - Anomaly Score (this work) \newline
(3) - Notes on the physical nature of the system \newline
(4) - Previous references to the system}
\label{tab: literature_examples}
\end{center}
\end{table*}

This work has exploited the ability of a particular unsupervised anomaly detection algorithm (the Unsupervised Random Forest, URF) to detect rare events in a set of PDC-processed \emph{TESS} light curves with a varied and diverse form and astrophysical origin. We have compiled a catalogue of the anomalies found.

We have identified a broad range of anomalous variability patterns on \emph{TESS} light curves. The method remains general and identifies many variability patterns without targeting a specific variability signature or astrophysical class. The resulting anomalies are also independent of whether the object is subject to previous studies or whether the astrophysical mechanisms are understood. Therefore, we have found a mix of unusual astrophysical behaviours (see \S~\ref{sec: giants}), objects of known classes that have extreme or rare physical parameters (e.g. \S~\ref{sec: binaries}), as well as processing artefacts (\S~\ref{sec: artefacts}) that are important to identify to produce a list of candidate astrophysical anomalies. The final catalogue can act as a repository of objects with outlying properties that can either form new stellar classes or inform astrophysical models of variability in know classes by adding examples of extreme cases.

Statistical analysis of the URF in Section \ref{sec: results} - \ref{sec: anomalous_objects} reveals a bimodal distribution of anomaly scores, with approximately 10\% of objects identified as anomalous. This population is not necessarily only astrophysical oddities, rather they possess a behaviour that is statistically distant from the general population. The anomalous objects include a population of known stellar classes that show extreme variability patterns (e.g. eclipsing binaries with particularly deep eclipses, instability strip objects and irregular light curves in YSOs, described respectively in \S~\ref{sec: binaries}, \ref{sec: instability_strip} \& \ref{sec: YSO}); non-astrophysical anomalies (e.g. non-linear detector response curves, described in \S~\ref{sec: artefacts}) and a fraction are the true unknown unknowns that demonstrate peculiar astrophysical behaviour, one such example is discussed in \S~\ref{sec: giants}. Analysis of individual sources across multiple sectors reveals that the identification of periodic anomalous signals is consistent across all observations while providing a strategy to identify rare events that occur during a specific observation (see \S~\ref{sec: high_sigma}). These objects may include catastrophic events (such as supernovae) and variability (periodic or not) with timescales exceeding the duration of a single sector observation, such as an eclipsing binary with a large orbital separation or long-period variables.

We find a link between certain anomalous behaviours and their stellar evolutionary stage as traced by \emph{Gaia} data, see  \S~\ref{sec: hr_diagram}. The population of \emph{TESS} anomalous objects differs from those found in \emph{Kepler} data in MG21. For example, red giants and supergiants score significantly higher in anomaly score in \emph{Kepler} with respect to \emph{TESS}. On the other hand, instability strip objects and blue supergiants appear anomalous in both datasets, while most of the main sequence stars have low anomaly scores for both \emph{Kepler} and \emph{TESS}. The comparison of both populations underscores an important aspect of anomaly detection; given a method for anomaly identification, the anomalous nature of a particular object is a function of the entire population and the specific observational parameters of the survey. Specifically, the high cadence \emph{TESS} population is composed of stars typically closer and intrinsically brighter than those in the \emph{Kepler} (see \S~\ref{sec: lc_data} for further details). This selection makes red giants less anomalous in the present work. Thus, even though, in both cases, we look at the periodograms and light curves as the features, the difference in the spectral range covered due to different cadences also affects the anomaly score of specific objects. Therefore, the analysis of anomaly scores in light curves should always be in the context of the population studied.

We also explored the link between the anomaly score and specific astrophysical configurations and found that a high anomaly score can indicate a rare or extreme configuration. For example, in \S~\ref{sec: binaries}, we show that the primary transit depth and orbital period in binaries dominate the final anomaly score. In the context of \emph{TESS} light curves, longer periods and deeper primary transits cause a higher anomaly score, which allows this work to increase the census of extreme examples within this object class. We also showed in \S~\ref{sec: YSO} that the irregular light curves resulting from obscuration in embedded YSOs have a high anomaly score and are the closest candidates to being analogues to the light curve of Boyajian's star.

We further confirm our ability to detect genuine astrophysical anomalies by comparison with recent literature. \autoref{tab: literature_examples} lists several anomalous objects recently identified. The majority of these \emph{"bona fide"} anomalies are assigned an anomaly score of 0.9 or higher, with 90\% of them having a score higher than 0.7. These objects range from quasi-periodic oscillations in a white dwarf to complex transits in multiple stellar systems. Only one of these objects is not assigned a high anomaly score: a rapidly oscillating Ap pulsator (TIC 264509538). This object has a pulsation period of 7.52 minutes \citep{Shi2020}, which is too short for our spectral analysis to pick it up.

Our work has also uncovered anomalous behaviour in well-studied sources that can be associated with their evolutionary stage or particular events during their evolution. One example is Betelgeuse, where we discovered an unusual variability pattern months before a major dimming event, see \S~\ref{sec: giants}. The anomaly score for the Betelgeuse light curve is 0.85, much higher than other stars at a similar evolutionary stage. It remains unclear whether the anomalous variability is related to the subsequent dimming event, but the fact it is present so close to the event may indicate abnormal photospheric activity ten months before the major dimming event started.

\section{Conclusion}
We have presented the results of applying the Unsupervised Random Forest (URF) anomaly detection method to a large amount of \emph{TESS} light curves characterised by their light curve points and frequency power spectrum, building on our previous work with \emph{Kepler} data. We have provided a catalogue of anomalous light curves, classified those anomalies according to their variability characteristics, and associated their anomalous nature to any particular evolutionary stage or astrophysical configuration. For anomalies belonging to known classes (e.g., eclipsing binaries), we have also investigated what physical parameters drive the anomaly score. By comparing our results with those from the \emph{Kepler} study, we have also studied how the anomalous nature of particular light curves depends on the characteristics of the general population itself. Here we summarise our main findings.

\begin{itemize}
    \item Nearly 10\% of the studied \emph{TESS} light curves form a separate group of objects with enhanced anomaly scores. They include a mix of previously unclassified objects and objects with previously assigned classes with outlying properties and/or configurations. 

    \item High amplitude variability, pulsations, rapid periodic patterns and long-term variability timescales that dominate the frequency spectrum are among the dominating properties that set the anomaly score for the different types of objects. As such, the method can serve as a detector of pulsating stars.

    \item Pulsating objects along the instability strip as well as M-dwarfs displaying flares, sunspots and other magnetic-related activity, are consistently among the most anomalous objects. Irregular light curves from young stars with significant circumstellar obscuration also rank high in the list of anomalies and constitute the closest analogues to Boyajian's star.

    \item TIC stars are on average cooler, older, and less massive than those targeted by \emph{Kepler}. This results in populations of giants, such as the red clump stars, being underrepresented in \emph{TESS} and white dwarfs overrepresented in comparison with \emph{Kepler} \citep{Berger20}. As a result, the distribution of anomalous objects over classes differs between the two populations. For example, the Giant Branch has a much lower average anomaly score in the \emph{TESS} dataset, despite the method to find these anomalies being very similar in both cases.

    \item Anomalous eclipsing binaries and exoplanet transits have significantly deeper transits and longer orbital periods than their "normal" counterparts. The method can also detect deep transits with orbital periods longer than the typical light curve duration.

    \item Among giant stars, only Blue Supergiants show a high anomaly score, mostly driven by more irregular, high amplitude light curves, likely associated with their young evolutionary stage. Additionally, white dwarfs are consistently identified as anomalies, with no apparent relation between anomaly score and surface temperature or other physical parameters other than the amplitude of their variability, which is consistent with unresolved, short-period pulsations as described in \citet{Corsico2019}.

    \item Instrumental and processing artefacts, such as single-pixel spikes and discontinuities in the light curve, are readily identified by our method, with the latter having a substantial impact on the anomaly score. They are sufficiently different from astrophysical phenomena to be easily identified, except for inverted transits that result from the inclusion of transiting objects in the background region of nearby targets, which could be misidentified as a lensing event.

\end{itemize}

\section*{Acknowledgements}

We would like to extend our gratitude to our collaborators Federica Bianco, Ashish Mahabal \& Matthew J. Graham for their input during conversations throughout this work. Furthermore, Nigel C. Hambly and Robert Mann for their guidance and input across the results. In carrying out this research we have used the scikit-learn Python package, \cite{Pedregosa11}. This paper includes data collected by the \emph{TESS} mission. Funding for the \emph{TESS} mission is provided by the NASA's Science Mission Directorate. D. A Crake is supported by the UK Science and Technology Facilities Council (STFC) through grant ST/S001948/1 and the award of a PhD studentship to DAC. For the purpose of open access, the author has applied a Creative Commons Attribution (CC BY) licence to any Author Accepted Manuscript version arising from this submission.




\bibliographystyle{mnras}
\bibliography{TESS_LCs}

\bsp	
https://www.overleaf.com/project/6048931c86886d3802d27ab6\label{lastpage}
\end{document}

%% file: Tables/results_split_1.tex
\begin{table*}
    \begin{tabular}{cccccc}
    \toprule
        tic\_id &                   designation &          ra &        dec &  main\_type  & anomaly\_score \\                             
        (1) & (2) & (3) & (4) & (5) & (6) \\
    \midrule
 TIC 471013946 &  Gaia DR2 2513021854931769344 &   33.821292 &   0.244569 &        WD* & 0.989290 \\
 TIC 166463090 &  Gaia DR2 6670392146960681984 &  298.835020 & -48.282248 &   gammaDor & 0.989262 \\
 TIC 177932603 &  Gaia DR2 3052092481686246016 &  107.658294 &  -7.122892 &   deltaCep & 0.987706 \\
 TIC 444535842 &   Gaia DR2 510991333155346688 &   19.837507 &  62.301401 &        NaN & 0.987691 \\
  TIC 30317282 &  Gaia DR2 4661514377893838464 &   74.272331 & -68.414734 &        EB* & 0.987162 \\
 TIC 439869954 &  Gaia DR2 2497895053130247040 &   43.384666 &  -0.562453 &        WD* & 0.987156 \\
 TIC 142867174 &  Gaia DR2 5054768627234519552 &   50.837600 & -32.270100 &        NaN & 0.986632 \\
 TIC 178366477 &  Gaia DR2 3071240270519385856 &  123.327106 &  -1.057903 &  CataclyV* & 0.986457 \\
 TIC 149894385 &  Gaia DR2 5587836255503080576 &  116.769000 & -34.372400 &        NaN & 0.986149 \\
 TIC 444000734 &                           NaN &  131.365000 &  10.914900 &        PM* & 0.986069 \\
    \bottomrule
    \end{tabular}
    \label{tab: data_table}
\end{table*}

%% file: Tables/results_split_2.tex
\begin{table*}
    \begin{tabular}{ccccccc}
        \toprule
        tic\_id & sectors & weirdness>0.9 & weirdness>0.6 & weirdness<0.6 &        Notes  & high\_variance \\
        (1) & (7) & (8) & (9) & (10) & (11) & (12) \\
        \midrule
 TIC 471013946 &      [4] &           [4] &            [] &            [] &            0  &   - \\
 TIC 166463090 &     [13] &          [13] &            [] &            [] &            0  &   - \\
 TIC 177932603 &      [7] &           [7] &            [] &            [] &            0  &   - \\
 TIC 444535842 &     [24] &          [24] &            [] &            [] &  [artefact ]  &   - \\
  TIC 30317282 &      [7] &           [7] &            [] &            [] &            0  &   - \\
 TIC 439869954 &      [4] &           [4] &            [] &            [] &            0  &   - \\
 TIC 142867174 &      [4] &           [4] &            [] &            [] &            0  &   - \\
 TIC 178366477 &      [7] &           [7] &            [] &            [] &            0  &   - \\
 TIC 149894385 &      [8] &           [8] &            [] &            [] &            0  &   - \\
 TIC 444000734 &      [7] &           [7] &            [] &            [] &            0  &   - \\
        \bottomrule
    \end{tabular}
    \caption{\newline
             (1) - TIC ID from \textit{Tess} catalogue\newline
             (2) - Designation ID from Gaia DR2 \newline
             (3) - Right Ascension \newline
             (4) - Declination \newline
             (5) - Main object type from SIMBAD \newline
             (6) - Average Weirdness Score from this work \newline
             (7) - Sectors the object is present \newline
             (8) - Sectors where the Weirdness Score > 0.9 \newline
             (9) - Sectors with 0.6 < Weirdness Score < 0.9 \newline
             (10) - Sectors with Weirdness Score < 0.6 \newline
             (11) - Notes from visual inspection, discussed in \S~\ref{sec: objects_found} \newline
             (12) - Flag for high variance objects, discussed in \S~\ref{sec: high_sigma}}
    \label{tab: datatable_cont}
\end{table*}

%% file: Tables/class_labels.tex
\begin{table}[ht]
    \centering
    \begin{tabular}{c|c|l}
    \hline
    Shorthand & Descriptive & Example \\
    \hline
    s    & sinusoidal            & \parbox[c]{1em}{\includegraphics[width=1in]{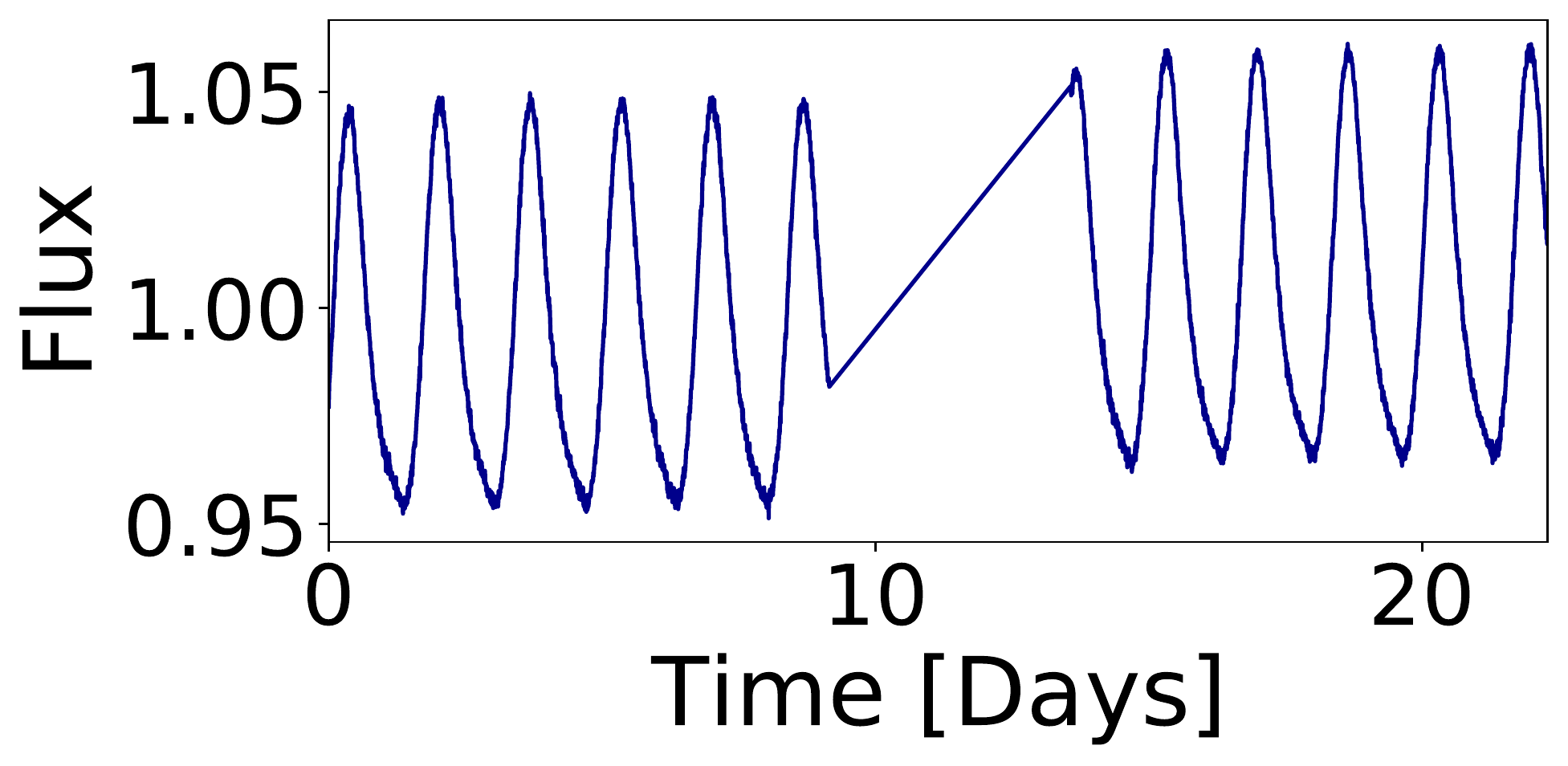}} \\
    m    & modulated             & \parbox[c]{1em}{\includegraphics[width=1in]{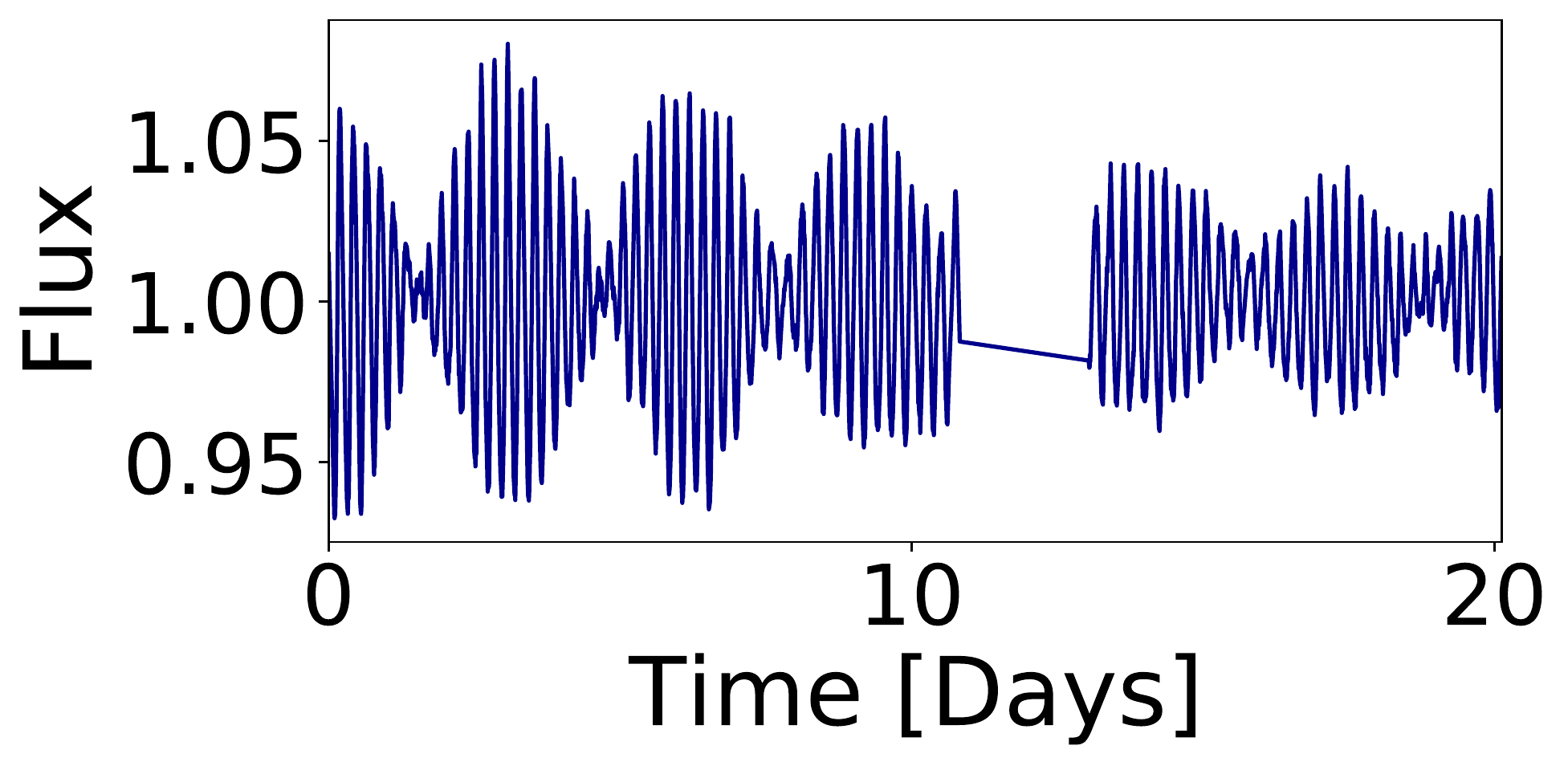}} \\
    e    & artefact              & \parbox[c]{1em}{\includegraphics[width=1in]{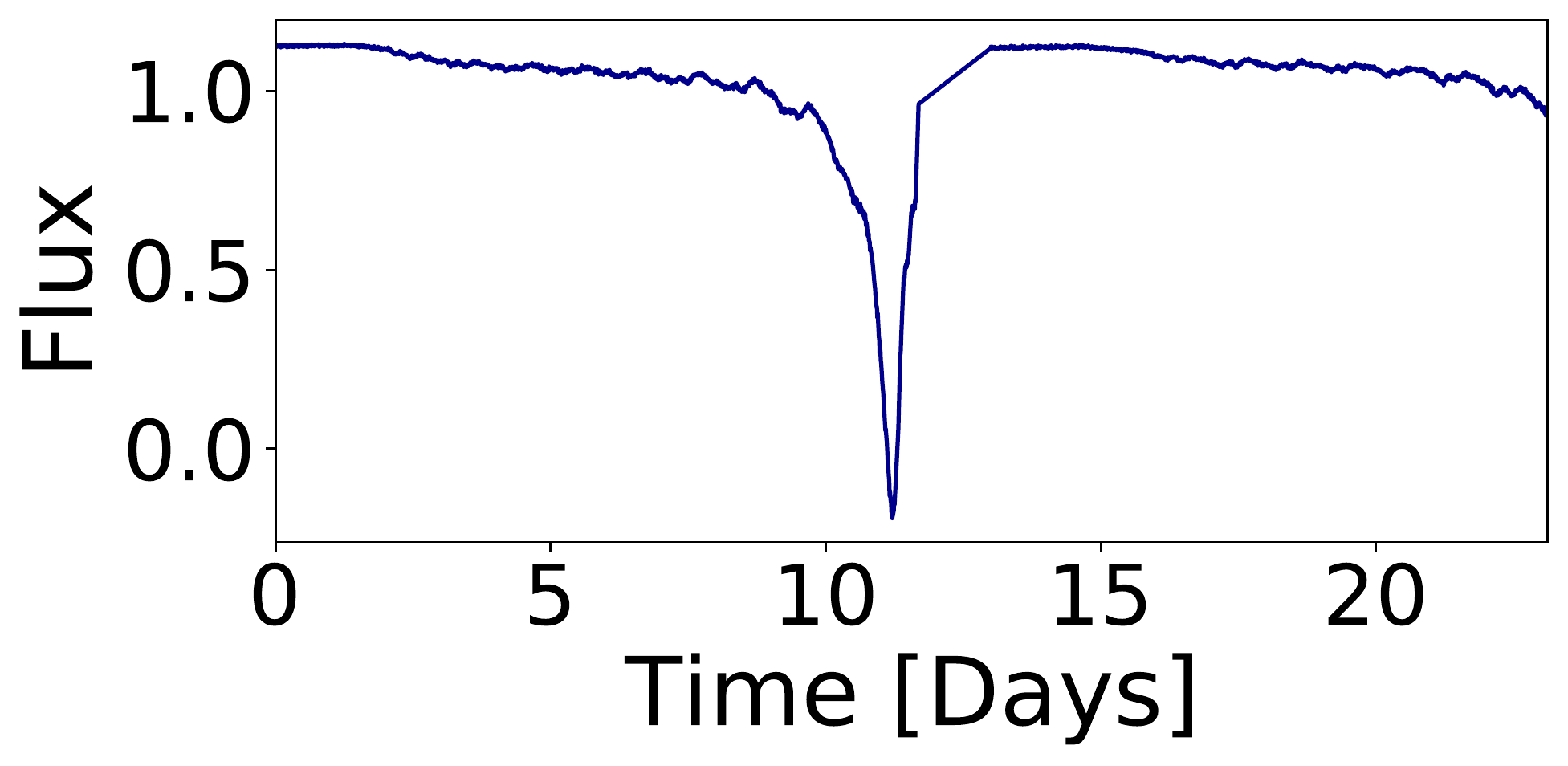}} \\
    p    & peaked                & \parbox[c]{1em}{\includegraphics[width=1in]{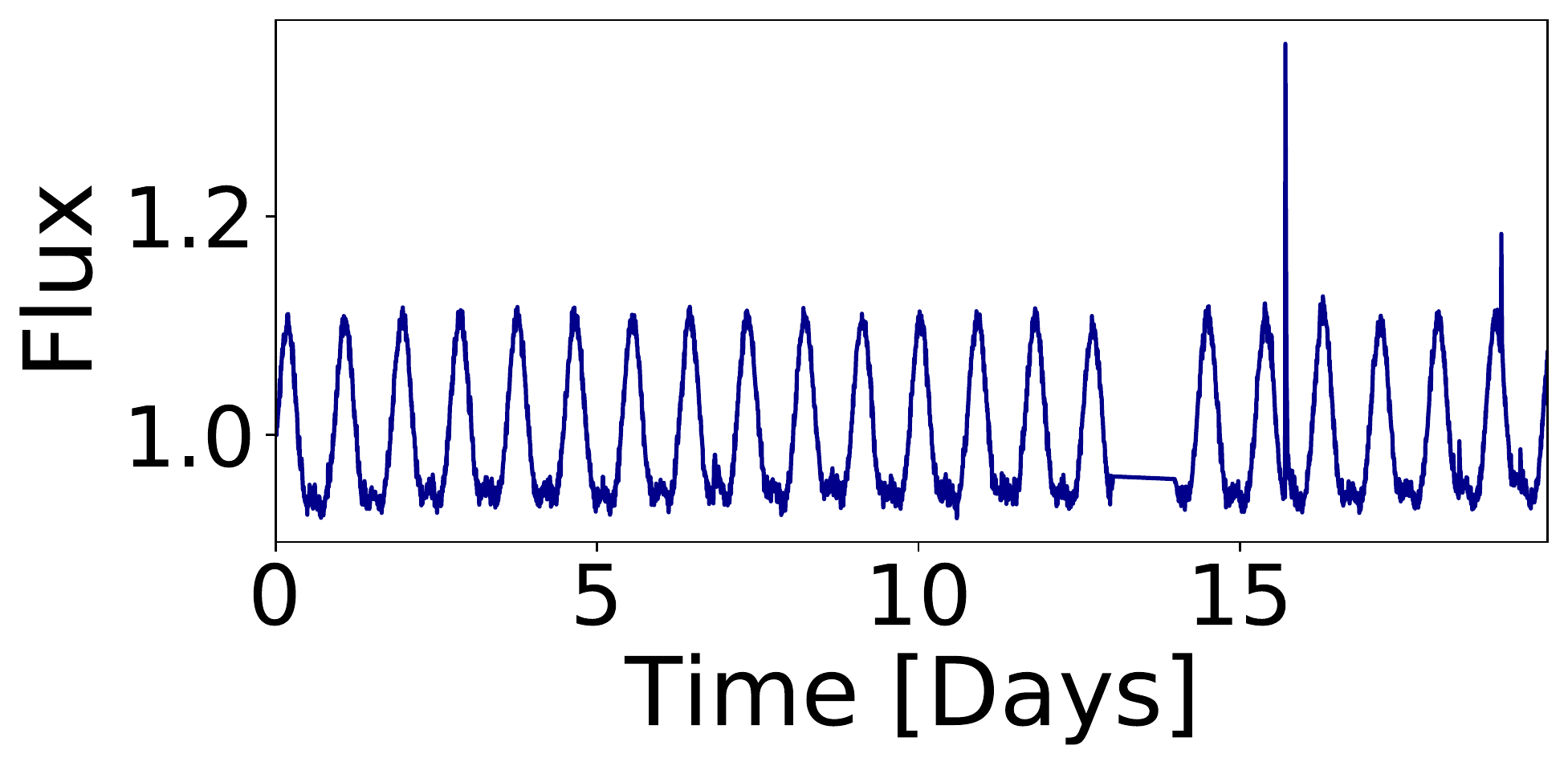}} \\
    dd   & dips                  & \parbox[c]{1em}{\includegraphics[width=1in]{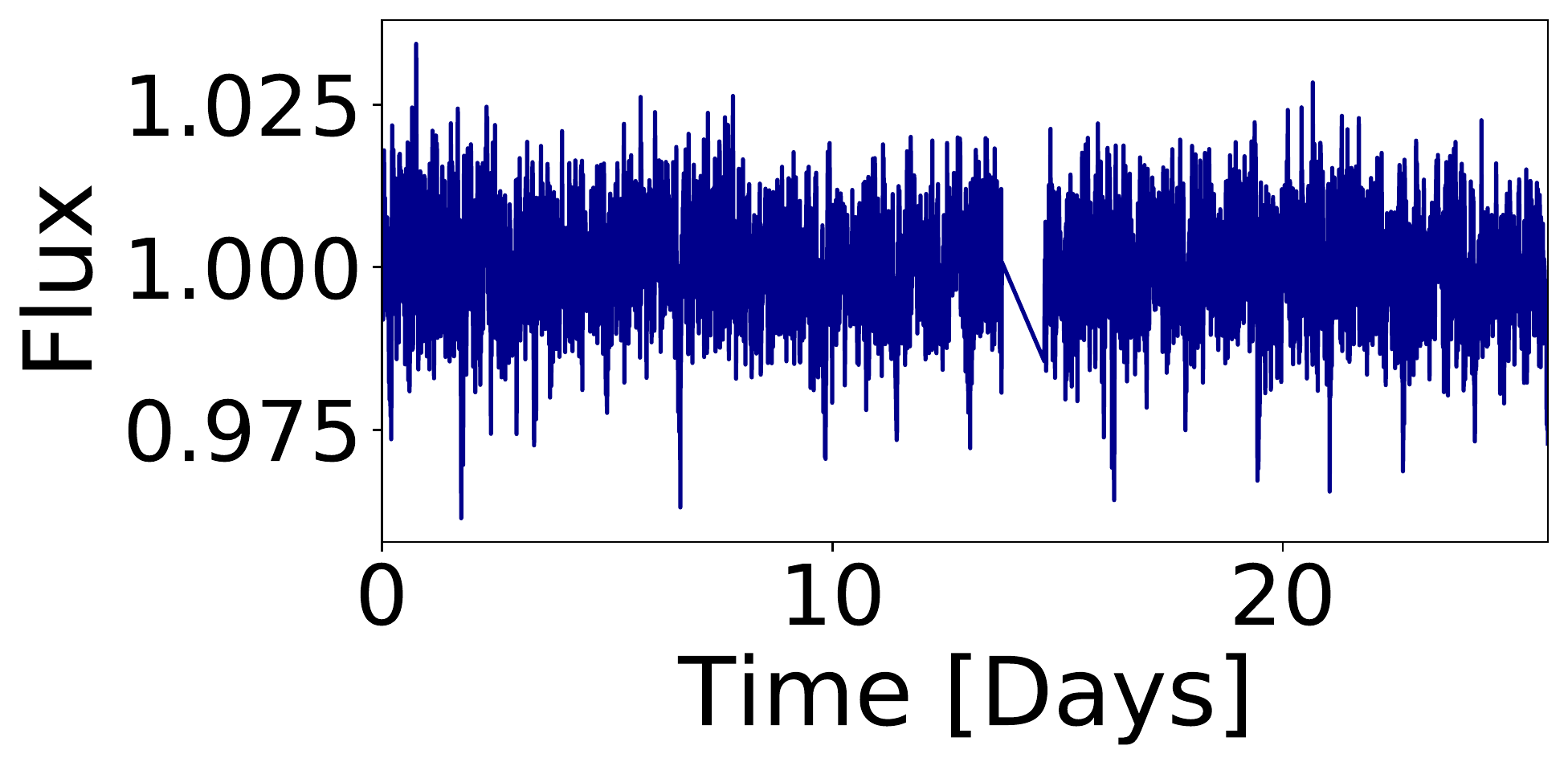}} \\
    i, I & irregular             & \parbox[c]{1em}{\includegraphics[width=1in]{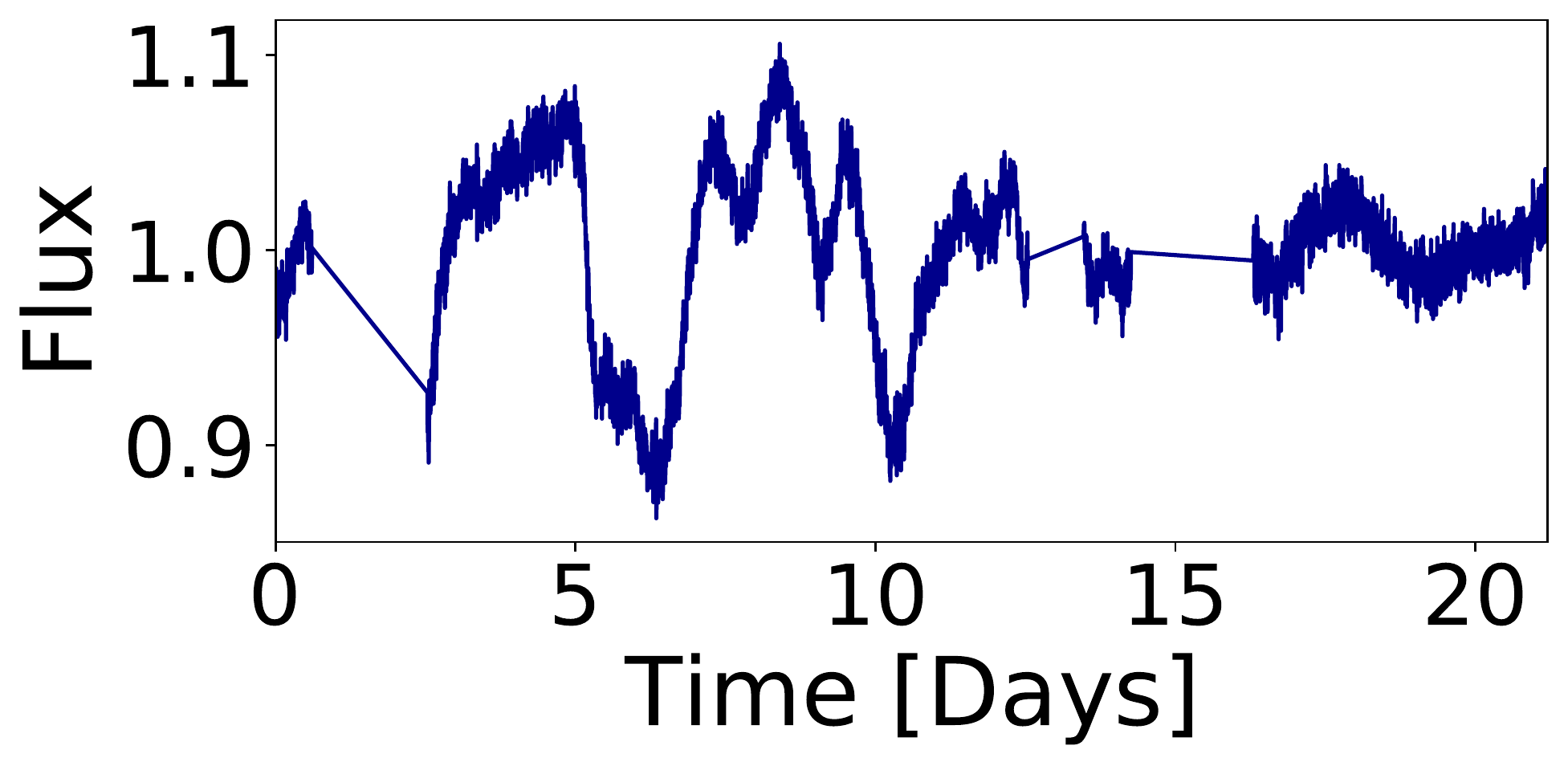}} \\
    mb   & multi-body system    & \parbox[c]{1em}{\includegraphics[width=1in]{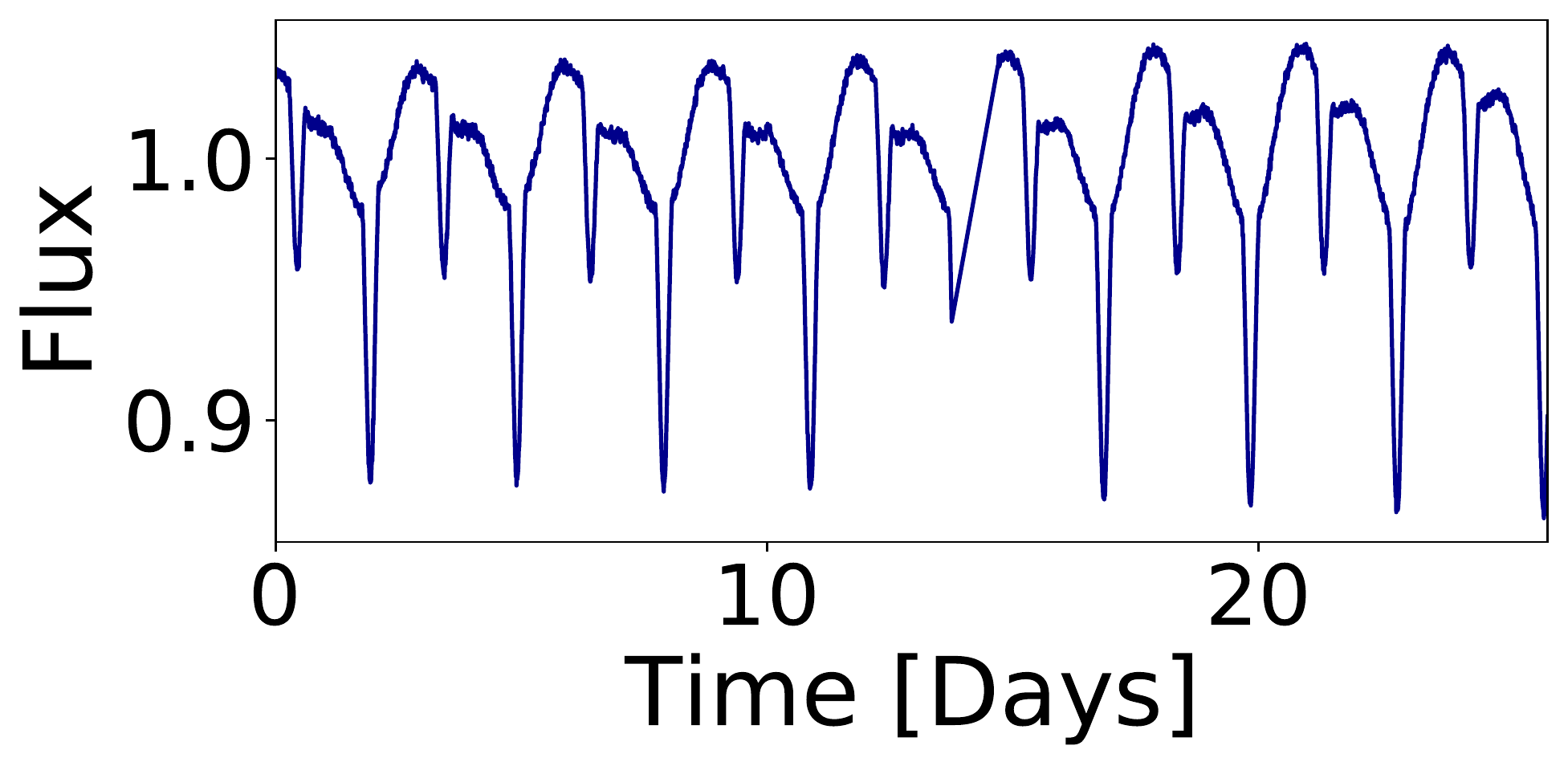}} \\
    t    & transit               & \parbox[c]{1em}{\includegraphics[width=1in]{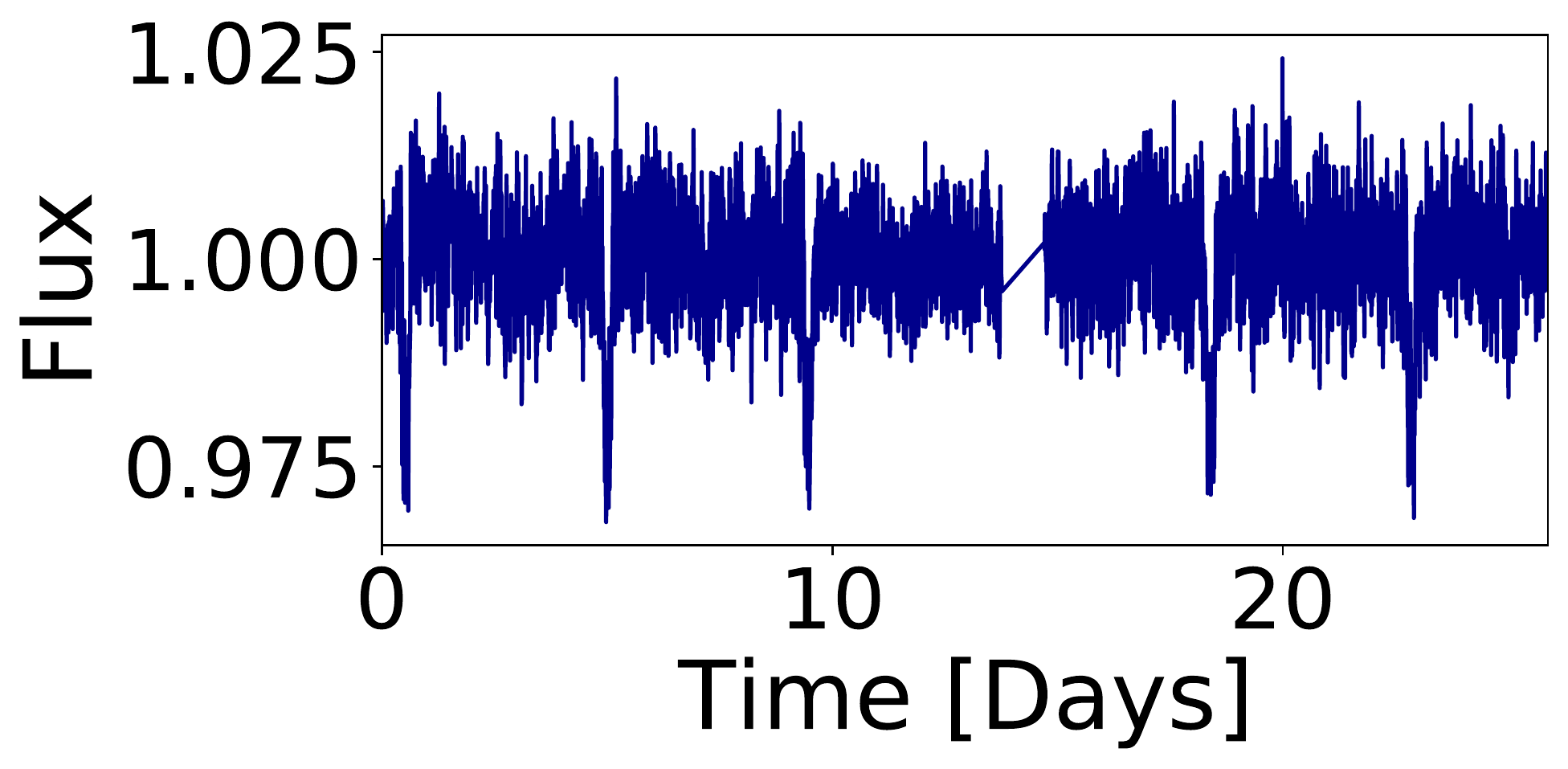}} \\
    LPV  & long period variable  & \parbox[c]{1em}{\includegraphics[width=1in]{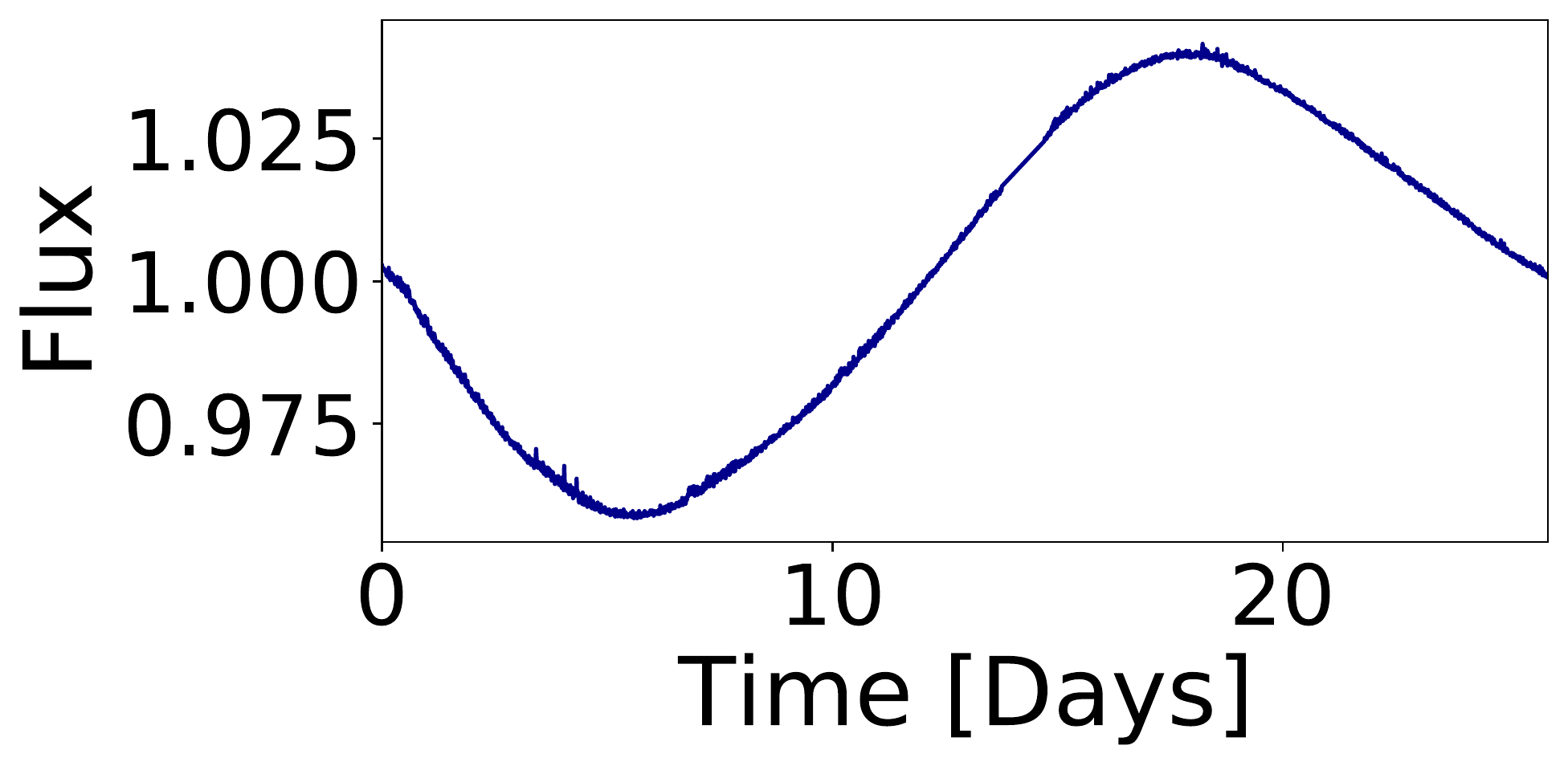}} \\
    -    & No Label              & \parbox[c]{1em}{\includegraphics[width=1in]{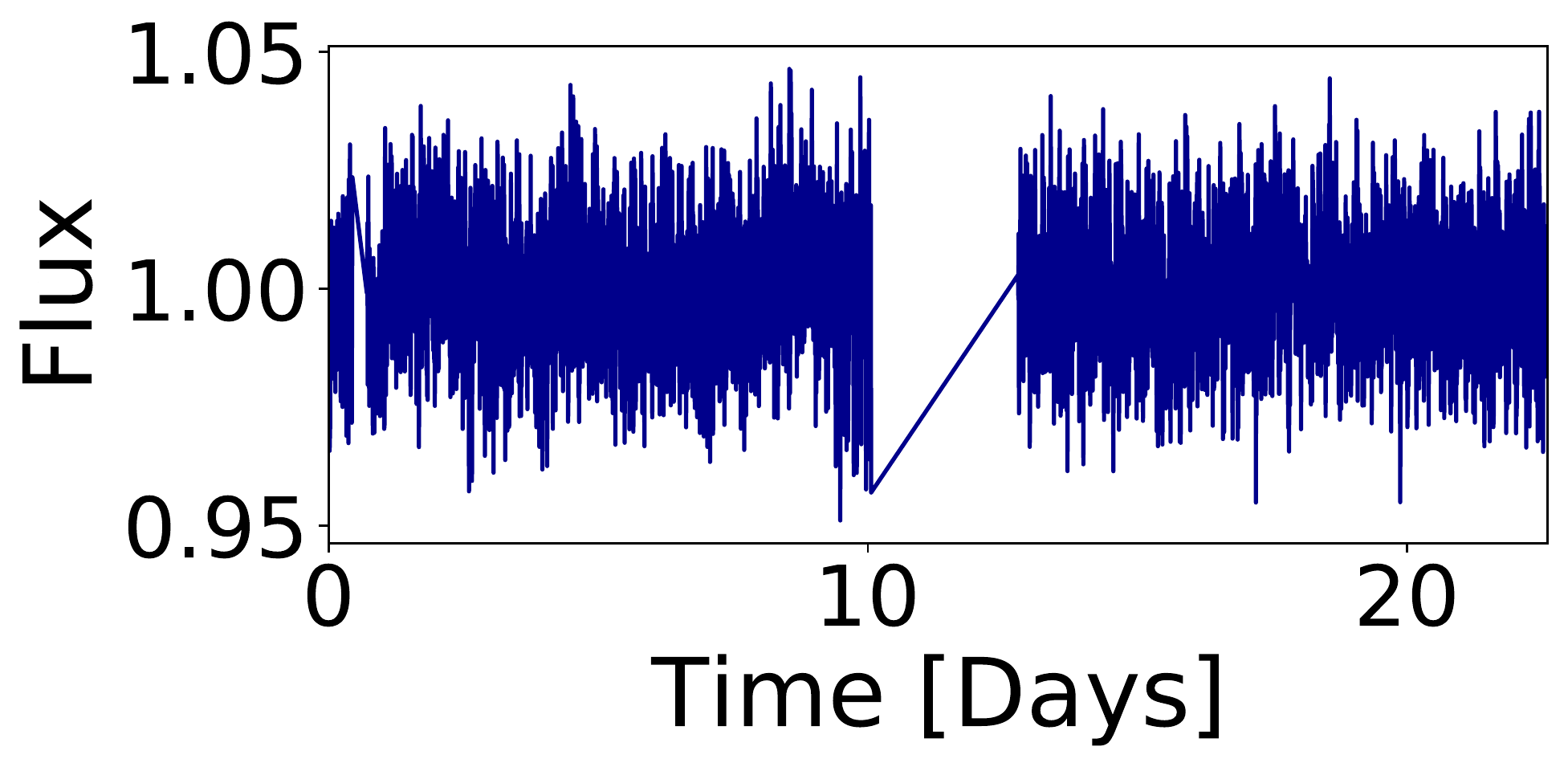}} \\
    \hline
    \multicolumn{3}{c}{Subcategories} \\
    \hline
    r    & rapid                 & * Additional label, descriptive. \\
    rr   & regular               & * Additional label, descriptive. \\
    d    & double                & * Additional label, descriptive. \\
    u    & unusual               & * Additional label, descriptive. \\
    a.   & asymmetrical          & * Additional label, descriptive. \\
    ?    & Indicates speculative & * Additional label, descriptive. \\
    \hline
    \end{tabular}
    \caption{List of the most common descriptors used in the classification of anomalous objects. The examples represent a typical object from these classes. In reality these labels cover a range of patterns and are often used in conjunction with each other, e.g. \textit{Modulated Sinusoidal}. Additional labels are used to separate the main labels (those with examples above) into subcategories.}
    \label{tab: my_labels}
\end{table}